\documentclass[12pt]{article}
\usepackage{amsmath,amssymb,amsthm,cite}
\usepackage[matrix,arrow]{xy}

\advance\topmargin by -\headheight
\advance\topmargin by -\headsep
\evensidemargin=\oddsidemargin
\renewcommand{\a}{\alpha}           
\newcommand{\alg}{{\mathrm{alg}}}   
\renewcommand{\b}{\beta}            
\newcommand{\barox}{\mathrel{\overline\otimes}} 
\newcommand{\bbfb}{Beatus Fa\`a di Bruno} 
\newcommand{\bfb}{Fa\`a di Bruno} 
\def\braket#1#2{\langle#1\mathbin|#2\rangle} 
\newcommand{\C}{\mathbb{C}}         
\newcommand{\Cc}{\mathcal{C}}       
\newcommand{\Ch}{{\mathrm{Ch}}}     
\newcommand{\CM}{{\mathrm{CM}}}     
\newcommand{\cocom}{{\mathrm{cocom}}} 
\newcommand{\cop}{{\mathrm{cop}}}   
\newcommand{\D}{\mathcal{D}}        
\DeclareMathOperator{\Der}{Der}     
\DeclareMathOperator{\Diff}{Diff}   
\newcommand{\Dl}{\Delta}            
\newcommand{\del}{\partial}         
\newcommand{\ddto}[1]{\frac{d}{d#1}\biggr|_{#1=0}} 
\newcommand{\dl}{\delta}            
\newcommand{\DS}{{\mathrm{DS}}}     
\newcommand{\dst}[2]{\langle#1,#2\rangle} 
\DeclareMathOperator{\End}{End}     
\newcommand{\eps}{\varepsilon}      
\newcommand{\F}{\mathcal{F}}        
\newcommand{\G}{\mathcal{G}}        
\newcommand{\Ga}{\Gamma}            
\newcommand{\g}{\mathfrak{g}}       
\newcommand{\ga}{\gamma}            
\newcommand{\gl}{\mathfrak{gl}}     
\renewcommand{\H}{\mathcal{H}}      
\newcommand{\Hl}{H_{\ell}}          
\DeclareMathOperator{\Hom}{Hom}     
\newcommand{\hideqed}{\renewcommand{\qed}{}} 
\newcommand{\hookto}{\hookrightarrow} 
\newcommand{\id}{{\mathrm{id}}}     
\DeclareMathOperator{\im}{im}       
\newcommand{\J}{\mathcal{J}}        
\renewcommand{\L}{\mathcal{L}}      
\renewcommand{\l}{\mathrm{l}}       
\newcommand{\La}{\Lambda}           
\newcommand{\la}{\lambda}           
\newcommand{\lt}{\triangleright}    
\newcommand{\marker}{\vspace{6pt}\noindent{$\rtri$}\enspace} 
\newcommand{\N}{\mathbb{N}}         
\newcommand{\nn}{\nonumber}         
\newcommand{\Om}{\Omega}            
\newcommand{\om}{\omega}            
\newcommand{\op}{\oplus}            
\newcommand{\opp}{{\mathrm{opp}}}   
\newcommand{\ox}{\otimes}           
\newcommand{\oxyox}{\otimes\cdots\otimes} 
\newcommand{\Pc}{\mathcal{P}}        
\newcommand{\Q}{\mathbb{Q}}         
\newcommand{\R}{\mathbb{R}}         
\def\roundbraket#1#2{(#1\mathbin|#2)} 
\newcommand{\row}[3]{{#1}_{#2},\dots,{#1}_{#3}} 
\newcommand{\Rr}{\mathcal{R}}       
\newcommand{\rt}{\triangleleft}     
\newcommand{\rtri}{\blacktriangleright} 
\newcommand{\Sf}{\mathbb{S}}        
\newcommand{\sepword}[1]{\quad\mbox{#1}\quad} 
\newcommand{\set}[1]{\{\,#1\,\}}    
\DeclareMathOperator{\T}{T}         
\newcommand{\tfifth}{\tfrac{1}{5}}  
\newcommand{\Th}{\Theta}            
\newcommand{\thalf}{\tfrac{1}{2}}   
\newcommand{\tquarter}{\tfrac{1}{4}}
\DeclareMathOperator{\Tr}{Tr}       
\newcommand{\tthird}{\tfrac{1}{3}}  
\newcommand{\twobytwo}[4]
{\begin{pmatrix}#1 & #2 \\#3 & #4\end{pmatrix}} 
\newcommand{\U}{\mathcal{U}}        
\newcommand{\uno}{\underline{1}}    
\newcommand{\vacin}{0_\mathrm{in}}  
\newcommand{\vacout}{0_\mathrm{out}}
\newcommand{\vacpersamp}{\dst{\vacin}{\vacout}} 
\newcommand{\vf}{\varphi}           
\newcommand{\vyv}{\vee\dots\vee}    
\newcommand{\w}{\wedge}             
\newcommand{\x}{\times}             
\newcommand{\7}{\dagger}            
\newcommand{\8}{\bullet}            
\renewcommand{\.}{\cdot}            
\renewcommand{\:}{:\,}              
\let\hash=\#
\renewcommand{\#}{\mathbin{\hash}}  
\def\<#1,#2>{\langle#1,#2\rangle}   
\def\wick:#1:{\mathopen:#1\mathclose:} 

\newcommand{\cross}{\parbox{2.5pc}{\begin{picture}(20,20)
\put(25,0){\line(-1,1){20}}
\put(5,0){\line(1,1){20}}
\end{picture}}}

\newcommand{\crossfatpt}{\parbox{2.5pc}{\begin{picture}(20,20)
\put(25,0){\line(-1,1){20}}
\put(5,0){\line(1,1){20}}
\put(12,7){$\bullet$}
\end{picture}}}

\newcommand{\fish}{\parbox{3pc}{\begin{picture}(20,20)
\put(4,10){\qbezier(0,-4)(16,20)(32,-4)}
\put(4,10){\qbezier(0,4)(16,-20)(32,4)}
\end{picture}}}

\newcommand{\winecup}{\parbox{2.5pc}{\begin{picture}(20,20)
\put(15,-4){\line(2,3){20}}
\put(25,-4){\line(-2,3){20}}
\put(10,15){\qbezier(0,3.75)(10,10)(20,3.75)}
\put(10,15){\qbezier(0,3.75)(10,-2.5)(20,3.75)}
\end{picture}}}

\newcommand{\bikini}{\parbox{4pc}{\begin{picture}(20,10)
\put(0,4){\qbezier(0,-3)(12,15)(22.8,0)}
\put(0,4){\qbezier(0,3)(12,-15)(22.8,0)}
\put(21.6,4){\qbezier(1.2,0)(12,15)(24,-3)}
\put(21.6,4){\qbezier(1.2,0)(12,-15)(24,3)}
\end{picture}}}

\makeatletter
\def\section{\@startsection{section}{1}{\z@}{-3.5ex plus -1ex minus
			  -.2ex}{2.3ex plus .2ex}{\large\bf}}
\def\subsection{\@startsection{subsection}{2}{\z@}{-3.25ex plus -1ex
			  minus -.2ex}{1.5ex plus .2ex}{\normalsize\bf}}
\renewcommand{\@dotsep}{200} 
\makeatother

\numberwithin{equation}{section}    

\theoremstyle{plain}
\newtheorem{thm}{Theorem}[section]  
\newtheorem{prop}[thm]{Proposition} 
\newtheorem{lema}[thm]{Lemma}       

\theoremstyle{definition}
\newtheorem{defn}{Definition}[section] 

\theoremstyle{remark}
\newtheorem{xmpl}{Example}[section] 

\begin{document}

\title{Combinatorial Hopf algebras in quantum field theory I}

\author{H\'ector Figueroa\dag\
\\[1pc]
\dag Departamento de Matem\'aticas, Universidad de Costa Rica,\\
San Pedro 2060, Costa Rica
\and
Jos\'e M. Gracia-Bond\'{\i}a\ddag\
\\[1pc]
\ddag Departamento de F\'{\i}sica Te\'orica I,
Universidad Complutense,\\
Madrid 28040, Spain}

\date{6 June 2005}

\maketitle

\begin{abstract}
This manuscript stands at the interface between combinatorial Hopf
algebra theory and renormalization theory.  Its plan is as follows:
Section~1 is the introduction, and contains as well an elementary
invitation to the subject.  The rest of part I, comprising Sections
2--6, is devoted to the basics of Hopf algebra theory and examples, in
ascending level of complexity.  Part II turns around the all-important
Fa\`a di Bruno Hopf algebra.  Section~7 contains a first, direct
approach to it.  Section~8 gives applications of the Fa\`a di Bruno
algebra to quantum field theory and Lagrange reversion.  Section~9
rederives the related Connes--Moscovici algebras.  In Part III we turn
to the Connes--Kreimer Hopf algebras of Feynman graphs and, more
generally, to incidence bialgebras.  In Section~10 we describe the
first.  Then in Section~11 we give a simple derivation of (the
properly combinatorial part of) Zimmermann's cancellation-free method,
in its original diagrammatic form.  In Section~12 general incidence
algebras are introduced, and the Fa\`a di Bruno bialgebras are
described as incidence bialgebras.  In Section~13, deeper lore on
Rota's incidence algebras allows us to reinterpret Connes--Kreimer
algebras in terms of distributive lattices.  Next, the general
algebraic-combinatorial proof of the cancellation-free formula for
antipodes is ascertained; this is the heart of the paper.  The
structure results for commutative Hopf algebras are found in
Sections~14 and~15.  An outlook section very briefly reviews the
coalgebraic aspects of quantization and the Rota--Baxter map in
renormalization.

\end{abstract}

\medskip
\noindent 2001 PACS: 11.10.Gh, 02.20.Uw, 02.40.Gh

\noindent Keywords: Hopf algebras, combinatorics, renormalization,
noncommutative geometry

\newpage

\tableofcontents

\newpage

\section*{Part I: Basic Combinatorial Hopf Algebra Theory}
\addcontentsline{toc}{section}{\protect\numberline{I}\enspace
BASIC HOPF ALGEBRA THEORY}

\section{Why Hopf algebras?}

Quantum field theory (QFT) aims to describe the fundamental phenomena
of physics at the shortest scales, that is, higher energies. In spite
of many practical successes, QFT is mathematically a problematic
construction. Many of its difficulties are related to the need for
{\it renormalization}. This complicated process is at present required
to make sense of quantities very naturally defined, that we are
however unable to calculate without incurring infinities. The
complications are of both analytical and combinatorial nature.

Since the work by Joni and Rota~\cite{JoniR} on incidence coalgebras,
the framework of Hopf algebras (a dual concept to groups in the spirit
of noncommutative geometry) has been recognized as a very
sophisticated one at our disposal, for formalizing the art of
combinatorics. Now, recent developments (from 1998 on) have placed
Hopf algebras at the heart of a noncommutative geometry approach to
physics. Rather unexpectedly, but quite naturally, akin Hopf algebras
appeared in two previously unrelated contexts: perturbative
renormalization in quantum field theories~\cite{KreimerOriginal,
ConnesKrRHI,ConnesKrRHII} and index formulae in noncommutative
geometry~\cite{ConnesMHopf}.

Even more recently, we have become aware of the neglected coalgebraic
side of the quantization procedure~\cite{BrouderG-24,Gang-of-Four}.
Thus, even leaving aside the role of ``quantum symmetry groups'' in
conformal field theory, Hopf algebra is invading QFT from both ends,
both at the foundational and the computational level. The
\textit{whole} development of QFT from principles to applications
might conceivably be subtended by Hopf algebra.

The approach from quantum theoretical first principles is still in its
first infancy. This is one reason why we have focused here on the
understanding of the contributions by Kreimer, Connes, and Moscovici
from the viewpoint of algebraic combinatorics ---in particular in
respect of incidence bialgebra theory. In other words (in contrast
with~\cite{DirkReport} for instance), we examine Rota's and Connes
and Kreimer's lines of thought in parallel. Time permitting, we will
return in another article to the perspectives broached
in~\cite{BrouderG-24,Gang-of-Four}, and try to point out ways from the
outposts into still unconquered territory.

In reference~\cite{BK} Broadhurst and Kreimer declare: ``[In
renormalization theory] combinations of diagrams\dots can provide
cancellations of poles, and may hence eliminate pole terms". The
practical interest of this is manifest. In fact, the ultimate goal of
tackling the Schwinger--Dyson equations in~QFT is linked to general
characterization problems for commutative Hopf
algebras~\cite{DirkWhisper}. This is one of the reasons why we have
provided a leisurely introduction to the subject, splashing pertinent
examples to chase dreariness away, and, we hope, leading the audience
almost imperceptibly from the outset towards the deeper structure
questions undertaken in Sections~14 and~15 of Part~IV.

The study of the more classical \bfb\ Hopf algebras, effected mostly
in Part~II, serves as a guiding thread of this survey. As well as
their applications, in particular to the Lagrange reversion formula
---a subject we find fascinating. The \bfb\ algebras, denoted~$\F(n)$,
are of the same general type as the Kreimer--Connes--Moscovici Hopf
algebras; they are in fact Hopf subalgebras of the Connes--Moscovici
Hopf algebras~$\H_\CM(n)$.

As hinted at above, the latter appeared in connection with the index
formula for transversally elliptic operators on a foliation. These
canonical Hopf algebras depend only on the codimension $n$ of the
foliation, and their action on the transverse frame bundle simplifies
decisively the associated computation of the index, that takes place
on the cyclic cohomology of~$\H_\CM(n)$. One of our main results is a
theorem describing $\H_\CM(n)$ as a kind of bicrossedproduct Hopf
algebra of~$\F(n)$ by the (action of/coaction on) the Lie algebra of
the affine transformation group. This is implicit
in~\cite{ConnesMHopf}, but there the construction, in the words of
G.~Skandalis~\cite{Yellow}, is performed ``by hand'', using the action
on the frame bundle. As the $\H_\CM(n)$ reappear in other contexts,
such as modular forms~\cite{ConnesMHecke}, a more abstract
characterization was on the order of the day.

Another focus of interest is the comprehensive investigation of the
range of validity of Zimmermann's combinatorial formula of
QFT~\cite{Zim}, in the algebraic-combinatorial context. It so happens
that the `natural' formulae for computing the antipodes of the
algebras we are dealing with are in some sense inefficient. This
inefficiency lies in the huge sets of cancellations that arise in the
final result.

A case in point is the alternating sum over chains characteristic of
incidence Hopf algebras. The Fa\`a di Bruno bialgebras are incidence
bialgebras, corresponding to the family of partially ordered sets
(posets) that are partition lattices of finite sets. In relation with
them at the end of his authoritative review~\cite{Schmitt1} on
antipodes of incidence bialgebras, Schmitt in 1987 wrote: ``[The
Lagrange reversion formula] can be viewed as a description of what
remains of the alternating sum [over chains] after all the
cancellations have been taken into account. We believe that
understanding exactly how these cancellations take place will not only
provide a direct combinatorial proof of the Lagrange inversion
formula, but may well yield analogous formulas for the antipodes of
Hopf algebras arising from classes of geometric lattices other than
partition lattices''.

Unbeknown, Zimmermann's formula had been performing this trick in
renormalization theory by then for almost twenty years. And already
in~\cite{Manoukian}, the Dyson--Salam method for renormalization was
thoroughly reexamined and corrected, and its equivalence (in the
context of the so-called BPHZ renormalization scheme) to Bogoliubov's
and Zimmermann's methods was studied. Inspired by the
book~\cite{Manoukian} and the work by Kreimer and Connes, the present
authors a couple of years ago investigated the Hopf algebra theory
underpinnings of these equivalences in~\cite{Ananke,Hektor}.

This is the punchline: on the one hand, the Connes--Kreimer algebras
of rooted trees and Feynman diagrams can be subsumed under the theory
of incidence algebras of distributive lattices (see~\cite{Stanley1}
for the latter); on the other, Zimmermann's formula can be
incorporated into the mainstream of combinatorial Hopf algebra theory
as the cancellation-free formula for the antipode of any such
incidence algebra. We show all this in Sections~11 and~13 of Part~III.
Haiman and Schmitt~\cite{Precursors} eventually found the equivalent
of Zimmermann's formula for \bfb\ algebras. This is also subsumed
here. Thus the trade between QFT and combinatorial Hopf algebra theory
is not one-way.

\marker
We did not want to assume that the readership is familiar with all the
notions of Hopf algebra theory; and some find a direct passage to the
starchy algebraist's diet too abrupt. This is why we start, in the
footsteps of~\cite{HSlim}, with a motivational discussion ---that
experts should probably skip. In the spirit of noncommutative
geometry, suppose we choose to study a set~$S$ via a commutative
algebra~$\F(S)$ of complex functions on it (we work with complex
numbers for definiteness only: nearly everything we have to say
applies when $\C$ is replaced by any unital commutative $\Q$-algebra;
also, in some parts of this paper it will be clear from the context
when we work with real numbers instead of~$\C$). The tensor product of
algebras $\F(S)\ox\F(S)$ has the algebra structure given by
$$
(f\ox g)(f'\ox g') = ff'\ox gg'.
$$
Also, there is a one-to-one map $\sigma:\F(S)\ox\F(S)\to\F(S\x S)$ given
by $f\ox g(s,s')\mapsto f(s)g(s')$. The image of~$\sigma$ consists of
just those functions $h$ of two variables for which the vector space
spanned by the partial maps $h_{s'}(s) := h(s,s')$ is of finite
dimension. Suppose now the set $S\equiv G$ is a group. Then there is
much more to~$\F(G)$ than its algebra structure. The multiplication
in~$G$ induces an algebra map $\rho : \F(G) \to \F(G\x G)$ given by
$$
\rho[f](x,y) = f(xy) =: y\lt f(x) =: f\rt x(y),
$$
where the functions $y\lt f$, $f\rt x$ are respectively the right
translate of $f$ by~$y$ and the left translate of $f$ by~$x$. Then
$\rho[f]\in\im\sigma$ iff $f$ is such that its translates span a
finite-dimensional space. An example is given by the space of
polynomials over the additive group~$\C$ acting on itself by
translation. A function with this property is called a
\textit{representative} function; representative functions clearly
form a subalgebra $\Rr(G)$ of~$\F(G)$. In summary, there is an algebra
map $\Dl=\sigma^{-1}\circ\rho$ from $\Rr(G)$ to $\Rr(G)\ox\Rr(G)$ that
we express by
$$
f(xy) = \sum_{j}f_{j(1)}(x)\ox f_{j(2)}(y) := \Dl f(x,y).
$$
This gives a linearized form of the group operation. Now, the
associative identity $f((xy)z) = f(x(yz))$ imposes
$$
(\id\ox\Dl)\circ\Dl = (\Dl\ox\id)\circ\Dl,
$$
where we denote $\id$ the identity map in the algebra of
representative functions. Moreover, $f\mapsto f(1_G)$ defines a map
$\eta: \Rr(G)\to\C$ (called the \textit{augmentation} map) that, in
view of~$f(x1_G) = f(1_Gx) = f(x)$ verifies
$$
(\id\ox\eta)\circ\Dl = (\eta\ox\id)\circ\Dl = \id.
$$
The basic concept of \textit{coalgebra} in Hopf algebra theory
abstracts the properties of the triple $(\Rr(G),\Dl,\eta)$.

\smallskip

Let us go a bit further and consider, for a complex vector space $V$,
representative maps $G\to V$ whose translates span a
finite-dimensional space of maps, say $\Rr_V(G)$. Let
$(f_1,\dots,f_n)$ be a basis for the space of translates
of~$f\in\Rr_V(G)$, and express
$$
y\lt f(x) = \sum_{i=1}^n c_i(y)f_i(x).
$$
In particular,
$$
y\lt f_j(x) = \sum_{i=1}^n c_{ij}(y)f_i(x)
$$
defines the $c_{ij}$; also $f=\sum_{i=1}^n\eta(f_i)c_i$. Now, the
$c_i$ are representative, since
$$
\sum_{i=1}^n c_i(zy)f_i(x) = z\lt(y\lt f)(x)
= \sum_{i=1}^n c_i(y)z\lt f_i(x)
= \sum_{i,j=1}^n c_j(y)c_{ij}(z)f_i(x),
$$
implying $y\lt c_i = \sum_{j=1}^n c_j(y)c_{ij}$. In consequence, the map
$$
v \ox f \mapsto f(.) v
$$
from $V\ox\Rr(G)$ to $\Rr_V(G)$ is bijective, and we identify these
two spaces. All this points to the importance of~$\Rr(G)$. But in what
wilderness are its elements found? A moment's reflection shows that
linear forms on any locally finite $G$-module provide representative
functions. Suppose $V$ is such a module, and let~$T$ denote the
representation of~$G$ on it. A map $\ga_T:V\to V\ox\Rr(G)$ is given by
$$
\ga_T(v)(x) = T(x)v.
$$
The fact $T(x)T(y)=T(xy)$ means that $\ga_T$ `interacts' with $\Dl$:
$$
(\ga_T\ox\id)\circ\ga_T = (\id_V\ox\Dl)\circ\ga_T;
$$
and the fact $T(1_G)=\id_V$ forces
$$
(\id_V\ox\eta) = \id_V.
$$
One says that $(V,\ga_T)$ is a \textit{comodule} for
$(\Rr(G),\Dl,\eta)$. A Martian versed in the `dual' way of thinking
could find this method a more congenial one to analyze group
representations.

Let us add that the coalgebraic aspect is decisive in the applications
of Hopf algebra theory to renormalization; those interested mainly in
this aspect might consult now our invitation at the beginning of
Section~10. Also, after extracting so much mileage from the
coalgebraic side of the group properties, one could ask, what is the
original algebra structure of~$\Rr(G)$ good for? This question we
shall answer in due course.

\section{Pr\'ecis of bialgebra theory}

We assume familiarity with (associative, unital)
algebras; but it is convenient here to rephrase the requirements for
an algebra $A$ in terms of its two defining maps, to wit $m : A \ox
A\to A$ and $u : \C \to A$, respectively given by $m(a, b) :=
ab;\;u(1) = 1_A$. They must satisfy:
\begin{itemize}
\item{}
Associativity: $m(m \ox \id) = m(\id \ox m) : A \ox A \to A$;
\item{}
Unity:
$m(u \ox \id) = m(\id \ox u) = \id : \C \ox A = A \ox \C = A \to A$.
\end{itemize}
In the following we omit the sign $\circ$ for composition of linear
maps.  These two properties correspond, respectively, to the
commutativity of the diagrams
$$
\vcenter{\hbox{
\xymatrix{
A \ox A \ox A \ar[r]^(.6){m \ox \id_A} \ar[d]_{\id_A \ox m}
& A \ox A \ar[d]^m   \\
A \ox A \ar[r]^m & A,}
}}
$$
and
$$
\vcenter{\hbox{
\xymatrix{
\C \ox A \ar[r]^{u \ox \id_A} \ar@{<->}[d] &  A \ox A \ar[d]^m  \\
  A \ar[r]^{\id_A} & A,}
\qquad\qquad
\xymatrix{
A \ox \C \ar[r]^{\id_A \ox u} \ar@{<->}[d] &  A \ox A \ar[d]^m  \\
  A \ar[r]^{\id_A} & A.}
}}
$$
The unnamed arrows denote the natural identifications $\C \ox A= A = A
\ox \C $ respectively given by $\la \ox a \mapsto \la a$ and $a \ox
\la \mapsto \la a$.

\begin{defn}
\label{df:coalgebra}
A \textit{coalgebra} $C$ is a vector space with the structure obtained
by reversing the arrows in the diagrams above characterizing an
algebra. A coalgebra is therefore also described by two linear maps:
the coproduct $\Dl : C\to C \ox C$ (also called diagonalization, or
``sharing"), and the counit (or augmentation) $\eta:C\to \C$, with
requirements:
\begin{itemize}
\item{}
Coassociativity: $(\Dl \ox \id)\Dl = (\id \ox \Dl)\Dl : C \to C \ox C
\ox C$;
\item{}
Counity: $(\eta \ox \id)\Dl = (\id \ox \eta)\Dl = \id : C \to C$.
\end{itemize}

These conditions, therefore, simply describe the commutativity of the
diagrams ``dual" to the above, namely
$$
\vcenter{\hbox{
\xymatrix{
C \ox C \ox C & C \ox C \ar[l]_(.4){\Dl\ox \id_C}  \\
C \ox C \ar[u]^{\id_C\ox\Dl} & C \ar[l]_\Dl \ar[u]_\Dl }
}}
$$
and
$$
\vcenter{\hbox{
\xymatrix{
\C \ox C & C \ox C \ar[l]_{\eta\ox \id_C}  \\
C \ar@{<->}[u]  & C, \ar[l]_{\id_C} \ar[u]_\Dl }
\qquad\qquad
\xymatrix{
C \ox \C & C \ox C \ar[l]_{\id_C\ox\eta}  \\
C \ar@{<->}[u] & C. \ar[l]_{\id_C} \ar[u]_\Dl }
}}
$$
\end{defn}

In general, the prefix `co' makes reference to the process of reversing
arrows in diagrams.

With $\Dl a := \sum_{j} a_{j(1)} \ox a_{j(2)}$, the second requirement is
explicitly given by:
\begin{equation}
\sum_j \eta(a_{j(1)})\,a_{j(2)} = \sum_j a_{j(1)}\,\eta(a_{j(2)}) = a.
\label{eq:counit}%
\end{equation}
Since it has a left inverse, $\Dl$ is always injective.

If $\Dl a_{j(1)} = \sum_{k} a_{jk(1)(1)} \ox a_{jk(1)(2)}$ and $\Dl
a_{j(2)} = \sum_{l} a_{jl(2)(1)} \ox a_{jl(2)(2)}$, the first
condition is
$$
\sum_{jk} a_{jk(1)(1)} \ox a_{jk(1)(2)} \ox a_{j(2)} =
\sum_{jl} a_{j(1)} \ox a_{jl(2)(1)} \ox a_{jl(2)(2)}.
$$
Thus coassociativity corresponds to the idea that, in decomposing the
`object' $a$ in sets of three pieces, the order in which the breakups
take place does not matter.

In what follows we alleviate the notation by writing simply
$\Dl a = \sum a_{(1)} \ox a_{(2)}$. Sometimes even the $\sum$ sign
will be omitted. Since $\sum a_{(1)(1)} \ox a_{(1)(2)} \ox
a_{(2)} = \sum a_{(1)} \ox a_{(2)(1)} \ox a_{(2)(2)}$, we can write
$$
\Dl^2 a = a_{(1)} \ox a_{(2)}\ox a_{(3)},\quad
\Dl^3 a = a_{(1)} \ox a_{(2)}\ox a_{(3)} \ox a_{(4)},
$$
and so on, for the $n$-fold coproducts. Notice that $(\eta \ox \id \ox
\cdots \ox \id)\Dl^n = \Dl^{n-1}$ ($n\,\id$ factors understood). And
similarly when $\eta$ is in any other position. We analogously
consider the $n$-fold products $m^n: A\ox\cdots\ox A\to A$, with $n+1$
factors $A$ understood.

Like algebras, coalgebras have a tensor product. The coalgebra $C\ox
D$ is the vector space $C\ox D$ endowed with the maps
\begin{equation}
\Dl_{\ox}(c\ox d) = \sum c_{(1)} \ox d_{(1)}\ox c_{(2)} \ox d_{(2)};
\qquad
\eta_{\ox}(c\ox d) = \eta(c)\eta(d).
\label{eq:coal-tensor}%
\end{equation}
That is $\Dl_{\ox} = (\id\ox\tau\ox\id)(\Dl_{C}\ox\Dl_{D})$, in
parallel with $m_{\ox} = (m_{A}\ox m_{B})(\id\ox\tau\ox\id)$ for
algebras.

A counital coalgebra (or comultiplicative) map $\ell: C\to D$ between
two coalgebras is a linear map that preserves the coalgebra structure,
$$
\Dl_D\ell = (\ell\ox \ell)\Dl_C : C \to D \ox D;
\qquad \eta_D\ell = \eta_C : C \to \C.
$$
Once more these properties correspond to the commutativity
of the diagrams
$$
\vcenter{\hbox{
\xymatrix{
C \ox C \ar[d]_{\ell\ox\ell} & C \ar[l]_(.4){\Dl_C} \ar[d]^\ell\\
D \ox D & D, \ar[l]_(.4){\Dl_D} }
\qquad\qquad
\xymatrix @C=9pt{
C \ar[rr]^\ell \ar[dr]_{\eta_C} && D \ar[dl]^{\eta_D}  \\
& \C, }
}}
$$
obtained by reversing arrows in the diagrams that express the
homomorphism properties of linear maps between unital algebras.

(Non)commutativity of the algebra and coalgebra operations is
formulated with the help of the ``flip map'' $\tau(a \ox b) := b \ox
a$. The algebra $A$ is \textit{commutative} if $m\tau = m : A \ox A\to
A$; likewise, the coalgebra $C$ is called \textit{cocommutative} if
$\tau\Dl = \Dl : C \to C \ox C$. These properties correspond,
respectively, to the commutativity of the diagrams
$$
\vcenter{\hbox{
\xymatrix @C=9pt{
A\ox A \ar[rr]^\tau \ar[dr]_{m} && A\ox A \ar[dl]^{m}  \\
& A, }
\qquad\qquad
\xymatrix @C=9pt{
C\ox C  && \ar[ll]_(.5){\tau} C\ox C    \\
& C.\ar[ur]_{\Dl} \ar[ul]^{\Dl} }
}}
$$

For commutative algebras, the map $m$
is a homomorphism, and similarly $\Dl$ is a coalgebra map for
cocommutative coalgebras. The same space $C$ with the coalgebra
structure given by $\tau\Dl$ is called the \textit{coopposite}
coalgebra $C^\cop$.

A \textit{subcoalgebra} of~$C$ is a subspace $Z$ such that
$$
\Dl Z \subseteq Z\ox Z.
$$
A coalgebra without nontrivial subcoalgebras is \textit{simple}. The
direct sum of all simple subcoalgebras of~$C$ is called its coradical
$R(C)$.

A very important concept is that of~\textit{coideal}. A subspace
$J$ is a coideal of~$C$ if
$$
\Dl J \subseteq J\ox C + C\ox J \sepword{and} \eta(J) = 0.
$$
The kernel of any coalgebra map $\ell$ is a coideal. In effect,
$$
\Dl_D\ell(\ker\ell) = 0 = (\ell\ox \ell)\Dl_C\ker\ell,
$$
forces $\Dl_C\ker\ell\subseteq\ker\ell\ox C + C\ox\ker\ell$, and
moreover
$$
\eta_C(\ker\ell) = \eta_D\ell(\ker\ell) = 0.
$$
If $J$ is a coideal, then $C/J$ has a (unique) coalgebra structure
such that the canonical projection $C\xrightarrow{q} C/J$ is a
coalgebra map ---see the example at the end of the section.

A coalgebra filtered as a vector space is called a \textit{filtered
coalgebra} when the filtering is compatible with the coalgebra
structure; that is, there exists a nested sequence of subspaces $C_n$
such that $C_{0}\subsetneq C_1\subsetneq\dots$ and
$\bigcup_{n\ge0}C_n=C$, and moreover
$$
\Dl C_n \subseteq \sum_{k=0}^n C_{n-k} \ox C_k.
$$

\marker
Given an algebra $A$ and a coalgebra $C$ over~$\C$, one can define the
\textit{convolution} of two elements $f,g$ of the vector space of
$\C$-linear maps $\Hom(C,A)$, as the map $f*g\in\Hom(C,A)$ given by
the composition
$$
C \xrightarrow{\Dl} C \ox C \xrightarrow{f\ox g} A \ox A
\xrightarrow{m} A.
$$
In other words,
$$
f*g(a) = \sum f(a_{(1)}) \, g(a_{(2)}).
$$

The triple $(\Hom(C,A),*,u_A\eta_C)$ is then a unital algebra. Indeed,
convolution is associative because of associativity of~$m$ and
coassociativity of~$\Dl$:
\begin{align*}
(f * g) * h
&= m ((f * g) \ox h) \Dl = m(m \ox \id)(f \ox g \ox h)(\Dl \ox \id)\Dl
\\
&= m(\id \ox m)(f \ox g \ox h)(\id \ox \Dl)\Dl = m (f \ox (g * h)) \Dl
           = f * (g * h).
\end{align*}

The unit is in effect $u_A\eta_C$:
\begin{align*}
f * u_A \eta_C &= m (f \ox u_A \eta_C) \Dl
           = m (\id_A \ox u_A)(f \ox \id_\C)(\id_C \ox \eta_C) \Dl
           = \id_A \, f \, \id_C = f,
\\
u_A\eta_C * f &= m (u_A\eta_C \ox f) \Dl
           = m (u_A \ox \id_A)(\id_\C \ox f)(\eta_C \ox \id_C) \Dl
           = \id_A \, f \, \id_C = f.
\end{align*}

Algebra morphisms $l: A\to B$ and coalgebra morphisms $\ell: D\to C$
respect convolution, in the following respective ways: if $f,g \in
\Hom(C,A)$ for some coalgebra $C$ and some algebra $A$, then
\begin{align}
l(f * g) &= lm_A(f \ox g)\Dl = m_B(l \ox l)(f \ox g)\Dl = m_B(lf \ox
lg)\Dl = lf * lg,
\label{eq:algprop}
\\
(f * g)\ell &= m(f \ox g)\Dl_C\ell = m(f \ox g)(\ell \ox \ell)\Dl_D =
m_A(f\ell\ox g\ell)\Dl_D = f\ell * g\ell.
\label{eq:coalgprop}%
\end{align}

\begin{defn}
\label{df:bialgebra}
To obtain a \textit{bialgebra}, say $H$, one considers on the vector
space $H$ both an algebra and a coalgebra structure, and further
stipulates the \textit{compatibility} condition that the algebra
structure maps $m$ and $u$ are counital coalgebra morphisms, when
$H\ox H$ is seen as the tensor product coalgebra.
\end{defn}

This leads (omitting the subscript from the unit in the algebra) to:
\begin{equation}
\Dl a\,\Dl b = \Dl(ab),  \quad  \Dl1 = 1\ox1,  \quad
\eta(a)\eta(b) = \eta(ab),  \quad  \eta(1) = 1;
\label{eq:mirror-symmetry}%
\end{equation}
in particular, $\eta$ is a one-dimensional representation of the
algebra~$H$.  For instance:
\begin{align*}
\Dl(ab) &= \Dl m(a\ox b) = (m\ox m)\Dl(a\ox b)
= (m\ox m)[a_{(1)} \ox b_{(1)}\ox a_{(2)}\ox b_{(2)}]
\\
&= a_{(1)}b_{(1)}\ox a_{(2)}b_{(2)}
= (a_{(1)}\ox a_{(2)})(b_{(1)}\ox b_{(2)})
= \Dl a\,\Dl b.
\end{align*}
Of course, it would be totally equivalent and (to our earthlings'
mind) looks simpler to postulate instead the
conditions~\eqref{eq:mirror-symmetry}: they just state that the
coalgebra structure maps $\Dl$ and $\eta$ are unital algebra
morphisms. But it is imperative that we familiarize ourselves with the
coalgebra operations. Note that the last equation
in~\eqref{eq:mirror-symmetry} is redundant.

The map $u\eta:H\to H$ is an idempotent, as $u\eta u\eta(a)=
\eta(a)u\eta(1_H)=\eta(a)1_H=u\eta(a)$. Therefore $H=\im u\eta\op\ker
u\eta=\im u\op\ker\eta=\C1_H\op\ker\eta$. A halfway house between
algebras or coalgebras and bialgebras is provided by the notions of
augmented algebra, which is a quadruple $(A,m,u,\eta)$, or augmented
coalgebra, which is a quadruple $(C,\Dl,u,\eta)$, with the known
properties in both cases.

A \textit{bialgebra morphism} is a linear map between two bialgebras,
which is both a unital algebra homomorphism and a counital coalgebra
map. A \textit{subbialgebra} of $H$ is a vector subspace $E$ that is
both a subalgebra and a subcoalgebra; in other words, $E$, together
with the restrictions of the product, coproduct and so on, is also a
bialgebra and the inclusion $E \hookto H$ is a bialgebra
morphism.

A \textit{biideal} $J$ of~$H$ is a linear subspace that is both an
ideal of the algebra~$H$ and a coideal of the coalgebra~$H$. The
quotient $H/J$ inherits a bialgebra structure.

Associated to any bialgebra $H$ there are the three bialgebras
$H^\opp$, $H^\cop$, $H^{\mathrm{copp}}$ obtained by taking opposite
either of the algebra structure or the coalgebra structure or both.

\medskip

Linear maps of a bialgebra $H$ into an algebra $A$ can in particular
be convolved; but if they are multiplicative, that is, algebra
homomorphisms, their convolution in general will be multiplicative
only if $A$ is commutative. In such a case if $f,g\in\Hom_\alg(H,A)$,
then
$$
f*g(ab) = f(a_{(1)})f(b_{(1)})g(a_{(2)})g(b_{(2)}) =
f(a_{(1)})g(a_{(2)})f(b_{(1)})g(b_{(2)}) = f*g(a)f*g(b).
$$
Similarly, the convolution of coalgebra maps of a coalgebra $C$ into a
bialgebra $H$ is comultiplicative when $H$ is cocommutative.

\begin{defn}
\label{df:filteredbialgebra}
A bialgebra $\N$-filtered as a vector space is called a
\textit{filtered bialgebra} when the filtering is compatible with both
the algebra and the coalgebra structures; that is, there exits a
nested sequence of subspaces $H_0\subsetneq H_1\subsetneq\dots$ such
that $\bigcup_{n\ge0}H_n=H$, and moreover
$$
\Dl H_n \subseteq \sum_{k = 0}^n H_{n-k} \ox H_k; \quad
H_nH_m \subseteq H_{n+m}.
$$
\textit{Connected} bialgebras are those filtered bialgebras for which
the first piece consists just of scalars: $H_{0} = u(\C)$; in that
case $R(H)=H_0$, and the augmentations $\eta,u$ are unique.
\end{defn}

\marker
Any coalgebra $C$ is filtered via a filtering, dubbed the
\textit{coradical filtering}, whose starting piece is the coradical.
When $H$ is a bialgebra, its coradical filtering is not necessarily
compatible with the algebra structure. Nevertheless, it is compatible
when $R:=R(H)$ is a subbialgebra of~$H$, in particular when $H_{0} =
u(\C)$. In short, the coradical filtering goes as follows: consider
the sums of tensor products
$$
H^1_R := R; \quad
H^2_R := R\ox H + H\ox R; \quad
H^3_R := R\ox H\ox H + H\ox R\ox H + H\ox H\ox R;
$$
and so on; then the subcoalgebras $H_n$ are defined by
$$
H_n = [\Dl^n]^{-1}(H^{n+1}_R).
$$
We refer the reader to~\cite{Montgomery,Puydedome} for more details on
the coradical filtering. Naturally, a bialgebra may have several
different filterings.

A bialgebra $H = \bigoplus^\infty_{n=0} H^{(n)}$ graded as a vector
space is called a \textit{graded bialgebra} when the grading~$\#$ is
compatible with both the algebra and the coalgebra structures:
$$
H^{(n)}H^{(m)} \subseteq H^{(n+m)} \sepword{and}
\Dl H^{(n)} \subseteq \bigoplus_{k = 0}^n H^{(n-k)} \ox H^{(k)}.
$$
That is, grading $H\ox H$ in the obvious way, $m$ and $\Dl$ are
homogeneous maps of degree zero. A graded bialgebra is filtered in
the obvious way. Most often we work with graded bialgebras
\textit{of finite type}, for which the $H^{(n)}$ are finite dimensional;
and, at any rate, all graded bialgebras are direct limits of subbialgebras
of finite type~\cite[Proposition~4.13]{MilnorM}.

\smallskip

Two main ``classical'' \textit{examples} of bialgebras, respectively
commutative and cocommutative, are the space of representative
functions on a compact group and the enveloping algebra of a Lie
algebra.

\begin{xmpl}
Let $G$ be a compact topological group (most often, a Lie group). The
Peter--Weyl theorem shows that any unitary irreducible representation
$\pi$ of~$G$ is finite-dimensional, any matrix element $f(x) :=
\<u,\pi(x)v>$ is a representative function on~$G$, and the vector
space $\Rr(G)$ generated by these matrix elements is a dense
$*$-subalgebra of $C(G)$. Elements of this space can be characterized
as those continuous functions $f: G \to \C$ whose translates $t\lt f:
x \mapsto f(xt)$, for all $t\in G$, generate a finite-dimensional
subspace of $C(G)$. The theorem says, in other words, that nothing of
consequence is lost in compact group theory by studying just~$\Rr(G)$.
As we know already, the commutative algebra $\Rr(G)$ is a coalgebra,
with operations
\begin{equation}
\Dl f(x,y) := f(xy); \quad \eta(f) := f(1_G).
\label{eq:coprod-repfns}%
\end{equation}
\end{xmpl}

\begin{xmpl}
The \textit{universal enveloping algebra} $\U(\g)$ of a Lie algebra
$\g$ is the quotient of the tensor algebra $T(\g)$ ---with
$T^0(\g)\simeq\C$--- by the two sided ideal $I$ generated by the
elements $XY - YX - [X,Y]$, for all $X,Y \in \g$. The word
``universal'' is appropriate because any Lie algebra homomorphism
$\psi: \g\to A$, where $A$ is a unital associative algebra, extends
uniquely to a unital algebra homomorphism $\U\psi:\U(\g)\to A$.

A coproduct and counit are defined first on elements of~$\g$ by
\begin{equation}
\Dl X := X \ox 1 + 1 \ox X,
\label{eq:precoprod-tensalg}%
\end{equation}
and $\eta(X) := 0$. These linear maps on $\g$ extend to homomorphisms
of~$T(\g)$; for instance,
$$
\Dl(XY) = \Dl X\Dl Y = XY \ox 1 + X \ox Y + Y \ox X + 1 \ox XY.
$$
It follows that the tensor algebra on any vector space is a (graded,
connected) bialgebra.
Now
$$
\Dl(XY - YX - [X,Y])
= (XY - YX - [X,Y]) \ox 1 + 1 \ox (XY - YX - [X,Y]),
$$
Thus $I$ is also a coideal (clearly $\eta(I) = 0$, too) and if $q:
T(\g) \to \U(\g)$ is the quotient map, then $I\subseteq\ker(q\ox
q)\Dl$; thus $(q\ox q)\Dl$ induces a coproduct on the quotient
$\U(\g)$, that becomes an irreducible bialgebra.
From~\eqref{eq:precoprod-tensalg} and the definition of~$I$, it is
seen that $\U(\g)$ is \textit{cocommutative}. Note also that it is a
graded coalgebra. The coradical filtering of $\U(\g)$ is just the
obvious filtering by degree~\cite{Montgomery}.
\end{xmpl}

When $\g$ is the Lie algebra of $G$, both previous constructions are
mutually dual in a sense that will be studied soon.

\section{Primitive and indecomposable elements}

\begin{defn}
An element $a$ in a bialgebra $H$ is said to be (1-)primitive when
$$
\Dl a = a \ox 1 + 1 \ox a.
$$
Primitive elements of~$H$ form a vector subspace $P(H)$, which is seen
at once to be a Lie subalgebra of $H$ with the ordinary bracket
$$
[a,b] := ab-ba.
$$
\end{defn}

For instance, elements of~$\g$ inside $\U(\g)$ are
primitive by definition; and there are no others.

Denote by $H_+ := \ker\eta$ the augmentation ideal of~$H$.
If for some $a \in H$, we can find $a_1,a_2 \in H_+$ with
$\Dl a = a_1 \ox 1 + 1 \ox a_2$, then by the counit property
$a = (\id \ox \eta) \Dl a = a_1$, and similarly $a = a_2$; so $a$ is
primitive. In other words,
\begin{equation}
P(H) = \Dl^{-1}(H_+ \ox 1 + 1 \ox H_+).
\label{eq:primitives}%
\end{equation}
By the counit property as well, $a\in\ker\eta$ is automatic for
primitive elements. If $\ell: H\to K$ is a bialgebra morphism, and $a
\in P(H)$ then $\Dl_{K} \ell(a) = (\ell \ox \ell) \Dl_H a = \ell(a)
\ox 1 + 1 \ox \ell(a)$; thus the restriction of $\ell$ to~$P(H)$
defines a Lie algebra map $P(\ell) \: P(H) \to P(K)$. If $\ell$ is
injective, obviously so is $P(\ell)$.

\begin{prop}
\label{pr:deltaFormula} 
Let $H$ be a connected bialgebra. Write ${H_+}_n := H_+\cap H_n$ for
all~$n$. If $a \in {H_+}_n$, then
\begin{equation}
\Dl a = a \ox 1 + 1 \ox a + y,
\sepword{where} y \in {H_+}_{n-1} \ox {H_+}_{n-1}.
\label{eq:gr-coprod}%
\end{equation}
Moreover, $H_1 \subseteq \C1 \op P(H)$.
\end{prop}

\begin{proof}
Write $\Dl a = a \ox 1 + 1 \ox a + y$. If $a\in H_+$, then
$(\id\ox\eta)(y) = (\id \ox \eta)\Dl a - a = 0$ and similarly
$(\eta\ox\id)(y) = 0$ by the counity properties~\eqref{eq:counit}.
Therefore $y\in H_+\ox H_+$, and thus
$$
y \in \sum_{i=0}^n {H_+}_i\ox {H_+}_{n-i}.
$$
As ${H_+}_0 = (0)$, the first conclusion follows.

Thus, when $H$ is a connected bialgebra, $H = H_0 \op \ker\eta$.  If
$c\in H_1$, we can write $c = \mu 1 + a$, with $\mu\in \C$ and
$a \in {H_+}_1$. Now, $\Dl a = a \ox 1 + 1 \ox a + \la(1 \ox 1)$ for
some $\la \in \C$; and $a + \la 1$ is primitive, thus
$c = (\mu-\la)1 + (a + \la 1)$ lies in $\C\,1 \op P(H)$.
\end{proof}

In a connected bialgebra, the primitive elements together with the
scalars constitute the ``second step'' of the coradical filtering. If
$H$ is graded, some $H^{(k)}$ might be zero; but if $k$ is the
smallest nonzero integer such that $H^{(k)}\ne 0$, compatibility of
the coproduct with the grading ensures $H^{(k)}\subseteq P(H)$.

In view of~\eqref{eq:gr-coprod}, it will be very convenient to
consider the reduced coproduct $\Dl'$ defined on $H_+$ by
\begin{equation}
\Dl'a := \Dl a - a \ox 1 - 1 \ox a =: \sum a'_{(1)}\ox a'_{(2)}.
\label{eq:gr-coprod-trunc}
\end{equation}
In other words, $\Dl'$ is the restriction of $\Dl$ as a map
$\ker\eta\to\ker\eta\ox\ker\eta$; and $a$ is primitive if and only if
it lies in the kernel of~$\Dl'$. Coassociativity of $\Dl'$ (and
cocommutativity, when it holds) is easily obtained from the
coassociativity of $\Dl$; and we write
$$
{\Dl'}^n a = \sum a'_{(1)}\ox\cdots\ox a'_{(n)}.
$$
The reduced coproduct is not an algebra homomorphism:
\begin{equation}
\Dl'(ab) = \Dl'a\,\Dl'b + (a\ox 1 + 1\ox a)\Dl'b +
\Dl'a\,(b\ox 1 + 1\ox b) + a\ox b + b\ox a.
\label{eq:purga-de-Benito}
\end{equation}

Let $p \: H \to H_+$ be the projection defined by~$p(a) := (\id -
u\eta) a = a - \eta(a)\,1$; then the previous
result~\eqref{eq:gr-coprod} is reformulated as $(p\ox p)\Dl = \Dl'p$;
more generally, it follows that
\begin{equation}
U_n := (p\ox p\ox\cdots \ox p)\Dl^n = {\Dl'}^n p,
\label{eq:eso-era}
\end{equation}
with $n+1$ factors in $p\ox p\ox\cdots\ox p$. This humble equality
plays a decisive role in this work.

\begin{thm}
\label{pr:tensor-and-primitives}
Consider now the tensor product $H\ox K$ of two connected graded
bialgebras $H$ and $K$, which is also a connected graded bialgebra,
with the coproduct~\eqref{eq:coal-tensor}. Then
\begin{equation}
P(H\ox K) = P(H) \ox 1 + 1\ox P(K) \cong P(H) \op P(K)
\label{eq:tensor-of-primitives}
\end{equation}
in $H\ox K$.
\end{thm}

\begin{proof}
The last identification comes from the obvious $P(H)\ox1\cap1\ox
P(K)=(0)$ in $H\ox K$. Let $a\in P(H)$ and $b \in P(K)$, then
\begin{align*}
\Dl_{\ox}(a \ox 1 + 1 \ox b)
&= (\id \ox \tau \ox \id)
\bigl((a \ox 1 + 1 \ox a) \ox (1 \ox 1)
+ (1 \ox 1) \ox (b \ox 1 + 1 \ox b) \bigr)
\\
&= (a \ox 1 + 1 \ox b) \ox (1\ox1) +
(1 \ox 1)(a \ox 1 + 1 \ox b),
\end{align*}
so $a \ox 1 + 1 \ox b \in P(H\ox K)$.

On the other hand, letting $x \in P(H\ox K)$:
\begin{equation}
\Dl_{\ox} x = x \ox 1 \ox 1 + 1 \ox 1 \ox x.
\label{eq:deltax}%
\end{equation}
Since $H\ox K$ is a graded bialgebra we can write
$$
x = a^0 \ox 1 + 1 \ox b^0 + \sum_{p,q \geq 1} a^p \ox b^q,
$$
where $a^0\in H_+$, $b^0\in K_+$, $a^p\in H^{(p)}$, $b^q\in K^{(q)}$.
By~\eqref{eq:gr-coprod-trunc} $\Dl_H a^i =
a^i \ox 1 + 1 \ox a^i + \sum {a^i}'_{(1)} \ox {a^i}'_{(2)}$ for
$i\geq 0$ and $\Dl_K b^j =
b^j \ox 1 + 1 \ox b^j + \sum {b^j}'_{(1)} \ox {b^j}'_{(2)}$ for
$j\geq 0$. Thus, after a careful book-keeping, it follows that
\begin{align*}
\Dl_{\ox} x
&= x \ox 1 \ox 1 + 1 \ox 1 \ox x +
\sum {a^0}'_{(1)} \ox 1 \ox {a^0}'_{(2)} \ox 1
+ \sum 1 \ox {b^0}'_{(1)} \ox 1 \ox {b^0}'_{(2)}
\\
&\quad + \sum 1 \ox b^q \ox a^p \ox 1
+ \sum a^p \ox 1 \ox 1 \ox b^q + R,
\end{align*}
where $R$ is a sum of terms of the form $c \ox d \ox e \ox f$ where at
least three of the following conditions hold: $c \in H_+$, $d \in
K_+$, $e \in H_+$, or $f \in K_+$.  A comparison
with~\eqref{eq:deltax} then gives $R=0$ and
$$
\sum {a^0}'_{(1)} \ox 1 \ox {a^0}'_{(2)} \ox 1
=  \sum 1 \ox {b^0}'_{(1)} \ox 1 \ox {b^0}'_{(2)}
= \sum a^p \ox 1 \ox 1 \ox b^q = 0.
$$
The vanishing of the third sum gives that $\sum a^p \ox b^q = 0$, so
$x = a^0 \ox 1 + 1 \ox b^0$, whereas the vanishing of the first and
second sums gives $\sum {a^0}'_{(1)} \ox {a^0}'_{(2)} = 0$ and $\sum
{b^0}'_{(1)} \ox {b^0}'_{(2)}= 0$, that is $a^0\in P(H)$ and $b^0\in
P(K)$.
\end{proof}

\begin{defn}
In a connected Hopf algebra $H_+$ is the unique maximal ideal, and
the~$H_+^m$ for $m\ge1$ form a descending sequence of ideals. The
graded algebra $Q(H) := \C1 \op H_+/H_+^2$ is called the set of
\textit{indecomposables} of~$H$.
\end{defn}

Algebraically, the $H$-module~$Q(H):=H_+/H_+^2$ is the tensor product
of~$H_+$ and~$\C$ by means of $\eta\: H \to \C$
\cite[Section~2.4]{Polaris}. We spell this out. Given $M$ and $N$,
respectively a right $H$-module by an action~$\phi_M$ and a left
$H$-module by an action~$\phi_N$, the vector space whose elements are
finite sums $\sum_j m_j\ox n_j$ with $m_j\in M$ and $n_j\in N$,
subject to the relations
$$
m\phi_M(a)\ox n = m\ox\phi_N(a)n, \sepword{for each} a\in H;
$$
is denoted $M\ox_H N$. Note now that $\C$ is a (left or right)
$H$-module by $a\lt\la=\la\rt a:=\eta(a)\la$. Also $H_+$ is a (right
or left) $H$-module. Thus, the tensor product $H_+\ox_H\C$ of~$H_+$
and~$\C$ over~$H$ by means of~$\eta$ is the graded vector space whose
elements are finite sums $\sum_j s_j\ox\b_j$ with $s_j\in H_+$ and
$\b_j\in\C$, subject to the relations
$$
sa\ox\b = s\ox\eta(a)\b, \sepword{for each} a\in A;
$$
if $s\in H_+^2$, then $s=0$ in $H_+\ox_H\C$. Similarly one defines
$\C\ox_H H_+\simeq H_+\ox_H\C$. Notice $\C\ox_H N\simeq N/H_+N$.

The quotient algebra morphism $H\to\C1\op H_+/H_+^2$ restricts to a
graded linear map $q_H\:P(H)\to H_+/H_+^2$; clearly this map will be
one-to-one iff $P(H)\cap H_+^2=0$, and onto iff $P(H) + H_+^2 = H_+$.

\begin{prop}
\label{pr:halfway}
If the relation
$$
P(H)\cap H_+^2 = (0),
$$
implying that primitive elements are all indecomposable, holds, then
$H$ is commutative.
\end{prop}

\begin{proof}
The commutator $[a,b]$ for $a,b$ primitive belongs both to~$P(H)$
and~$H_+^2$; therefore it must vanish. Proceeding by induction on the
degree, one sees that the bracket vanishes for general elements
of~$H$: indeed, let $[a,b]=0$ for all $a\in H_p,b\in H_q$ and consider
$[a,b]$ for, say, $a\in H_p,b\in H_{q+1}$. A straightforward
computation, using~\eqref{eq:purga-de-Benito}, shows that
$\Dl'[a,b]=0$; so $[a,b]$ is primitive; and hence zero.
\end{proof}

If $\ell\: H\to K$ is an onto bialgebra
map, then the induced map $Q(\ell)\: Q(H)\to Q(K)$ is onto.

(The terminology ``indecomposable elements" used for instance in this
section, is somewhat sloppy, as in fact the indecomposables are
defined only modulo $H_+^2$. However, to avoid circumlocutions, we
shall still use it often, trusting the reader not be confused.)

\section{Dualities}

Consider the space $C^*$ of all linear functionals on a coalgebra $C$.
One identifies $C^*\ox C^*$ with a subspace of $(C\ox C)^*$ by
defining
\begin{equation}
f\ox g(a\ox b) = f(a)g(b),
\label{eq:ovo-di-Colombo}
\end{equation}
where $a,b\in C;f,g\in C^*$. Then $C^*$ becomes an algebra with
product the restriction of $\Dl^{\!t}$ to $C^* \ox C^*$; with $^t$
denoting transposed maps. We have already seen this, as this product
is just the convolution product:
$$
fg(a) = \sum f(a_{(1)})g(a_{(2)});
$$
and $u_{C^*}1=\eta$ means the unit is~$\eta^t$.

It is a bit harder to obtain a coalgebra by dualization of an
algebra~$A$ ---and thus a bialgebra by a dualization
of another.  The reason is that $m^t$ takes $A^*$ to $(A \ox A)^*$ and
there is no natural mapping from $(A \ox A)^*$ to $A^* \ox A^*$; if
$A$ is not finite dimensional, the inverse of the identification
in~\eqref{eq:ovo-di-Colombo} does not exist, as the first of these
spaces is larger than the second.

In view of these difficulties, the pragmatic approach is to focus on
the (strict) pairing of two bialgebras $H$ and $K$, where each may be
regarded as included in the dual of the other. That is to say, we
write down a bilinear form $\dst{a}{f} := f(a)$ for $a\in H$ and $f
\in K$ with implicit inclusions $K \hookto H^*,H \hookto K^*$. The
transposing of operations between the two bialgebras boils down to the
following four relations, for $a,b \in H$ and $f,g \in K$:
\begin{gather}
\dst{ab}{f} = \dst{a \ox b}{\Dl_{K}f},  \qquad
\dst{a}{fg} = \dst{\Dl_H a}{f \ox g},
\notag \\
\eta_H(a) = \dst{a}{1},  \sepword{and}  \eta_{K}(f) = \dst{1}{f}.
\label{eq:nada-sin-ti}
\end{gather}
The nondegeneracy conditions allowing us to assume that $H \hookto
K^*$ and $K \hookto H^*$ are: (i)~$\dst{a}{f} = 0$ for all $f \in K$
implies $a = 0$, and (ii) $\dst{a}{f} = 0$ for all $a \in H$ implies
$f = 0$. It is plain that $H$ is a commutative bialgebra iff $K$ is
cocommutative.

The two examples at the end of Section~3 are tied up by duality as
follows. Let $G$ be a compact connected Lie group whose Lie algebra
is~$\g$. The function algebra $\Rr(G)$ is a commutative bialgebra,
whereas $\U(\g)$ is a cocommutative bialgebra. Moreover, representative
functions are smooth~\cite{alemanesbuenos}. On identifying
$\g$ with the space of left-invariant vector fields on the group
manifold~$G$, we can realize $\U(\g)$ as the algebra of left-invariant
differential operators on~$G$. If $X \in \g$, and $f \in \Rr(G)$, we
define
$$
\dst{X}{f} := Xf(1) = \ddto{t} f(\exp tX),
$$
and more generally, $\dst{X_1\dots X_n}{f} :=
X_1(\cdots(X_nf)\cdots)(1)$; we also set $\dst{1}{f} := f(1)$. This
yields a duality between~$\Rr(G)$ and~$\U(\g)$. Indeed, the Leibniz
rule for vector fields, namely $X(fh) = (Xf)h + f(Xh)$, gives
\begin{align}
\dst{X}{fh} &= Xf(1) h(1) + f(1) Xh(1)
          = (X \ox 1 + 1 \ox X)(f \ox h)(1 \ox 1)
\notag \\
&= \Dl X(f \ox h)(1 \ox 1) = \dst{\Dl X}{f \ox h};
\label{eq:Leibniz-Ug}%
\end{align}
while
\begin{align*}
\dst{X \ox Y}{\Dl f} &= \ddto{t} \ddto{s} (\Dl f)(\exp tX \ox \exp sY)
          = \ddto{t} \ddto{s} f(\exp tX \exp sY)
\\
&= \ddto{t} (Yf)(\exp tX) = X(Yf)(1) = \dst{XY}{f}.
\end{align*}
The necessary properties are easily checked. Relation
\eqref{eq:Leibniz-Ug} shows that $\Dl X = X \ox 1 + 1 \ox X$ encodes
the Leibniz rule for vector fields.

\marker
A more normative approach to duality is to consider instead the
subspace $A^\circ$ of~$A^*$ made of functionals whose kernels contain
an ideal of finite codimension in~$A$. Alternatively, $A^\circ$ can be
defined as the set of all functionals $f\in A^*$ for which there are
functionals $g_1,\dots,g_r; h_1,\dots h_r$ in $A^*$ such that $f(ab) =
\sum_{j=1}^rg_j(a)h_j(b)$; that is to say, $A^\circ$ is the set of
functions on the monoid of~$A$ that are both linear and
representative. It can be checked that $m^t$ maps $A^\circ$ to
$A^\circ\ox A^\circ$, and so $(A^\circ, m^t|_{A^\circ},u^t)$ defines a
coalgebra structure on~$A^\circ$, with $\eta_{A^\circ}(f) = f(1_A)$.

Given a bialgebra $(H,m,u,\Dl,\eta)$, one then sees that
$(H^\circ, \Dl^{\!t}, \eta^t, m^t,u^t)$ is again a bialgebra,
called the \textit{finite dual} or Sweedler dual of~$H$; the
contravariant functor $H\mapsto H^\circ$ defines a duality of the
category of bialgebras into itself. In the previous case of a dual
pair $(H,K)$, we actually have $K \hookto H^\circ$ and
$H \hookto K^\circ$.

If $G$ is a group, $\C G$ denotes the group algebra of~$G$, that is,
the complex vector space freely generated by~$G$ as a basis, with
product defined by extending linearly the group multiplication of $G$,
so $1_G$ is the unit in $\C G$. Endowed with the coalgebra structure
given by (the linear extensions of) $x\to x\ox x$ and $\eta(x):= 1$,
it is a cocommutative bialgebra. In view of the discussion of
Section~1, $\Rr(G)$ is the Sweedler dual of~$\C G$.

In a general bialgebra $H$, a nonzero element $g$ is called
\textit{grouplike} if $\Dl g := g \ox g$; for it $\eta(g) = 1$. The
product of grouplike elements is grouplike.

The \textit{characters} of a bialgebra~$H$ (further discussed in the
following section) are by definition the multiplicative elements
of~$H^*$. They belong to~$H^\circ$, as for them $m^tf = f\ox f$. Then,
the set~$G(H^\circ)$ of grouplike elements of $H^\circ$ coincides with
the set of characters of~$H$. If $H=\Rr(G)$, the map $x\to x^0$ given
by $x^0(f) =f(x)$ gives a map $G\to G(\Rr^0(G))$; it will become clear
soon that it is a homomorphism of groups.

Among the interesting elements of the dual of a bialgebra, there are
also the derivations or \textit{infinitesimal characters}: these are
linear functionals $\dl$ satisfying
$$
\dl(ab) = \dl(a)\eta(b) + \eta(a)\dl(b) \sepword{for all} a,b \in H.
$$
This entails $\dl(1) = 0$. The previous relation can also be written
as $m^t(\dl) = \dl \ox \eta + \eta \ox \dl$, which shows that
infinitesimal characters belong to $H^\circ$ as well, and are
primitive there. Thus the Lie algebra of primitive elements
of~$H^\circ$ coincides with the Lie algebra $\Der_\eta H$ of
infinitesimal characters.

\marker
When $A$ is a graded algebra of finite type, one can consider the
space
$$
A' := \bigoplus_{n\ge0}{A^{(n)}}^*,
$$
where $\dst{{A^{(n)}}^*}{A^{(m)}} = 0$ for $n\ne m$, and there is
certainly no obstacle to define the graded coproduct on homogeneous
components of~$A'$ as the transpose of
$$
m: \sum_{k=0}^n A^{(k)}\ox A^{(n-k)} \to A^{(n)}.
$$
If the algebra above is a bialgebra $H$, one obtains in this way a
subbialgebra~$H'$ of $H^\circ$, called the \textit{graded dual}
of~$H$. Certainly $(H,H')$ form a nondegenerate dual pair. Note $H''=H$.

If $I$ is a linear subspace of~$H$ graded of finite type, we denote
by~$I^\perp$ its orthogonal in~$H'$. Naturally $I^{\perp\perp}=I$.

\begin{prop}
\label{pr:ortogonal-of-primitives}
For a graded connected bialgebra of finite type~$H$,
\begin{equation}
P(H')^\perp = \C1 \op H_+^2.
\label{eq:new-number}
\end{equation}
\end{prop}

\begin{proof}
Let $p\in P(H')$. Using~\eqref{eq:nada-sin-ti} we obtain
$\dst{p}{1}=\eta_{H'}(p)=0$. Also, for $a_1,a_2\in H_+$
$$
\dst{p}{a_1a_2} = \dst{p\ox1+1\ox p}{a_1\ox a_2}
= \eta_H(a_2)\dst{p}{a_1} + \eta_H(a_1)\dst{p}{a_2}=0.
$$
Thus $\C1\op H_+^2\subseteq P(H')^\perp$. Use of~\eqref{eq:nada-sin-ti}
again easily gives $(\C1\op H_+^2)^\perp\subseteq P(H')$. Then
$P(H')^\perp\subseteq\C1\op H_+^2$, and therefore $P(H')^\perp =
\C1\op H_+^2$.
\end{proof}

In general $H'\subsetneq H^\circ$. As an example, let the polynomial
algebra $H = \C[Y]$ be endowed with the coproduct associated to
translation,
\begin{equation}
\Dl Y^n = \sum_{k=0}^n\binom{n}{k}Y^k\ox Y^{n-k},
\label{eq:binom}
\end{equation}
which is a homomorphism in view of the Vandermonde identity; this is
the so-called \textit{binomial bialgebra}. Consider the elements
$f^{(n)}$ of~$H^*$ defined by
$$
\dst{f^{(n)}} {Y^m} = \dl_{nm}.
$$
Obviously any $\phi \in H^*$ can be written as
$$
\phi = \sum_{n\ge 0} c_n f^{(n)},
$$
where the complex numbers $c_n$ are given by $c_n=\dst{\phi}{Y^n}$.
Now, write $f:= f^{(1)}$; notice that $f$ is primitive since
$$
\dst{\Dl^t f}{Y^n \ox Y^m} = \dst{f}{Y^n \,Y^m}
= \dst{f}{Y^{n+m}} = \dl_{1n}\dl_{0m} + \dl_{0n}\dl_{1m}
= \dst{f\ox \eta + \eta \ox f}{Y^n \ox Y^m}.
$$
On the other hand,
$$
\dst{f^2}{Y^n} = \dst{f\ox f}{\Dl Y^n}
= \sum_{k=0}^n \binom{n}{k} f(Y^{n-k}) f(Y^k).
$$
Since $f(Y^k)= 0$ unless $k=1$, then $f^2 = 2! f^{(2)}$. A simple
induction entails $f^n = n! f^{(n)}$. Thus, $\phi$ can be written as
$\phi = \sum_{n\ge 0} d_n f^n$ with $d_n=\frac{c_n}{n!}$; so $H^*
\simeq \C[[f]]$, the algebra of formal (exponential, if you wish)
power series in $f$.

It is also rather clear what $\C[Y]'$ is: in terms of the $f^{(n)}$ it
is the divided powers bialgebra, namely the bialgebra with basis
$f^{(n)}$ for $n\ge0$, where the product and coproduct are,
respectively, given by
$$
f^{(n)}f^{(m)} = \binom{n+m}{n}f^{(n+m)}
\sepword{and}
\Dl f^{(n)} = \sum_{k=0}^n f^{(n-k)} \ox f^{(k)}.
$$
We can conclude that $\C[Y]' = \C[f]$.

Consider now $\phi$ in the Sweedler dual $H^\circ$. By definition
there exists some (principal) ideal $I=(p(Y))$, with $p$ a (monic)
polynomial, such that $\phi(I) =0$. Therefore we shall first describe
all the $\phi$ that vanish on a given ideal~$I$. We start with the
case $p(Y) = (Y - \la)^r$ for some $\la \in \C$ and $r\in\N$. Let
$\phi_\la := \sum_{n\ge0} \la^n f^n = \exp(\la f)$. The set $\set{(Y -
\la)^m\: m\ge0}$ is also a basis of $H$. As before, one can consider
the elements $g^{(m)}_\la$ of $H^*$ defined by $\dst{g^{(m)}_\la}{(Y
- \la)^l} =\dl_{lm}$. We are going to prove that $g^{(m)}_\la =
f^{(m)}\, \phi_\la$. Indeed
\begin{align*}
\dst{f^{(m)}\phi_\la}{(Y - \la)^l}
&= \dst{f^{(m)}\phi_\la}{\sum_{k=0}^l \binom{l}{k}(-\la)^{l-k} Y^k} \\
&= \sum_{k=0}^l \binom{l}{k}(-\la)^{l-k}
         \dst{f^{(m)} \ox\phi_\la}{\Dl Y^k}  \\
&= \sum_{k=0}^l \sum_{j=0}^k\binom{l}{k}\binom{k}{j}(-\la)^{l-k}
         \dst{f^{(m)} \ox\phi_\la}{Y^j \ox Y^{k-j}}  \\
&= \sum_{k=0}^l \sum_{j=0}^k\binom{l}{k}\binom{k}{j}(-\la)^{l-k}
          f^{(m)}(Y^j) \phi_\la(Y^{k-j}).
\end{align*}
Since $f^{(m)}(Y^j)$ vanish if $m>j$, it is clear that
$\dst{f^{(m)}\phi_\la}{(Y - \la)^l} =0$ if $m>l$. If $m=l$ only one
term survives and $\dst{f^{(m)}\phi_\la}{(Y - \la)^m} =1$. On the
other hand, if $m<l$ then
\begin{align*}
\dst{f^{(m)}\phi_\la}{(Y - \la)^l}
&= \sum_{k=m}^l \binom{l}{k}\binom{k}{m}(-1)^{l-k} \la^{l-m} \\
&= \frac{\la^{l-m}}{m!} \sum_{k=m}^l \binom{l}{k} (-1)^{l-k}
          k(k-1) \cdots (k-m+1).
\end{align*}
Successive derivatives of the binomial identity give
$$
l(l-1) \cdots (l-m+1) (x -1)^{l-m}
= \sum_{k=m}^l \binom{l}{k} (-1)^{l-k} k(k-1) \cdots (k-m+1) x^{k-m},
$$
therefore
$$
0= \sum_{k=m}^l \binom{l}{k} (-1)^{l-k} k(k-1) \cdots (k-m+1),
$$
and we conclude that $\dst{f^{(m)}\phi_\la}{(Y - \la)^l} = \dl_{lm}$.
Any $\phi \in H^\circ$ can be written as
$$
\phi = \sum_{m\ge0} e_m f^{(m)} \, \phi_\la.
$$
It follows that those $\phi$ satisfying $\dst{\phi}
{\bigl((Y-\la)^r\bigr)} =0$ are of the form
\begin{equation}
\phi = \sum_{m=0}^{r-1} e_m f^{(m)} \,\phi_\la;
\label{eq:finite-exp}
\end{equation}
we can think of them as linear recursive sequences~\cite{Bienhechor}.
In general, $p(Y) = \prod_{i=1}^s (Y - \la_i)^{r_i}$, and the $\phi$
satisfying $\dst{\phi}{(p(Y))} =0$ will be linear combinations of
terms as in~\eqref{eq:finite-exp}. Thus
$$
H^\circ = \biggl\{\sum e_{ij}\, f^{(i)}\, \phi_{\la_j} \:
e_{ij}, \, \la_j \in \C\biggr\}.
$$

Furthermore, since $\phi_\la(Y^n) = \la^n = \phi_\la(Y)^n$, it ensues
that $\phi_\la \in \Hom_\alg(H,\C) = G(H^\circ)$. Conversely, if
$\phi$ is grouplike and $\phi(Y) =\la$, then
$$
\dst{\phi}{Y^2} = \dst{\Dl^t\phi}{Y\ox Y}
= \dst{\phi\ox \phi}{Y\ox Y} = \bigl(\phi(Y)\bigr)^2 = \la^2,
$$
and so on. It follows that $\phi = \phi_\la$, in other words the
grouplike elements of $H^\circ$ are precisely the exponentials
$\phi_\la=\exp(\la f)$, and since $\phi_\la\phi_\mu=\phi_{\la+\mu}$,
we conclude that $G(\C[Y]^\circ) \cong(\C,+)$. In summary, $H^\circ$
can be rewritten as
$$
\C[Y]^\circ = H' \ox G(\C[Y]^\circ).
$$
An interpretation of $\C[Y]^\circ$ as the space of proper rational
functions was given in~\cite{VerdeSta}.

In general, when $H$ is commutative, $H'=\U\bigl(P(H^\circ)\bigr)
=\U(\Der_\eta H)$. This is an instance of the Milnor--Moore
theorem~\cite{MilnorM}, on which we shall dwell a bit in Section~14.
There are no grouplike elements in~$H'$, apart from~$\eta$. The
characters of a graded connected commutative bialgebra~$H$ can be
recovered as the set of grouplike elements in the completion $H^\circ$
of the algebra $H'=\U(\Der_\eta H)$. The sets $\sum_{k\ge m}(\Der_\eta
H)^k$, for $m=1,2,\ldots$ form a basis of neighbourhoods of~0 for a
vector space topology on $\U(\Der_\eta H)$; the grading properties
mean that the Hopf algebra operations are continuous for this
$H_+$-adic topology. An element of the completion of~$\U(\Der_\eta H)$
is a series $\sum_{k\ge0}z_k$ with $z_k\in(\Der_\eta H)^k$ for each
$k$. As $\dl^{m+1}(a)=0$ if $a\in(\ker\eta)^m$, the element
$\exp(\dl)\in H^\circ$ makes sense for each $\dl\in\Der_\eta H$; and
$\exp\dl\in G(H^\circ)$ since
$$
\Dl\exp\dl = \exp\Dl\dl = \exp(\eta\ox\dl+\dl\ox\eta) =
\exp\dl\ox\exp\dl,
$$
by continuity of~$\Dl$. This exponential map is a bijection between
$\Der_\eta H$ and $G(H^\circ)$~\cite{HHfather,Calypso}, with inverse
$$
\log\mu = -\sum_{n=1}^\infty\frac{(\eta-\mu)^n}{n}.
$$
That is, the group $G(H^\circ)$ is a pro-unipotent Lie group, and one
regards the commutative Hopf algebra~$H$ as an algebra of affine
coordinates on that group. As a general paradigm, the dual
of~$\Rr(G)$, where $G$ is a Lie group with Lie algebra~$\g$, is of the
form $\U(\g)\ox\C G$ as a coalgebra; as an algebra, it is the smash
product of $\U(\g)$ and $\C G$. Smash products will be briefly
discussed in Section~9; here they are just a fancy name for the
adjoint action of $G$ on~$\g$.

\marker
Finally we can come back to our introductory remarks in the second
half of Section~1. The whole idea of harmonic analysis is to
``linearize'' the action of a group ---or a Lie algebra--- on a space
$X$ by switching the attention, as it is done in noncommutative
geometry, from $X$ itself to spaces of functions on~$X$, and the
corresponding operator algebras. Now, linear representations of groups
and of Lie algebras can be tensor multiplied, whereas general
representations of algebras cannot. Thus from the standpoint of
representation theory, the main role of the coproduct is to ensure
that the action on $H$-modules propagates to their tensor products. To
wit, if a bialgebra $H$ acts on~$V$ and~$W$, then it will also act
on~$V\ox W$ by
$$
h_\ox\.(v\ox w) = h_{(1)}\.v \ox h_{(2)}\.w,
$$
for all $h\in H, v\in V, w\in W$. In other words, if $\phi_V:H\ox V\to
V,\phi_W:H\ox W\to W$ denote the actions, then
\begin{equation}
\phi_{V\ox W} := (\phi_V\ox\phi_W)(\id\ox\tau\ox\id)(\Dl\ox\id\ox\id).
\label{eq:tensorrep}
\end{equation}
Indeed,
$$
h_\ox\.(k_\ox\.(v\ox w))) = (h_{(1)}k_{(1)})\.v \ox
(h_{(2)}k_{(2)})\.w = (hk)_\ox\.(v\ox w),
$$
and moreover
$$
1_\ox\.(v\ox w) := 1\.v\ox 1\.w = v\ox w
$$
as it should. In this view, the product structure on the module of all
representations of a group comes from the comultiplication:
$g \mapsto g \ox g$, for $g\in G$; in the case of representations of
Lie algebras, where a $\g$-module is the same as a module for the
bialgebra~$\U(\g)$, there is analogously a product. Note that
$V \ox W \not\simeq W \ox V$ in an arbitrary $H$-module category.

\begin{defn}
\label{df:corepresentation}
Dually, we envisage \textit{corepresentations} of coalgebras~$C$. A
right corepresentation or coaction of~$C$ on a vector space $V$ is a
linear map $\ga: V\to V\ox C$ such that $(\id\ox\Dl)\ga=
(\ga\ox\id)\ga$ and $\id=(\id\ox\eta)\ga$. These conditions are
expressed by the commutativity of the diagrams
$$
\vcenter{\hbox{
\xymatrix{
V \ox C \ox C &  V \ox C \ar[l]_(.4){\id\ox\Dl}  \\
V \ox C \ar[u]^{\ga\ox\id} & V,\ar[l]_(.4){\ga} \ar[u]_{\ga} }
\qquad\qquad
\xymatrix{
V  &  V \ox \C \ar@{<->}[l]  \\
V \ar[u]^{\id} \ar[r]^(.3){\ga} & V \ox C, \ar[u]_{\id\ox\eta} }
}}
$$
obtained by the process of reversing arrows from the axioms of a
left representation of an algebra $\phi\: A\ox V \to V$:
$$
\vcenter{\hbox{
\xymatrix{
A \ox A \ox V \ar[r]^(.6){m\ox\id} \ar[d]_{\id\ox\phi}
& A \ox V \ar[d]^{\phi}   \\
A \ox V \ar[r]^{\phi} & V, }
\qquad\qquad
\xymatrix{
V  &  \C \ox V \ar@{<->}[l] \ar[d]^{u\ox\id}  \\
V \ar[u]^{\id}  & A \ox V. \ar[l]_(.6){\phi} }
}}
$$
\end{defn}

We use the convenient notation
$$
\ga(v) = \sum v^{(\uno)}\ox v^{(2)},
$$
with $v^{(\uno)}\in V$, $v^{(2)}\in C$.
So, for instance, the first defining relation becomes
$$
\sum v^{(\uno)(\uno)} \ox v^{(\uno)(2)} \ox v^{(2)}
= \sum v^{(\uno)} \ox v^{(2)}_{(1)} \ox v^{(2)}_{(2)}.
$$

Representations of algebras come from corepresentations of their
predual coalgebras: if $\ga$ is a corepresentation as above, then
$$
h\.v := (\id\ox h)\ga(v) \sepword{or} h\.v := \sum v^{(\uno)} h(v^{(2)})
$$
defines a representation of~$C^*$. Indeed
\begin{align*}
h_1\.(h_2\.v)
&= h_1\.\biggl( \sum v^{(\uno)} h_2(v^{(2)})\biggr)
= \sum v^{(\uno)(\uno)} h_1(v^{(\uno)(2)}) h_2(v^{(2)})
\\
&= \sum v^{(\uno)} h_1(v^{(2)}_{(1)}) h_2(v^{(2)}_{(2)})
=  \sum v^{(\uno)} h_1 h_2(v^{(2)})
= (h_1h_2)\.v.
\end{align*}

Now we use the product structure: if a bialgebra $H$ coacts on~$V$
and~$W$, it coacts on the tensor product $V\ox W$ by
$$
\ga_\ox(v\ox w) = \sum v^{(\uno)}\ox w^{(\uno)}\ox v^{(2)}w^{(2)}, \quad
v\in V,w\in W;
$$
that is
$$
\ga_{V\ox W} := (\id\ox\id\ox m)(\id\ox\tau\ox\id)(\ga_V \ox \ga_W),
$$
in complete parallel to~\eqref{eq:tensorrep}. The required
corepresentation properties are easily checked as well. For instance,
\begin{align*}
\sum &(v\ox w)^{(\uno)(\uno)} \ox (v\ox w)^{(\uno)(2)} \ox
(v\ox w)^{(2)}
\\
&= \sum v^{(\uno)(\uno)} \ox w^{(\uno)(\uno)} \ox
v^{(\uno)(2)} w^{(\uno)(2)} \ox v^{(2)} w^{(2)}
\\
&= \sum \bigl(v^{(\uno)(\uno)} \ox 1 \ox v^{(\uno)(2)} \ox
v^{(2)}\bigr)\, \.\,\bigl(1 \ox w^{(\uno)(\uno)} \ox w^{(\uno)(2)} \ox
w^{(2)}\bigr)
\\
&= \sum \bigl(v^{(\uno)} \ox 1 \ox v^{(2)}_{(1)} \ox
v^{(2)}_{(2)}\bigr)\, \.\,\bigl(1 \ox w^{(\uno)} \ox w^{(2)}_{(1)} \ox
w^{(2)}_{(2)}\bigr)
\\
&= \sum v^{(\uno)} \ox w^{(\uno)} \ox v^{(2)}_{(1)} w^{(2)}_{(1)} \ox
v^{(2)}_{(2)} w^{(2)}_{(2)}
\\
&= \sum (v\ox w)^{(\uno)} \ox (v\ox w)^{(2)}_{(1)} \ox
(v\ox w)^{(2)}_{(2)}.
\end{align*}

These simple observations prove decisive to our reconstruction of the
Connes--Moscovici Hopf algebra in Section~9.

\section{Antipodes}

\begin{defn}
\label{df:hopfalgebra}
A \textit{skewgroup} or \textit{Hopf algebra} is a bialgebra~$H$
together with a (necessarily unique) convolution inverse~$S$ for the
identity map~$\id$. Thus,
$$
\id * S = m (\id \ox S) \Dl = u\eta,  \qquad
S * \id = m (S \ox \id) \Dl = u\eta,
$$
which boils down to the commutativity of the diagram
$$
\vcenter{\hbox{
\xymatrix{
H \ox H  \ar[d]_{\id\ox S} & H \ar[l]_(.4)\Dl \ar[r]^(.4)\Dl
& H \ox H \ar[d]^{S\ox \id}  \\
H \ox H \ar[r]^(.6)m  & H \ar[u]_{u\circ\eta} & H \ox H \ar[l]_(.6)m.}
}}
$$
In terms of elements this means
$$
\sum a_{(1)}Sa_{(2)} = \eta(a)  \sepword{and}
\sum Sa_{(1)}\,a_{(2)} = \eta(a).
$$
The map $S$ is usually called the \textit{antipode} or
\textit{coinverse} of~$H$. The notion of Hopf algebra occurred first
in the work by Hopf in algebraic topology~\cite{Skewclassical}.
\end{defn}

Uniqueness of the antipode can be seen as follows. Let $S,S'$ be two
antipodes on a bialgebra. Then
$$
S'a = S'a_{(1)}\eta(a_{(2)})
= S'a_{(1)}\,a_{(2)(1)}Sa_{(2)(2)}
= S'a_{(1)(1)}\,a_{(1)(2)}Sa_{(2)}
= \eta(a_{(1)})Sa_{(2)} = Sa.
$$
We have used counity in the first equality, and successively the
antipode property for~$S$, coassociativity, the antipode property
for~$S'$ and counity again.

A bialgebra morphism between two Hopf algebras $H,K$ is automatically
a Hopf algebra morphism, i.e., it exchanges the antipodes: $\ell S_H =
S_K\ell$. For that, it is enough to prove that these maps are
one-sided convolution inverses for $\ell$ in $\Hom(H,K)$. Indeed,
since the identity in $\Hom(H,K)$ is $u_K\eta_H$, it is enough to
notice that
\begin{equation}
\ell S_H * \ell = \ell(S_H * \id_H) = \ell u_H\eta_H = u_K\eta_H
= u_K\eta_K\ell = (\id_K * S_K)\ell = \ell * S_K\ell;
\label{eq:more-than-meets-the-eye}
\end{equation}
associativity of convolution then yields
$$
S_K\ell = u_K\eta_H * S_K\ell = \ell S_H * \ell * S_K\ell
= \ell S_H * u_K\eta_H = \ell S_H.
$$

The antipode is an antimultiplicative and anticomultiplicative map of
$H$. This means
$$
Sm = m\tau(S \ox S), \quad S1 = 1 \sepword{and}
\tau\Dl S = (S \ox S)\Dl, \quad \eta S = S.
$$
The first relation, evaluated on $a \ox b$, becomes the familiar
antihomomorphism property $S(ab) = SbSa$. For the proof of it we
refer to~\cite[Lemma~1.26]{Polaris}; the second relation is a similar
exercise.

A grouplike element $g$ of a Hopf algebra $H$ is always invertible
with $g^{-1} = Sg$. Indeed,
$$
1 = u\eta(g) = m(\id\ox S)\Dl g = gSg = m(S\ox\id )\Dl g = (Sg)g.
$$

Often the antipode $S$ is \textit{involutive} (thus invertible); that
is, $S^2 = \id_H$.

\begin{prop}
\label{pr:involutiveness-of-S} 
$S$ is involutive if and only if
\begin{equation}
Sa_{(2)}a_{(1)} = a_{(2)}Sa_{(1)} = \eta(a).
\label{eq:coinv-invol}%
\end{equation}
\end{prop}

\begin{proof}
The relation $Sa_{(2)}\,a_{(1)} = \eta(a)$ implies
$S*S^2a = Sa_{(1)}\,S^2a_{(2)} = S\bigl(Sa_{(2)}\,a_{(1)}\bigr) =
S\eta(a) =\eta(a)$. Hence $S*S^2=S^2*S=u\eta$, which entails
$S^2=\id$. Reciprocally, if $S^2 = \id$, then
$$
Sa_{(2)}\,a_{(1)} = Sa_{(2)}\,S^2a_{(1)}
= S\bigl(Sa_{(1)}\,a_{(2)}\bigr) = S\eta(a) = \eta(a),
$$
and analogously $a_{(2)}Sa_{(1)} = \eta(a)$.
\end{proof}

In other words, the coinverse $S$ is involutive when it is still the
inverse of the identity for the new operation obtained from~$*$ by
twisting it with the flip map. Property~\eqref{eq:coinv-invol} clearly
holds true for Hopf algebras that are commutative or cocommutative.
The antipode for a commutative and cocommutative Hopf algebra is an
involutive bialgebra morphism.

A \textit{Hopf subalgebra} of $H$ is a vector subspace $E$ that is a
Hopf algebra with the restrictions of the antipode, product, coproduct
and so on, the inclusion $E \hookto H$ being a bialgebra
morphism.  A Hopf ideal is a biideal $J$ such that $SJ\subseteq J$;
the quotient $H/J$ gives a Hopf algebra.

\medskip

A glance at the defining conditions for the antipode shows that,
if~$H$ is a Hopf algebra, then $H^{\mathrm{copp}}$ is also a Hopf
algebra with the same antipode. However, the bialgebras $H^\opp$,
$H^\cop$ are Hopf algebras if and only if $S$ is invertible, and then
the antipode is precisely $S^{-1}$~\cite[Section 1.2.4]{KlimykS}. We
prove this for $H^\cop$. Assume $S^{-1}$ exists. It will be an algebra
antihomomorphism. Hence
$$
S^{-1}a_{(2)}\,a_{(1)} = S^{-1}(Sa_{(1)}\,a_{(2)}) = S^{-1}\eta(a)1 =
\eta(a)1;
$$
similarly $m(\id\ox S^{-1})\Dl_\cop=u\eta$.  Reciprocally, if
$H^\cop$ has an antipode $S'$, then
$$
SS' * Sa = \sum SS'a_{(1)}\,Sa_{(2)}
= \sum S(a_{(2)}\,S'a_{(1)}) = S\bigl(\eta(a) 1\bigr) = u\eta(a).
$$
Therefore $SS' = \id$, so $S'$ is the inverse of~$S$ under composition.

\medskip

The duality nonsense of Section~4 is immediately lifted to the Hopf
algebra category. The dual of $(H,m,u,\Dl,\eta,S)$ becomes $(H^\circ,
\Dl^{\!t}, \eta^t, m^t,u^t,S^t)$; to equations~\eqref{eq:nada-sin-ti}
we add the condition
$$
\dst{S_Ha}{f} = \dst{a}{S_Kf},
$$
which is actually redundant.

As for examples, the bialgebra $\C G$ is a Hopf algebra with coinverse
$Sx=x^{-1}$; the bialgebra $\Rr(G)$ is a Hopf algebra with coinverse
\begin{equation}
Sf(x) = f(x^{-1}).
\label{eq:paradigm}
\end{equation}

\marker
The antipode is a powerful inversion machine. If~$H$ is a Hopf
algebra, both algebra homomorphisms on an algebra~$A$ and coalgebra
morphisms on a coalgebra~$C$ can be inverted in the convolution
algebra. In fact, by going back to
reinspect~\eqref{eq:more-than-meets-the-eye}, we see already that in
the case of algebra morphisms $fS$ is a left inverse for $f$; also,
using~\eqref{eq:algprop},
$$
f*fS = f(\id*S) = fu_H\eta_H = u_A\eta_H.
$$
In the case of coalgebra maps in~\eqref{eq:more-than-meets-the-eye} we
see that $Sf$ is a right inverse for $f$; and similarly it is a left
inverse. Recall that $\Hom_\alg(H,A)$ denotes the convolution monoid
of multiplicative morphisms on an algebra $A$ with neutral
element~$u_A\eta_H$. The `catch' is that $fS$ does not belong to
$\Hom_\alg(H,A)$ in general; as remarked before, it will if~$A$ is
commutative (a moment's reflection reassures us that although $S$ is
antimultiplicative, $fS$ indeed is multiplicative). In that case
$\Hom_\alg(H,A)$ becomes a group (an abelian one if~$H$ is
cocommutative). In particular, that is the case of the set
$\Hom_\alg(H,\C)$ of multiplicative functions or characters, and
of~$\Hom_\alg(H,H)$, when $H$ is commutative. (One also gets a group
when $H$ is cocommutative, and considers the coalgebra morphisms
from~$H$ to a coalgebra~$C$.)

In the first of the examples given in Section~2, the group of real
characters of~$\Rr(G)$ reconstructs $G$ in its full topological glory:
this is Tannaka--Kre\u{\i}n duality ---see~\cite[Ch.~1]{Polaris}
and~\cite{HHfather}. Characters of connected graded Hopf
algebras have special properties, exhaustively studied in~\cite{ABS}.

\medskip

A pillar of wisdom in Hopf algebra theory: connected bialgebras
are \textit{always} Hopf. There are at least two `grand strategies' to
produce the antipode $S: H \to H$ in a connected bialgebra. One is to
exploit its very definition as the convolution inverse of the identity
in $H$, via a geometric series:
$$
S := \id^{*-1} = S_G := (u \eta - (u \eta -\id))^{*-1}:= u \eta + (u
\eta -\id) + (u \eta -\id)^{*2} + \cdots
$$

\begin{prop}
\label{pr:geometricseries}
The geometric series expansion of $Sa$ with $a \in H_n$ has at
most $n+1$ terms.
\end{prop}

\begin{proof}
If $a \in H_0$ the claim holds since $(u \eta - \id)1 = 0$. It also
holds in ${H_+}_1$ because these elements are primitive. Assume
that the statement holds for elements in ${H_+}_{n-1}$, and let
$a\in{H_+}_n$; then
\begin{align*}
(u \eta -\id)^{*(n+1)}(a)
&= (u \eta -\id) * (u \eta -\id)^{*n}(a)
\\
&= m [(u \eta -\id) \ox (u \eta -\id)^{*n}] \Dl a
\\
&= m [(u \eta -\id) \ox (u \eta -\id)^{*n}]
( a \ox 1 + 1 \ox a + \Dl'a ).
\end{align*}
The first two terms vanish because $(u \eta - \id)1 = 0$. By the
induction hypothesis each of the summands of the third term are also
zero.
\end{proof}

In view of~\eqref{eq:eso-era}, on $H_+$ we can write for $k\geq1$
\begin{equation}
(u\eta -\id)^{*k+1} = (-1)^{k+1} m^k {\Dl'}^k.
\label{eq:quasicojo-formula}%
\end{equation}
There is then a fully explicit expression for the antipode for
elements without degree~0 components (recall $S1=1$), in terms of the
product and the reduced coproduct
\begin{equation}
S_G = -\id + \sum^\infty_{k=1}(-1)^{k+1} m^k {\Dl'}^k.
\label{eq:cojo-formula}%
\end{equation}
All this was remarked in~\cite{Ananke}.

\medskip

The second canonical way to show that a connected bialgebra is a Hopf
algebra amounts to take advantage of the equation $m (S\ox\id) \Dl a =
0$ for $a\in H_+$. For $a\in{H_+}_n$ and $n \geq 1$, one ushers in the
recursive formula:
\begin{equation}
S_B(a) := - a - \sum S_Ba'_{(1)}a'_{(2)},
\label{eq:Bogol-recur}%
\end{equation}
using the notation in~\eqref{eq:gr-coprod-trunc}.

\begin{prop}
\label{pr:bogo-equals-series}
If $H$ is a connected bialgebra, then $S_Ga = S_Ba$.
\end{prop}

\begin{proof}
The statement holds, by a direct check, if $a\in{H_+}_1$. Assume that
$S_Gb=S_Bb$ whenever $b\in{H_+}_n$ , and let $a\in{H_+}_{n+1}$. Then
\begin{align*}
S_Ga &= (u\eta-\id)a + \sum_{i=1}^n(u\eta-\id)^{*i} * (u\eta-\id)a
\\
&= -a + m \biggl( \sum_{i=1}^n (u\eta-\id)^{*i} \ox (u\eta-\id)
\biggr) \Dl a
\\
&= -a + m \sum_{i=1}^n (u\eta-\id)^{*i} \ox (u\eta-\id)(a\ox 1 + 1\ox
a + \Dl'a)
\\
&= -a + \sum_{i=1}^n (u\eta-\id)^{*i} a'_{(1)} \, (u\eta-\id)a'_{(2)}
= -a - \sum_{i=1}^n (u\eta-\id)^{*i} a'_{(1)} \, a'_{(2)}
\\
&=  -a - S_Ba'_{(1)} \,a'_{(2)} = S_Ba,
\end{align*}
where the penultimate equality uses the induction hypothesis.
\end{proof}

Taking into account the alternative expression
$$
S_Ga = (u\eta-\id)a +
\sum_{i=1}^n (u\eta-\id) * (u\eta-\id)^{*i}a,
$$
it follows that the twin formula
$$
S'_Ba := - a - \sum a'_{(1)} S'_Ba'_{(2)}
$$
provides as well a formula for the antipode. The subindices $B$
in~$S_B,S'_B$ reminds us that this second strategy corresponds
precisely to Bogoliubov's formula for renormalization in quantum field
theory.

\medskip

The geometric series leading to the antipode can be generalized as
follows. Consider, for~$H$ a connected bialgebra and~$A$ an arbitrary
algebra, the set of elements $f\in\Hom(H,A)$ fulfilling $f(1)=1$. They
form a convolution monoid, with neutral element~$u_A\eta_H$, as
$f*g(1)=1$, if both $f$ and~$g$ are of this type. Moreover we can
repeat the inversion procedure:
$$
f^{*-1} := (u\eta - (u\eta-f))^{*-1}
:= u\eta + (u\eta-f) + (u\eta-f)^{*2} + \cdots
$$
Then $f^{*-1}(1)=1$, and for any $a\in H_+$
$$
(u\eta-f)^{*k+1}a = (-1)^{k+1}m^k(f\ox\cdots\ox f){\Dl'}^k a,
$$
vanish for $a\in H_n$ when $n\le k$. Therefore the series stops. The
convolution monoid then becomes a group; of which, as we already know,
the set $\Hom_\alg(H,A)$ of multiplicative morphisms is a subgroup
when $A$ is commutative.

\marker
The foregoing indicates that, associated to any connected bialgebra,
there is a natural filtering ---we call it \textit{depth}--- where the
order $\dl(a)$ of a generator $a\in H_+$ is $k>0$ when $H$ is the
smallest integer such that $a\in\ker({\Dl'}^k p)$; we then say $a$ is
$k$-primitive. Whenever $a\in H_n$, it holds $\dl(a)\le n$. On account
of~\eqref{eq:eso-era}, $\ker(U_k) :=\ker({\Dl'}^k p)=
(\Dl^k)^{-1}(H^{k+1}_R)$. In other words, for those bialgebras, depth
is the coradical filtering. Antipodes are automatically filtered.

On $H_+$ one has $(u\eta -\id)^{*k+1}a=0$, in view
of~\eqref{eq:quasicojo-formula}, if ${\Dl'}^k a=0$; but of course the
converse is not true.

\begin{defn}
\label{df:quasiprimitive}
For $H$ commutative, we say $a$ indecomposable is
\textit{quasiprimitive} if
\begin{equation}
Sa = -a; \sepword{this implies} (u\eta - \id)^{*2} a = 0.
\label{eq:dirty-trick}
\end{equation}
\end{defn}
Obviously primitive elements are quasiprimitive. In Section~15 we give
examples of elements that are quasiprimitive, but not primitive, and
discuss their relevance. We will also see that a basis for~$H$ can be
found such that, for any element~$b$ of it, $Sb=-b$ or $b$.

\medskip

Following~\cite{godgift}, we conclude this section with a relatively
short proof of that~$\dl$ indeed is a filtering, in the framework of
connected graded bialgebras of finite type, comprising our main
examples. Let us denote in the reminder of this section
$H_k=\C1\op\ker{\Dl'}^k =\set{a\in H:\dl(a)\le k}$, for $k\ge0$. This
is certainly a linear filtering.

\begin{prop}
\label{pr:Hperp}
Let $H'_+ := \bigoplus_{l\ge1}{H^{(l)}}^* \subset H'$. Then
$H_k^\perp = {H'_+}^{k+1}$ ---thus $H_k = ({H'_+}^{k+1})^\perp$.
\end{prop}

\begin{proof}
(Derivations are typical elements of~$1^\perp$.) The assertion is true
and obvious for $k=0$. Consider $(\id-u\eta)^t$, the projection
on~$H'_+$ with kernel~$1_{H'}$. Let $\kappa_1,\dots,\kappa_{k+1}\in
H'_+$ and $a\in H_k$. We have
\begin{align*}
\dst{\kappa_1\cdots\kappa_{k+1}}{a}
&=\dst{\kappa_1\ox\cdots\ox\kappa_{k+1}} {\Dl^k a}
\\
&= \dst{{(\id-u\eta)^t}^{\ox(k+1)}
(\kappa_1\ox\cdots\ox\kappa_{k+1})}{\Dl^k a}
\\
&= \dst{\kappa_1\ox\cdots\ox\kappa_{k+1}}{U_k a} = 0.
\end{align*}
Therefore ${H'_+}^{k+1}\subseteq H_k^\perp$. Now, let
$a\in({H'_+}^{k+1})^\perp$. Then again
$$
\dst{\kappa_1\ox\cdots\ox\kappa_{k+1}}{U_k a} =
\dst{\kappa_1\cdots\kappa_{k+1}}{a} = 0.
$$
Therefore $U_k a\in H_+^{\ox(k+1)}\cap\bigl({H'_+}^{\ox(k+1)})^\perp =
(0)$. Consequently $({H'_+}^{k+1})^\perp\subseteq H_k$, which
implies $H_k^\perp\subseteq {H'_+}^{k+1}$.
\end{proof}

Given any augmented graded algebra~$A$, connected in the sense that
$A^{(0)}\simeq0$ with $A_+=\bigoplus_{i\ge1}A^{(i)}$, it is not too
hard to see that
$$
\sum_{i+j>k} A_+^i\ox A_+^j
= \bigcap_{l+m=k} \bigl(A_+^{l+1}\ox A + A\ox A_+^{m+1}\bigr).
$$

As a corollary:

\begin{prop}
\label{pr:coalgebra-filtering}
The filtering by the~$H_k$ is a coalgebra filtering.
\end{prop}

\begin{proof}
We have
\begin{align*}
\sum_{l+m=k}H_l\ox H_m &=
\sum_{l+m=k}({H'_+}^{l+1})^\perp\ox({H'_+}^{m+1})^\perp
= \sum_{l+m=k}({H'_+}^{l+1}\ox H' + H'\ox{H'_+}^{m+1})^\perp
\\
&=\biggl(\bigcap_{l+m=k}\bigl({H'_+}^{l+1}\ox
H'+H'\ox{H'_+}^{m+1}\bigr)\biggr)^\perp =
\biggl(\sum_{i+j>k}{H'_+}^i\ox{H'_+}^j\biggr)^\perp.
\end{align*}
Now, let $a\in H_k$ and $\kappa_1\in{H'_+}^i,\kappa_2\in{H'_+}^j,
i+j>k$. Then $\kappa_1\kappa_2\in{H'_+}^{k+1}=H_k^\perp$. Thus
$$
0 = \dst{a}{\kappa_1\kappa_2} = \dst{\Dl a}{\kappa_1\ox\kappa_2}.
$$
This means $\Dl a\in \sum_{l+m=k}H_l\ox H_m$, and so $H_k$ is a coalgebra
filtering as claimed.
\end{proof}

\begin{prop}
\label{pr:calgebra-filtering}
The filtering by the~$H_k$ is an algebra filtering.
\end{prop}

\begin{proof}
If $a\in H_k$, then ${\Dl'}^k a=0$, and the counit property entails
${\Dl'}^{k-1}a\in P(H)^{\ox k}$. Let now $a\in H_l,b\in H_m$ with
$$
{\Dl'}^{l-1}a=
\sum a'_{(1)}\ox a'_{(2)}\ox\cdots\ox a'_{(l)}, \quad
{\Dl'}^{m-1}b=
\sum b'_{(1)}\ox\cdots\ox b'_{(m)}.
$$
By~\eqref{eq:eso-era} once again
\begin{align}
{\Dl'}^{l+m-1}(ab) &= U_{l+m-1}(ab) =
(\id-u\eta)^{\ox\ l+m}(\Dl^{l+m-1}a\Dl^{l+m-1}b)
\notag \\
&= \sum\sum_{\sigma \in S_{l+m,l}}
a'_{(\sigma(1))} \ox\cdots\ox a'_{(\sigma(l))} \ox b'_{(\sigma(l+1))}
\ox\cdots\ox b'_{(\sigma(l+m))}
\notag \\
&=: \sum\sum_{\sigma \in S_{l+m,l}}\sigma\.
(a'_{(1)}\ox\cdots\ox a'_{(l)}\ox b'_{(1)}\ox\cdots\ox b'_{(m)}),
\label{eq:powerful-minister}
\end{align}
where $S_{n,p}$ denotes the set of $(p,n-p)$-shuffles; a
$(p,q)$-shuffle is an element of the group of permutations $S_{p+q}$
of $\{1, 2,\dots p+q\}$ in which $\sigma(1)<\sigma(2)<\cdots
<\sigma(p)$ and $\sigma(p+1)<\cdots<\sigma(p+q)$.

Equation~\eqref{eq:powerful-minister} implies that
${\Dl'}^{l+m-1}(ab)\in P(H)^{\ox\ l+m}$, and hence ${\Dl'}^{l+m}(ab)
=0$.
\end{proof}

We retain as well the following piece of information:
$$
{\Dl'}^{n-1}(p_1\dots p_n) =
\sum_{\sigma\in S_n}p_{\sigma(1)}\ox\cdots\ox p_{\sigma(n)},
$$
for primitive elements $p_1,\dots,p_n$. The proof is by induction.
Obviously we have in particular:
$$
\Dl'(p_1p_2) = p_1\ox p_2 + p_2\ox p_1.
$$
Assuming
$$
{\Dl'}^{n-2}(p_1\dots p_{n-1}) =
\sum_{\sigma\in S_{n-1}}p_{\sigma(1)}\ox\cdots\ox p_{\sigma(n-1)},
$$
equation~\eqref{eq:powerful-minister} tells us that
$$
{\Dl'}^{n-1}(p_1\cdots p_n) = \sum_{\tau\in S_{n,1}}\sum_{\sigma\in
S_{n-1}} \tau\.(p_{\sigma(1)}\ox\cdots\ox p_{\sigma(n-1)}\ox p_n) =
\sum_{\sigma\in S_n}p_{\sigma(1)}\ox\cdots\ox p_{\sigma(n)}.
$$

\section{Symmetric Algebras}

In our second example in Section~2, we know $SX = -X$ for $X\in\g$,
since $X$ is primitive, and $S(XY)= YX$; but the concrete expression
in terms of a basis given a priori can be quite involved. Consider,
however, the universal enveloping algebra corresponding to the trivial
Lie algebra structure on~$V$. This is clearly a commutative and
cocommutative Hopf algebra, nothing else than the familiar symmetric,
free commutative algebra or, in physics, boson algebra~$B(V)$ over $V$
of quantum field theory. Given~$V$, a complex vector space, $B(V)$ is
defined as $\bigoplus_{n=0}^\infty V^{\vee n}$, where $V^{\vee n}$ is
the complex vector space algebraically generated by the symmetric
products
$$
v_1 \vee v_2 \vee\cdots\vee v_n
:= \frac{1}{n!} \sum_{\sigma \in S_n}
v_{\sigma(1)} \ox v_{\sigma(2)} \ox\cdots\ox v_{\sigma(n)},
$$
with $V^{\vee 0} = \C$ by convention. On $B(V)$ a coproduct and counit
are defined respectively by $\Dl v := v\ox1 + 1\ox v$ and $\eta(v) = 0$,
for $v\in V$, and then extended by the homomorphism property. In general,
$$
\Dl(a^1 \vee a^2) = \sum a^1_{(1)}\vee a^2_{(1)} \ox a^1_{(2)}\vee
a^2_{(2)},
$$
for $a^1,a^2\in B(V)$.  Another formula is:
\begin{equation}
\Dl(v_1 \vee v_2 \vee \cdots \vee v_n) = \sum_I U(I) \ox U(I^c),
\label{eq:shuffles}
\end{equation}
with sum over all the subsets $I\subseteq\{v_1, v_2, \dots v_n\},
I^c = \{v_1, v_2, \dots v_n\} \setminus I$, and $U(I)$ denotes the
$\vee$ product of the elements in $I$. Thus, if $u = v_1\vee v_2
\vee\dots\vee v_n$, then
$$
\Dl'u = \sum_{p=1}^{n-1}  \sum_{\sigma \in S_{n,p}}
v_{\sigma(1)} \vyv v_{\sigma(p)} \ox v_{\sigma(p+1)}
\vyv v_{\sigma(n)}.
$$
Here we are practically repeating the calculations at the end of the
previous section. In the particularly simple case when $V$ is one
dimensional, the Hopf algebra $B(V)\simeq\U(\C)$ is just the binomial
bialgebra. Finally $Sa=-a$ for elements of $B(V)$ of odd degree and
$Sa=a$ for even elements; the reader can amuse himself checking
how~\eqref{eq:cojo-formula} works here.

It should be clear that an element of $B(V)$ is primitive iff it
belongs to $V$. For a direct proof, let $\{e_i\}$ be a basis for $V$,
any $a \in B(V)$ can be represented as
$$
a = \a 1 + \sum_{k\ge 1} \sum_{i_1 \leq \dots \leq i_k}
\a_{i_1, \dots, i_k} e_{i_1} \vee\dots\vee e_{i_k}.
$$
for some complex numbers $\a_{i_1, \dots, i_k}$.  Now, if $a$ is
primitive then $\a = \eta(a) = 0$, and
$$
\Dl a = a\ox1 + 1\ox a
= \sum \a_{i_1, \dots, i_k} (e_{i_1} \vee\dots\vee e_{i_k} \ox 1
+ 1 \ox e_{i_1} \vee\dots\vee e_{i_k}).
$$
but also
\begin{align*}
\Dl a
&= \sum_k \sum \a_{i_1, \dots, i_k}
(e_{i_1} \ox 1 + 1 \ox e_{i_1}) \cdots
(e_{i_k} \ox 1 + 1 \ox e_{i_k}) \\
&= \sum_k \sum \a_{i_1, \dots, i_k}
(e_{i_1} \vee\dots\vee e_{i_k} \ox 1
+ 1 \ox e_{i_1} \vee\dots\vee e_{i_k} + \sum_J e_J \ox e_{J^c}),
\end{align*}
where the last sum runs over all nonempty subsets
$J = \{i_1,\dots,i_l\}$ of $[k]:= \set{1,\dots,k}$ with at most
$k-1$ elements, $J^c$ denotes the complement of $J$ in $[k]$, and
$e_J := e_{i_1} \vee\dots\vee e_{i_l}$. A comparison of the two
expressions gives
$$
\sum_{k \geq 2} \sum_{i_1\leq\dots\leq i_k}
\a_{i_1, \dots, i_k} \sum_{\emptyset \neq
J \subsetneq [k]} e_J \ox e_{J^c} = 0.
$$
This forces $\a_{i_1, \dots, i_k} = 0$ for $k \geq 2$, so $a \in V$.

\begin{defn}
\label{df:primitively-generated}
A Hopf algebra $H$ is said to be \textit{primitively generated} when
the smallest subalgebra of $H$ containing all its primitive elements
is~$H$ itself. Cocommutativity of~$H$ is clearly a necessary condition
for that. It is plain that~$B(V)$ is primitively generated.
\end{defn}

\marker
Symmetric algebras have the following universal property: any morphism
of (graded, if you wish) vector spaces $\psi: V \to H$, where
$V^{(0)}=(0)$ and~$H$ is a unital graded connected
\textit{commutative} algebra, extends uniquely to a unital graded
algebra homomorphism $B\psi: B(V) \to H$; one says $B(V)$ is free
over~$V$. Note the isomorphism $B(V\op{\tilde V})\simeq B(V)\ox
B(\tilde V)$, implemented by $V\op\tilde V\ni(v, \tilde v)\mapsto v\ox
1+1\ox\tilde v$, extended by linear combinations of products. In this
perspective, the comultiplication~$\Dl$ on~$B(V)$ is the extension
$Bd$ induced by the diagonal map $d\: V\mapsto V\op V$.

\begin{prop}
\label{pr:hopf-morphism}
Let $H$ be a graded connected commutative Hopf algebra, and denote by
$\psi_H$ the inclusion $P(H) \hookto H$. The universal property
gives a graded algebra map $B\psi_H\: B(P(H)) \to H$. This is a
graded Hopf algebra morphism.
\end{prop}

\begin{proof}
The coproduct $\Dl: H \to H \ox H$ gives a linear map $P\Dl: P(H) \to
P(H \ox H)$ such that $\Dl \psi_H = \psi_{H\ox H}\, P\Dl$. The
universal property of symmetric algebras then gives us maps
$B\psi_H,B\psi_{H\ox H}$ and $BP\Dl$ such that $\Dl B\psi_H =
B\psi_{H\ox H}B P\Dl$. By~\eqref{eq:tensor-of-primitives},
$$
B\bigl(P(H\ox H)\bigr) \cong B\bigl(P(H) \op P(H)\bigr) \cong
B\bigl(P(H)\bigr) \ox B\bigl(P(H)\bigr); \quad B\psi_{H\ox H}\simeq
B\psi_H\ox B\psi_H.
$$
Therefore $B P\Dl$ is identified to the coproduct of~$B(P(H))$ and
$$
\Dl B\psi_H = (B\psi_{H}\ox B\psi_{H})\Dl_{B(P(H))}.
$$
In a similar fashion it is seen that $B\psi_H$ respects counity and
coinverse.
\end{proof}

If $V \subseteq P(H)$, then the subalgebra $\C[V]$ generated by $V$ is
a primitively generated Hopf subalgebra of~$H$. The inclusion $\iota :
V \hookto H$ induces a morphism of graded algebras, indeed of Hopf
algebras, $B\iota: B(V) \to H$ whose range is $\C[V]$. In particular,
$\C[P(H)]$ is the largest primitively generated subalgebra of $H$ and
$B\psi_H: B(P(H)) \to H$ is a morphism of Hopf algebras \textit{onto}
$\C[P(H)]$; therefore $H$ is primitively generated only if the
underlying algebra is generated by $P(H)$. All these statements follow
from the previous proof and the simple observation that $v\in V$ is
primitive in both $\C[V]$ and $B(V)$.

\begin{prop}
\label{pr:mophism-is-injective}
The morphism $B\psi_H : B(P(H)) \to H$ is injective. In particular, if
$H$ is primitively generated, then $H = B(P(H))$.
\end{prop}

\begin{proof}
The vector space $P(H)$ can be regarded as the direct limit of all its
finite-dimensional subspaces $V$, hence $B\bigl(P(H)\bigr)$ is the
direct limit of all $B(V)$ ---tensor products commute with direct
limits--- and the map $B\psi_H$ is injective if and only if its
restriction to each of the algebras $B(V)$ is injective. Thus it is
enough to prove the proposition for $V$ a finite-dimensional subspace
of~$P(H)$.

We do that by induction on $m = \dim V$. For $m = 0$, there is nothing
to prove. Assume, then, that the claim holds for all subspaces $W$
of~$P(H)$ with $\dim W \leq m-1$, and let $V$ be a subspace of~$P(H)$
with $\dim V = m$. Let $W$ be an $(m-1)$-dimensional subspace of~$V$,
then $B \psi : B(W) \to \C[W]$ is an isomorphism of Hopf algebras. Take
$Y \in V \setminus \C[W]$ homogeneous of minimal degree, so $V = W \op
\C Y$. Then $B(V) \cong B(W) \ox B(\C Y) \cong \C[W]\ox B(\C Y)$. Now,
$\C[V]\cong\C[W\op\C Y]\cong\C[W]\ox \C[Y]$. The remarks prior to the
statement imply that
$B\bigl(P(\C[Y])\bigr)\cong\C\bigl[P(\C[Y])\bigr]$, and since $Y\in
V\subset P(H)$, clearly $P(\C[Y]) = \C Y$, so $B(\C Y)\cong\C[Y]$. It
follows that
$$
B(V) \cong \C[W] \ox B(\C Y) \cong \C[W] \ox \C[Y] \cong \C[V],
$$
which completes the induction.
\end{proof}

A similar argument allows us to take up some unfinished business: the
converse of Proposition~\ref{pr:halfway}.

\begin{thm}
\label{pr:the-other-half}
The relation
\begin{equation}
P(H)\cap H_+^2 = (0),
\label{eq:prim-indecomp}
\end{equation}
or equivalently $q_H: P(H)\to Q(H)$ is one-to-one, holds if $H$ is
commutative.
\end{thm}

\begin{proof}
Suppose $H$ commutative has a unique generator. Then a moment's
reflection shows that $H$ has to be the binomial algebra, and then
$P(H)\simeq Q(H)$. Suppose now that the proposition is proved for
algebras with less than or equal to $n$ generators. Let the elements
$a_1,\dots,a_{n+1}$ be such that their images by the canonical
projection $H\to Q(H)$ form a basis of~$Q(H)$. Leave out the element
of highest degree among them, and consider $B$, the Hopf subalgebra of
$H$ generated by the other $n$ elements. We form $\C\ox_B H=H/B_+ H$.
This is seen to be a Hopf algebra with one generator. Moreover,
$H\simeq B\ox\C\ox_B H$. Then~\eqref{eq:tensor-of-primitives} implies
$P(H)= P(B)\op P(\C\ox_B H)$. By the induction hypothesis, $q_H=q_B\op
q_{\C\ox_B H}$ is injective. The proposition is then proved for
finitely generated Hopf algebras. Hopf algebras of finite type are
clearly direct limits of finitely generated Hopf algebras; and direct
limits preserve the functors $P,Q$ and injective maps;
so~\eqref{eq:prim-indecomp} holds true for Hopf algebras of finite
type as well. Finally, by the result of~\cite{MilnorM} already invoked
in Section~3, the proposition holds for all commutative connected
graded Hopf algebras without exception.
\end{proof}

\medskip

Yet another commutative and cocommutative Hopf algebra is the
polynomial algebra $\C[Y_1,Y_2,Y_3,\dots]$ with coproduct given by
$$
\Dl Y_n = \sum_{j=0}^n Y_j \ox Y_{n-j};
$$
to be studied later, using symmetric algebra theory. We denote this
algebra by~$\Hl$, and call it the \textit{ladder Hopf algebra}; it is
related to nested Feynman graphs and to the Hopf algebra of rooted
trees, considered in Section~12.

\marker
To finish, we come back to the universal property of~$B(V)$ as an
algebra. For that, the coalgebra structure previously considered does
not come into play, and one could ask whether other coproducts are
available, that also give a Hopf algebra structure on~$B(V)$; the
answer is yes, but the one exhibited here is the only one that makes
$B(V)$ a \textit{graded} Hopf algebra. More to the point: dually,
given graded cocommutative coalgebras and linear maps into vector
spaces $\vf \: H\to V$, there is a universal `cofree cocommutative'
coalgebra, say $Q_\cocom(V)$, `cogenerated' by $V$, together with a
unique coalgebra map $Q_\cocom\vf \: H \to Q_\cocom(V)$ restricting
to~$\vf$ by the projection $Q_\cocom(V) \to V$. Now, $Q_\cocom(V)$ is
nothing but the vector space underlying $B(V)$ with the coalgebra
structure already given here; and there is a unique algebra structure
on $Q_\cocom(V)$ making it a graded Hopf algebra.

In many contexts it is important to waive the (co)commutativity
requisite on~$B$ and on~$Q_\cocom$; this respectively leads to the
tensor or free graded algebra (already touched upon in Example~2.2)
and to the cotensor or cofree graded coalgebra, for which the most
natural Hopf algebra structure is Ree's~\cite{Anticipating} and
Chen's~\cite{Catecismo} \textit{shuffle} algebra. In some detail:
already in Section~2 we touched upon the tensor algebra $T(V)$, which
is made into a (graded, connected, cocommutative) Hopf algebra by the
algebra morphisms
$$
\Dl 1 = 1 \ox 1, \quad  \Dl X := X \ox 1 + 1 \ox X; \qquad
\eta(X) := 0, \quad \eta(1) = 1,
$$
for $X \in V$. It is useful to think of the basis of~$V$ as an
alphabet, whose elements are letters, and to write the tensor product
simply as a \textit{concatenation}. Then we have, as before,
\begin{equation}
\Dl (X_1 X_2 \cdots X_n) = \sum_{p=0}^{n}\sum_{\sigma\in S_{n,p}}
X_{\sigma(1)} \cdots X_{\sigma(p)} \ox
X_{\sigma(p+1)} \cdots X_{\sigma(n)},
\label{eq:shuffles-strike-again}
\end{equation}
with $S_{n,p}$ denoting the $(p,n-p)$-shuffles. It should be clear
that primitive elements of $T(V)$ constitute a free Lie
algebra~\cite{Reutenauer}. For instance, if there are two letters
$\{X,Y\}$ in~$V$, a basis is made of
$$
X,\  Y,\  [X,Y],\  [X,[X,Y]],\  [[X,Y],Y],\  [X,[X,[X,Y]]],\ 
[X,[[X,Y],Y]],
$$
and so on. The graded dual of~$T(V)$ ---which at least in the finite
type case is isomorphic to~$T(V)$ as a graded vector space--- is the
shuffle Hopf algebra~$Sh(V)$, a most interesting object that appears
in many contexts. We find its (necessarily associative and
commutative) product, denoted by $\pitchfork$, by
dualizing~\eqref{eq:shuffles-strike-again}: for the words
$u = d_1d_2\cdots d_p$ and $v = d_{p+1}d_{p+2}\cdots d_n$ with each
$d_i$ in~$V$, their product is
$$
u \pitchfork v
= \sum_{\sigma \in S_{n,p}} d_{\sigma(1)} \cdots d_{\sigma(n)}.
$$
For instance:
$$
1 \pitchfork d = d \pitchfork 1 = d;\
d_1 \pitchfork d_2 = d_2 \pitchfork d_1 = d_1d_2 + d_2d_1; \ 
d_1d_2 \pitchfork d_3 = d_3d_1d_2 + d_1d_3d_2 + d_1d_2d_3.
$$
There is a recursive definition given by
$$
d_1d_2\cdots d_p \pitchfork d_{p+1}d_{p+2}\cdots d_n =
d_1\bigl(d_2\cdots d_p \pitchfork d_{p+1}d_{p+2}\cdots d_n\bigr)
+ d_{p+1}\bigl(d_1d_2\cdots d_p \pitchfork d_{p+2}\cdots d_n\bigr).
$$
We have indeed
$$
\dst{u \pitchfork v}{X_1X_2\cdots X_n}
= \dst{u \ox v}{\Dl(X_1X_2\cdots X_n)}.
$$
On can easily show, using the universal property, that $Sh(V)$ is the
unique commutative graded Hopf algebra structure on the cofree graded
subalgebra~$Q(V)$~\cite{MaestrosSutiles}. The coproduct on the latter
is given by
$$
\dst{\Dl u}{X_1X_2\cdots X_l \ox Y_1Y_2\cdots Y_k} =
\dst{u}{X_1X_2\cdots X_lY_1Y_2\cdots Y_k},
$$
that is, \textit{deconcatenation}:
$$
\Dl u = \sum_{u=vw} v \ox w.
$$

\newpage

\section*{Part II: The Fa\`a di Bruno Bialgebras}
\addcontentsline{toc}{section}{\protect\numberline{II}\enspace
THE FA\`A di BRUNO BIALGEBRAS}

\section{Partitions, Bell polynomials and the \bfb\ algebras}

In a ``moral" sense, epitomized by Example~1 of Section~2, the
discussion around equation~\eqref{eq:paradigm} and the consideration
of~$G(H^\circ)$ in Section~4, commutative skewgroups are equivalent to
groups. Now, we would like to deal with relatively complicated groups,
like diffeomorphism groups. Variants of Hopf algebra theory
generalizing categories of (noncompact in general) topological groups
do exist~\cite{KustermansV}. It is still unclear how to handle
diffeomorphism groups globally, though: the interplay between topology
and algebra becomes too delicate~\cite{Leslie}. We settle for a
``perturbative version". Locally, one can think of orientation
preserving diffeomorphisms of~$\R$ leaving a fixed point as given by
formal power series like
\begin{equation}
f(t) = \sum_{n=0}^\infty f_n\frac{t^n}{n!},
\label{eq:expl-ser}%
\end{equation}
with $f_{0}=0,f_1>0$. (Orientation preserving diffeomorphisms of the
circle are just periodic ones of~$\R$, and locally there is no
difference.) Among the functions on the group $G$ of diffeomorphisms
the coordinate functions on jets
$$
a_n(f):=f_n=f^{(n)}(0), \quad n\ge1,
$$
single out themselves. The product of two diffeomorphisms is
expressed by series composition; to which just like in Example~1 we
expect to correspond a coproduct for the $a_n$ elements.  As the $a_n$
are representative, it is unlikely that this reasoning will lead us
astray.

Let us then work out $f\circ g$, for $f$ as above and $g$ of the same
form with coordinates $g_n=a_n(g)$.  This old problem is solved with
the help of the \textit{Bell polynomials}.  The (partial, exponential)
Bell polynomials $B_{n,k}(x_1,\dots,x_{n+1-k})$ for $n\ge1,\,1\le k\le
n$ are defined by the series expansion:
\begin{equation}
\exp\biggl(u\sum_{m\geq1}x_m\frac{t^m}{m!}\biggr) = 1 +
\sum_{n\geq1}\frac{t^n}{n!}\biggl[\sum_{k=1}^nu^k
B_{n,k}(x_1,\dots,x_{n+1-k})\biggr].
\label{eq:ring-a-Bell}%
\end{equation}
The first Bell polynomials are readily found:
$B_{n,1} = x_n$,
$B_{n,n} = x_1^n$;
$B_{2,1} = x_2$,
$B_{3,2} = 3x_1x_2$,
$B_{4,2} = 3x_2^2 + 4x_1x_3$,
$B_{4,3} = 6x_1^2x_2$,
$B_{5,2} = 10x_2x_3 + 5x_1x_4$,
$B_{5,3} = 10x_1^2x_3 + 15x_1x_2^2$,
$B_{5,4} = 10x_1^3 x_2$, \dots\
Each Bell polynomial $B_{n,k}$ is homogeneous of degree~$k$.

We claim the following: if $h(t)=f\circ g(t)$, then
\begin{equation}
\qquad h_n = \sum_{k=1}^nf_kB_{n,k}(g_1,\dots,g_{n+1-k}).
\label{eq:wonder-series}%
\end{equation}
One can actually allow here for $f_{0}\ne0$; then the same result
holds, together with $h_0 = f_0$.

The proof is quite easy: it is clear that the $h_n$ are linear in the
$f_n$:
$$
h_n = \sum_{k=1}^nf_kA_{n,k}(g).
$$
In order to determine the $A_{n,k}$ we choose the series
$f(t)=e^{ut}$.  This entails $f_k=u^k$ and
$$
h = f\circ g = e^{ug}
= \exp\biggl(u\sum_{m\geq1}g_m\frac{t^m}{m!}\biggr)
= 1 + \sum_{n\geq1}\frac{t^n}{n!}\biggl[\sum_{k=1}^nu^k
B_{n,k}(g_1,\dots,g_{n+1-k})\biggr],
$$
from which at once there follows
\begin{equation}
A_{n,k}(g) = B_{n,k}(g_1,\dots,g_{n+1-k}).
\label{eq:faa-diB}%
\end{equation}
So $h_1 = f_1g_1, h_2 = f_1g_2 + f_2g_1^2, h_3 = f_1g_3 + 3f_2g_1g_2 +
f_3g_1^3$, and so on. Francesco Fa\`a di Bruno (beatified in 1988)
gave a formula equivalent to~\eqref{eq:faa-diB} about a hundred and
fifty years ago~\cite{FaadiBruno}. There are older instances of it:
see our comment at the beginning of Section~16. Lest the reader think
we are dealing with purely formal results, we remark that, if $g$ is
real analytic on an open interval~$I_1$ of the real line and takes
values on another open interval~$I_2$, on which $f$ is analytic as
well, then $f\circ g$ given by~\eqref{eq:wonder-series} is analytic
on~$I_1$, too~\cite{Krank}. To obtain explicit formulae for the
$B_{n,k}$, one can proceed directly from the definition. We shall only
need the multinomial identity
$$
(\b_1+\b_2+\dots+\b_r)^k =
\sum_{c_1+c_2+\dots+c_r=k}\frac{k!}{c_1!c_2!\cdots c_r!}
\b_1^{c_1}\b_2^{c_2}\cdots\b_r^{c_r},
$$
that generalizes directly the binomial identity. To see that, note
that if $c_1+c_2+\dots +c_r=k$ then the multilinear coefficient
$\binom{k}{c_1,c_2,\dots,c_r}$ of
$\b_1^{c_1}\b_2^{c_2}\cdots\b_r^{c_r}$ is the number of ordered
$r$-tuples of mutually disjoint subsets $(S_1,S_2,\dots S_r)$ with
$|S_i|=c_i$ whose union is $\{1,2,\dots,k\}$. Then, since $S_1$ can be
filled in $\binom{k}{c_1}$ different ways, and once $S_1$ is filled,
$S_2$ can be filled in $\binom{n-n_1}{n_2}$ ways, and so on:
$$
\binom{k}{\row c1r}
= \binom{k}{c_1} \binom{k-c_1}{c_2} \binom{k-c_1-c_2}{c_3}
\cdots \binom{c_r}{c_r}
= \frac{k!}{c_1!\,c_2!\cdots c_r!}\,.
$$

Now, we can expand
\begin{align*}
\sum_{k\geq0} \frac{u^k}{k!}
\biggl(\sum_{m\geq1}x_m\frac{t^m}{m!}\biggr)^k
&=\sum_{k\geq0}\frac{u^k}{k!}
\biggl(\;\sum_{c_1+c_2+\cdots=k}\frac{k!}{c_1!c_2!\cdots}
\bigl(x_1t\bigr)^{c_1}\bigl(x_2t^2/2!\bigr)^{c_2}\cdots\biggr)
\\
&=\sum_{c_1,c_2,\dots\geq0}
\frac{u^{c_1+c_2+c_3+\cdots}\,t^{c_1+2c_2+3c_3+\cdots}}
{c_1!c_2!\cdots(1!)^{c_1}(2!)^{c_2}\cdots}x_1^{c_1}x_2^{c_2}\cdots.
\end{align*}
Taking the coefficients of $u^kt^n/n!$ in view
of~\eqref{eq:ring-a-Bell}, it follows that
\begin{equation}
B_{n,k}(x_1,\dots,x_{n+1-k}) =
\sum\frac{n!}{c_1!c_2!c_3!\cdots(1!)^{c_1}(2!)^{c_2}(3!)^{c_3}\cdots}
x_1^{c_1}x_2^{c_2}x_3^{c_3}\cdots
\label{eq:wonder-formula}%
\end{equation}
where the sum is over the sets of positive integers $c_1,c_2,\dots,c_n$
such that $c_1+c_2+c_3+\dots+c_n=k$ and $c_1+2c_2+3c_3+\dots+nc_n=n$.

It is convenient to introduce the notations
$$
\binom{n}{\la;k} := \frac{n!}
{\la_1!\la_2!\cdots\la_n!(1!)^{\la_1}(2!)^{\la_2}\dots(n!)^{\la_n}},
$$
where $\la$ is the sequence $(1,1,\dots;2,2,\dots;\dots)$, better
written $(1^{\la_1},2^{\la_2},3^{\la_3}\dots)$, of~$\la_1$~1's,
$\la_2$~2's and so on; and $x^\la:=x_1^{\la_1}x_2^{\la_2}x_3^{\la_3}
\cdots$; obviously some of the $\la_i$ may vanish, and certainly
$\la_n$ is at most~1.

\marker
The coefficients $\binom{n}{\la;k}$ also have a combinatorial meaning.
We have already employed the concept of \textit{partition} of a
\textit{set}: if~$S$ is a finite set, with $|S|=n$, a partition
$\{\row A1k\}$ is a collection of $k\le n$ nonempty, pairwise disjoint
subsets of $S$, called \textit{blocks}, whose union is~$S$. It is
often convenient to think of~$S$ as of $[n]:= \{1,2,\dots,n\}$.
Suppose that in a partition of $[n]$ into $k$ blocks there are $\la_1$
singletons, $\la_2$ two-element subsets, and so on, thereby precisely
$\la_1+\la_2+\la_3+\dots=k$ and $\la_1+2\la_2+3\la_3 +\dots=n$;
sometimes~$k$ is called the length of the partition and~$n$ its
weight. We just saw that the number of \textit{ordered} $\la_1,\dots,
\la_r$-tuples of subsets partitioning $[n]$ is
$$
\frac{n!}{(1!)^{\la_1}(2!)^{\la_2}\dots(r!)^{\la_r}}.
$$
Making the necessary permutations, we conclude that $[n]$ possesses
$\binom{n}{\la;k}$ partitions of class~$\la$. Also
$$
B_{n,k}(1,\dots,1) = \sum_\la\binom{n}{\la;k} = |\Pi_{n,k}|,
$$
with $\Pi_{n,k}$ standing for the set of all partitions of
$[n]$ into $k$ subsets; the $|\Pi_{n,k}|$ are the
so-called Stirling numbers of the second kind.

Later, it will be convenient to consider partitions of an
\textit{integer} $n$, a concept that should not be confused with
partitions of the set $[n]$. A partition of~$n$ is a sequence of
positive integers $(f_1,f_2,\dots,f_k)$ such that $f_1\ge f_2\ge
f_3\ge\dots$ and $\sum_{i=1}^kf_i=n$. The number of partitions of~$n$
is denoted $p(n)$. Now, consider a partition~$\pi$ of~$[n]$ of type
$\la(\pi)=(1^{\la_1},2^{\la_2}, 3^{\la_3}\dots)$, and let~$m$ be the
largest number for which $\la_m$ does not vanish; we put
$f_1=f_2=\cdots=f_{\la_m}=m$; then we take for $f_{\la_m+1}$ the
largest number such that $\la_{f_{\la_m+1}}$ among the remaining
$\la$'s does not vanish, and so on. The procedure can be inverted, and
it is clear that partitions of~$n$ can be indexed by the sequence
$(f_1,f_2,\dots,f_k)$ of their definition, or by $\la$. The number of
partitions $\pi$ of~$[n]$ for which $\la(\pi)$ represents a partition
of~$n$ is precisely $\binom{n}{\la;k}$. To take a simple example, let
$n=4$. There are the following partitions of~4: (4)$\equiv(4^1)$,
corresponding to one partition of~$[4]$; (3,1)$\equiv(1^1,3^1)$,
corresponding to four partitions of~$[4]$; (2,2)$\equiv (2^2)$,
corresponding to three partitions of~$[4]$; (2,1,1)$\equiv(1^2,2^1)$,
corresponding to six partitions of~$[4]$; (1,1,1,1)$\equiv(1^4)$,
corresponding to one partition of~$[4]$. In all $p(4)=5$, whereas the
number $B_4$ of partitions of~$[4]$ is~15. We have $p(5)=7$, whereas
the number of partitions of~$[5]$ is $B_5=52$.

\medskip

The results~\eqref{eq:wonder-series} and~\eqref{eq:wonder-formula} are
so important that to record a slightly differently worded argument to
recover them will do no harm: let $f,g,h$ be power series as above;
notice that
$$
h(t) = \sum_{k=0}^\infty \frac{f_k}{k!}
\biggl(\sum_{l=1}^\infty \frac{g_l}{l!}t^l\biggr)^k.
$$
To compute the $n$-th coefficient $h_n$ of $h(t)$ we only need to
consider the partial sum up to $k=n$, since the other products contain
powers of $t$ higher than $n$, on account of $g_0 = 0$. Then
for~$n\ge1$, from Cauchy's product formula
$$
h_n = \sum_{k=1}^n \frac{f_k}{k!}
\sum_{l_1 + \cdots +l_k = n, \, 1 \le l_i}
\frac{n! \, g_{l_1} \cdots g_{l_k}}{l_1! \cdots l_k!}.
$$
Now, each sum $l_1 + \cdots + l_k = n$ can be written in the form
$\a_1 +2\a_2 + \cdots + n\a_n =n$ for a unique vector $(\row{\a}1n)$,
satisfying $\a_1 + \cdots +\a_n = k$; and since there are
$k!/\a_1!\cdots \a_n!$ ways to order the $g_l$ of each term, it again
follows that
$$
h_n = \sum_{k=1}^n \frac{f_k}{k!}
\sum_\a \frac{n!k!}{\a_1! \cdots \a_n!}
\frac{g_1^{\a_1} \cdots g_n^{\a_n}}
{(1!)^{\a_1} \, (2!)^{\a_2} \cdots (n!)^{\a_n}}
= \sum_{k=1}^n f_k \, B_{n,k}(g_1,\dots,g_{n+1-k}),
$$
where the second sum runs over the vectors fulfilling the conditions
just mentioned.

The (complete, exponential) Bell polynomials $Y_n$ are defined
by~$Y_0=1$ and
$$
Y_n(x_1,\dots,x_n)=\sum_{k=1}^nB_{n,k}(x_1,\dots,x_{n+1-k});
$$
that is, taking $u=1$ in~\eqref{eq:ring-a-Bell}:
$$
\exp\biggl(\sum_{m\geq1}x_m\frac{t^m}{m!}\biggr) =
\sum_{n\geq0}\frac{t^n}{n!}Y_n(x_1,\dots,x_n);
$$
and the \textit{Bell numbers} by $B_n:=Y_n(1,\dots,1)$.  It is clear
that the Bell numbers coincide with cardinality of the set~$\Pi_n$ of
all partitions of~$\{1,2,\dots,n\}$; a fact already registered in our
notation.  Some amusing properties can be now derived: in
formula~\eqref{eq:ring-a-Bell}, take
$u=x_m=1$.  We get
$$
\sum_{n=0}^\infty\frac{B_nt^n}{n!}= \exp(e^t-1) \sepword{or}
\log\sum_{n=0}^\infty\frac{B_nt^n}{n!}= e^t-1.
$$
Differentiating $n+1$ times both sides, it ensues the recurrence
relation:
$$
B_{n+1} = \sum_{k=0}^n\binom{n}{k}B_k.
$$
The same relation is of course established by combinatorial arguments.
For consider the partitions of $[n]$ as starting point to determine
the number of partitions of $[n+1]$.  The number $n+1$ must lie in a
block of size $k+1$ with $0\leq k\leq n$, and there are $\binom{n}{k}$
choices for such a block.  Once the block is chosen, the remaining
$n-k$ numbers can be partitioned in $B_{n-k}$ ways.  Summing over $k$,
one sees
$$
B_{n+1} = \sum_{k=0}^n\binom{n}{k}B_{n-k},
$$
which is the same formula.

\marker
The analytical smoke has cleared, and now we put the paradigm
of~\eqref{eq:coprod-repfns} in Example~1 to work.  Taking our cue from
Example~1, we have the right to expect that the formula
$$
\Dl a_n(g,f) := a_n(f\circ g) = a_{n(1)}(f)\,a_{n(2)}(g)
$$
give rise to a coproduct for the polynomial algebra generated by the
coordinates $a_n$ for $n$ going from 1 to $\infty$. In other words,
\begin{equation}
\Dl a_n = \sum_{k=1}^n\sum_{\la}\binom{n}{\la;k}a^{\la} \ox a_k =
\sum_{k=1}^nB_{n,k}(a_1,\dots,a_{n+1-k})\ox a_k
\label{eq:peace-and-strength}
\end{equation}
must yield a bialgebra, which is commutative but clearly not
cocommutative. The quirk in defining $\Dl a_n(g,f)$ by~$a_n(f\circ g)$
rather than by~$a_n(g\circ f)$ owns to the wish of having the linear
part of the coproduct standing on the right of the $\ox$ sign, and not
on the left.

The first few values for the coproduct will be
\begin{align}
\Dl a_1 &= a_1 \ox a_1,
\notag \\
\Dl a_2 &= a_2 \ox a_1 + a^2_1 \ox a_2,
\notag \\
\Dl a_3 &= a_3 \ox a_1 + a_1^3 \ox a_3 + 3a_2a_1 \ox a_2,
\label{eq:deltaex}%
\\
\Dl a_4 &= a_4 \ox a_1 + a_1^4 \ox a_4 + 6a_2a_1^2 \ox a_3
+ (3a_2^2 + 4a_3a_1) \ox a_2,
\notag \\
\Dl a_5 &= a_5 \ox a_1 + a_1^5 \ox a_5 + 10a_2a_1^3 \ox a_4
+ (10a_3a_1^2 + 15a_2^2a_1) \ox a_3 + (5a_4a_1 + 10a_2a_3) \ox a_2.
\notag 
\end{align}

The Hopf algebras of rooted trees and of Feynman graphs introduced in
QFT by Kreimer and Connes~\cite{ConnesKrTrees, ConnesKrRHI}, as well
as the Connes--Moscovici Hopf algebra~\cite{ConnesMHopf}, are of the
same general type, with a linear part of the coproduct standing on the
right of the $\ox$ sign and a polynomial one on the left. The kinship
is also manifest in that, as conclusively shown in~\cite{ConnesKrRHII}
---see also~\cite{SierraNevada}--- one can use Feynman diagrams to
obtain formulae of the type of~\eqref{eq:peace-and-strength}. In what
follows, we shall clarify the relations, and show how all those
bialgebras fit in the framework and machinery of incidence bialgebras.
But before doing that, we plan to explore at leisure the obtained
bialgebra and some of its applications.

We do not have a connected Hopf algebra here.  Indeed, since $a_1$ is
\textit{grouplike}, it ought to be invertible, with inverse $Sa_1$.
Besides, if $f^{(-1)}$ denotes the reciprocal series of~$f$, then,
according to the paradigm followed, $S$ should be given
by~\eqref{eq:paradigm}:
$$
Sa_1 = a_1(f^{(-1)}) = a_1^{-1}(f).
$$
To obtain a connected Hopf algebra it is necessary to set $a_1 = 1$;
in other words, to consider only formal power
series~\eqref{eq:expl-ser} of the form $f(t) = t + \sum_{n\geq 2}
f_n\,t^n/n!$. The resulting graded connected bialgebra is hereinafter
denoted $\F$ and called the \bfb\ algebra (terminology due to Joni and
Rota~\cite{JoniR}); the degree is given by $\#(a_n)=n-1$, with the
degree of a product given by definition as the sum of the degrees of
the factors. If $G_1$ is the subgroup of diffeomorphisms of $\R$ such
that $f(0)=0$ and $df(0)=\id$, we could denote $\F$
by~$\Rr^\cop(G_1)$. The coproduct formula is accordingly simplified as
follows:
$$
\Dl a_n = \sum_{k=1}^n\sum_{\la}\binom{n}{\la;k}
a_2^{\la_2}a_3^{\la_3}\cdots \ox a_k
= \sum_{k=1}^nB_{n,k}(1,\dots,a_{n+1-k})\ox a_k.
$$

\medskip

Now we go for the antipode in~$\F$. Formula~\eqref{eq:cojo-formula}
applies, and in this context reduces to
\begin{equation}
Sa_n = - a_n + \sum_{j=2}^{n-1} (-1)^j \sum_{1 < k_{j-1} < \cdots <
k_1 < n} B_{n,k_1} \, B_{k_1,k_2} \, \cdots \, B_{k_{j-2},k_{j-1}} \,
a_{k_{j-1}}.
\label{eq:fdb-ant}
\end{equation}
In~\eqref{eq:fdb-ant} the arguments of the Bell polynomials have been
suppressed for concision.  In particular,
\begin{align*}
Sa_2 &= -a_2, \\
Sa_3 &= -a_3 + 3a_2^2,  \\
Sa_4 &= -a_4 - 15a_2^3 + 10a_2a_3.
\end{align*}
Let us give the details in computing $Sa_5$:
\begin{align}
Sa_5 &= -a_5 + (B_{5,2}a_2 + B_{5,3}a_3 + B_{5,4}a_4)
-(B_{5,4}B_{4,3}a_3 + B_{5,4}B_{4,2}a_2 + B_{5,3}B_{3,2}a_2)
\notag \\
& \qquad + B_{5,4}B_{4,3}B_{3,2}a_2
\notag \\
&= -a_5 + 15a_2a_4 + 10a_3^2 + 25a_2^2a_3
-(130a_2^2a_3 + 75a_2^4) + 180a_2^4
\notag \\
&= -a_5 + 15a_2a_4 + 10a_3^2 -105a_2^2a_3 + 105a_2^4.
\label{eq:trust-but-verify}
\end{align}
The computation using instead $S_B$ runs as follows
\begin{align*}
Sa_5 &= -a_5 -10a_2 S_Ba_4 - (10a_3 + 15a_2^2) S_Ba_3
-(5a_4 + 10a_2a_3)S_Ba_2 \\
&= -a_5 + 10a_2(a_4 + 6a_2 S_Ba_3 + (3a_2^2 +4a_3)S_Ba_2)
+ (10a_3 +15a_2^2)(a_3 + 3a_2 S_Ba_2) \\
& \quad + 5a_2a_4 + 10a_2^2a_3 \\
&= -a_5 + 10a_2a_4 - 60a_2^2(a_3 + 3a_2 S_Ba_2) - 30a_2^4
-40a_2^2a_3 + 10a_3^2 -30a_2^2a_3 + 15a_2^2a_3  \\
& \quad - 45a_2^4 + 5a_2a_4 + 10a_2^2a_3 \\
&= -a_5 + 10a_2a_4 - 60a_2^2a_3 + 180a_2^4 - 30a_2^4
-40a_2^2a_3 + 10a_3^2 -30a_2^2a_3 + 15a_2^2a_3 \\
& \quad - 45a_2^4 + 5a_2a_4 + 10a_2^2a_3 \\
&= -a_5 + 15a_2a_4 + 10a_3^2 -105a_2^2a_3 + 105a_2^4.
\end{align*}

Note that in both procedures there are the same cancellations,
although the expansions do not coincide term-by-term. However, since
this \bfb\ algebra (vaguely) looks of the same general type as a Hopf
algebra of Feynman graphs, this is a case where we would expect a
formula \`a la Zimmermann, that is, without cancellations, to compute
the antipode. Such a formula indeed exists, as mentioned in the
introduction~\cite{Precursors}. It leads to:
$$
Sa_n = \sum_{k=1}^{n-1}(-1)^kB_{n-1+k,k}(0,a_2,a_3 \dots).
$$
The elegance of this equation is immediately appealing. Using a
standard identity of the Bell polynomials it can be further simplified:
\begin{equation}
Sa_n = \sum_{k=1}^{n-1}(-1)^k(n-1+k)\cdots n\,
B_{n-1,k}\Bigl(\frac{a_2}{2},\frac{a_3}{3}, \dots\Bigr).
\label{eq:reverted}
\end{equation}
For instance,~\eqref{eq:trust-but-verify} is recovered at once with no
cancellations, the coincidence of~\eqref{eq:fdb-ant} and the last
formula actually gives nonstandard identities for Bell polynomials.

\marker
We cannot omit a description, however brief, of the graded dual~$\F'$.
Consider the primitive elements $a_n'\in\F'$ given by $\dst{a_n'}{a_m} =
\dl_{nm}$, then also
$$
\dst{a_n'}{a_pa_q} =\dst{\Dl a_n'}{a_p \ox a_q}
= \dst{a_n'\ox1+1\ox a_n'}{a_p \ox a_q} = 0,
$$
if $p\geq 1$ and $q\geq 1$ ---the $a_n'$ kill nontrivial products of the
$a_q$ generators. On the other hand,
\begin{align*}
\dst{a_n'a_m'}{a_q}
&= \dst{a_n'\ox a_m'}{\Dl a_q} \\
&= \sum_{k=1}^q \dst{a_n'\ox a_m'}
{B_{q,k}(1,a_2,\dots,a_{q+1-k}) \ox a_k}  \\
&= \sum_{k=1}^q \dst{a_n'}{B_{q,k}(1,a_2,\dots,a_{q+1-k})}
\dst{a_m'}{a_k} \\
&=  \dst{a_n'}{B_{q,m}(1,a_2,\dots,a_{q+1-m})}.
\end{align*}
The polynomial $B_{q,m}$ is homogeneous of degree $m$, and the only
monomial in it giving a nonvanishing contribution is the
$x_1^{m-1}x_n$ term. Its coefficient is $\binom{q}{\la;m}$ with $\la$
the sequence $(1^{m-1},0, \dots,n^1,0,\dots)$, satisfying
$q=\la_1+2\la_2+\cdots+q\la_q=m-1+n$. Thus
$$
\dst{a_n'a_m'}{a_q} =
\begin{cases}
\binom{m+n-1}{n} & \text{if } q = m+n-1
\\
0 & \text{otherwise}.
\end{cases}
$$
On the other hand,
$$
\Dl(a_q a_r) = a_q a_r \ox 1 + 1\ox a_q a_r + a_q \ox a_r +
a_r \ox a_q + R,
$$
where $R$ is either vanishing or a sum of terms of the form $b\ox c$
with $b$ or $c$ a monomial in $a_2, a_3, \dots$ of degree greater
than~1. Therefore
$$
\dst{a_n'a_m'}{a_q a_r}  = \dst{a_n' \ox a_m'}{\Dl(a_q a_r)}
=\begin{cases}
1 &\text{if } n = q \neq m = r \text{ or } n = r \neq m = q,
\\
2 &\text{if } m = n = q = r,
\\
0 &\text{otherwise}.
\end{cases}
$$
Furthermore, it is clear that all the terms of the coproduct of three
or more $a$'s are the tensor product of two monomials where at least
one of them is of order greater than~1. So
$\dst{a'_n a'_m}{a_{q_1}a_{q_2}a_{q_3}\cdots}=0$. All together
gives
$$
a'_n a'_m
= \binom{m-1+n}{n} a'_{n+m-1} + \bigl(1 + \dl_{nm}\bigr)(a_n a_m)'.
$$
In particular,
$$
[a'_n, a'_m] := a'_n a'_m - a'_m a'_n
= (m-n) \frac{(n+m-1)!}{n!m!} a_{n+m-1}'.
$$
Therefore, taking $b'_n := (n+1)!a'_{n+1}$, we get the simpler looking
\begin{equation}
[b'_n, b'_m] = (m-n)b'_{n+m}.
\label{eq:formula-after-all}
\end{equation}
The Milnor--Moore theorem implies that $\F'$ is isomorphic to the
universal enveloping algebra of the Lie algebra defined by the last
equation. The algebra $\F'$ can be realized by vector
fields~\cite{ConnesMHopf}. As we saw in Chapter~4, $\F^\circ$ is
bigger and contains grouplike elements $f,g\ldots$ with~$g*f=f\circ
g$.

\smallskip

We shall come back to the structure of the \bfb\ algebra in
Section~12. Also, primitivity in the \bfb\ algebra is thoroughly
examined in Section~15.

\section{Working with the \bfb\ Hopf algebra}

The \bfb\ formulae~\eqref{eq:ring-a-Bell},~\eqref{eq:wonder-series}
and~\eqref{eq:wonder-formula}, and the algebra $\F$ are ubiquitous in
quantum field theory.  We give a couple of examples and then we turn
to a famous example in a combinatorial-algebraic context.

\begin{xmpl}
Consider charged fermions in an external field.  The complex space
$\H$ of classical solutions of the corresponding Dirac equation is
graded by $E_+ - E_-$, where $E_+,E_-$ project on the particle and
antiparticle subspaces; we write $\H^\pm := E_\pm\H$.  Operators on
$\H = \H_+ \op \H^-$ can be presented in block form,
$$
A = \twobytwo{A_{++}}{A_{+-}}{A_{-+}}{A_{--}}.
$$
In particular, a unitary operator
$$
S = \twobytwo{S_{++}}{S_{+-}}{S_{-+}}{S_{--}}
$$
corresponds to a classical scattering matrix if and only if
\begin{align*}
S_{++}S_{++}^\7 + S_{+-}S_{+-}^\7 & = E_+, \quad S_{--}S_{--}^\7 +
S_{-+}S_{-+}^\7 = E_-, \notag \\
S_{++}S_{-+}^\7 + S_{+-}S_{--}^\7 &= S_{--}S_{+-}^\7 +
S_{-+}S_{++}^\7 = 0.
\end{align*}
By the Shale--Stinespring theorem~\cite[Ch.~6]{Polaris}, this operator
is implemented in Fock space iff $S_{+-},S_{-+}$ are
Hilbert--Schmidt.  Assuming this is the case, the quantum scattering
matrix $\Sf$ is obtained through the spin representation~\cite{Rhea}
and it can be proved that (the square of the absolute value of) the
vacuum persistence amplitude is of the form
$$
|\vacpersamp|^2 = |\dst{\vacin}{\Sf\vacin}|^2 = \det(1 -
S_{+-}S_{+-}^\7).
$$
The determinant exists because of the Shale--Stinespring condition.
With $A$ of trace class and of norm less than~$1$ one can use the
development
$$
\det(1 - A) = \exp(\Tr\log(1 - A)) =: \exp\biggl( -\sum_{k=1}^\infty
\frac{\sigma_k}{k} \biggr),
$$
with $\sigma_k:=\Tr A^k$. We may want to reorganize this series, for
instance to expand the exact result for $\dst{\vacin}{\Sf\vacin}$ in
terms of coupling constants. Say
$$
1 + \sum_{n=1}^\infty b_n
:= \exp\biggl( -\sum_{k=1}^\infty \frac{\sigma_k}{k} \biggr).
$$
Our formula \eqref{eq:ring-a-Bell} gives
\begin{equation}
b_n = \frac{1}{n!}\sum_{k=1}^n(-1)^k
B_{n,k}(\sigma_1,\sigma_2,2\sigma_3,\ldots,(n-k)!\sigma_{n+1-k});
\label{eq:appalling}
\end{equation}
so, for example,
$$
b_1 = -\sigma_1,  \quad  b_2 = \thalf(\sigma_1^2 - \sigma_2),  \quad
b_3 = -\tfrac{1}{6} (\sigma_1^3 - 3\sigma_1\sigma_2 + 2\sigma_3),  \dots
$$
One finds~\cite{SalamM} that
$$
b_n = \frac{(-1)^n}{n!}
\det \begin{pmatrix}
\sigma_1 & n-1 & 0 & \dots & 0                   \\
\sigma_2 & \sigma_1 & n-2 & \dots & 0               \\
\vdots & \vdots & \vdots & \ddots & \vdots    \\
\sigma_{n-1} & \sigma_{n-2} & \sigma_{n-3} & \dots & 1 \\
\sigma_n & \sigma_{n-1} & \sigma_{n-2} & \dots & \sigma_1 \end{pmatrix}.
$$
\end{xmpl}

\begin{xmpl}
The first generating functional in QFT is usually defined by the
expression
$$
Z[j] = 1 + \sum_{n=1}^\infty\frac{1}{n!}\biggl(\frac{i}{\hbar}\biggr)^n
\int d^4x_1\cdots\int d^4x_n\,G(x_1,\dots,x_n)
j(x_1)\cdots j(x_n).
$$
This is often taken to be a formal expression, but actually it makes
sense perturbatively if by the Green function $G(x_1,\dots,x_n)$ we
understand a renormalized chronological product \`a la
Epstein--Glaser. The second generating functional is defined:
$$
W[j]:= -i\hbar\log Z[j].
$$
We assert that
$$
W[j] =
\sum_{m=1}^\infty\frac{1}{m!}\biggl(\frac{i}{\hbar}\biggr)^{m-1}
\int d^4x_1\cdots\int d^4x_m\,G_c(x_1,\dots,x_m)\,
j(x_1)\cdots j(x_m);
$$
where $G_c$ denotes the \textit{connected} Green
functions~\cite[Ch.~5]{IZ}:
$$
G(x_1,\dots,x_n) := \sum_{k=1}^n\sum_{\la\in\Pi_{n,k}}G^{\la_1}_c(x_1)
G^{\la_2}_c(x_1,x_2)\cdots
$$
with $(1^{\la_1},2^{\la_2},\dots)$ as before a partition of $[n]$ in
$k$ blocks.  To prove our assertion directly, one can use the \bfb\
formula
$$
\exp\biggl(\sum_{m\geq1}\frac{x_m}{m!}\biggr) =
\sum_{n\geq1}\frac{1}{n!}\biggl[\sum_{k=1}^n\sum_{\la_1,\la_2\dots}
\frac{n!}{\la_1!\la_2!\la_3!\cdots(1!)^{\la_1}(2!)^{\la_2}(3!)^{\la_3}
\cdots} x_1^{\la_1}x_2^{\la_2}x_3^{\la_3}\cdots\biggr]
$$
Therefore
\begin{eqnarray*}
\exp\bigl(\frac{i}{\hbar}W[j]\bigr) &=&
\sum^\infty_{c_1=0}\frac{1}{c_1!}\biggl[\frac{i}{\hbar}\int d^4x_1\,
G_c(x_1)j(x_1)\biggr]^{c_1}
\\
&& \x\sum^\infty_{c_2=0}\frac{1}{c_2!}\biggl[
\biggl(\frac{i}{\hbar}\biggr)^2
\frac{1}{2!}\int d^4x_1\,d^4x_2G_c(x_1,x_2)j(x_1)j(x_2)\biggr]^{c_2}
\\
&& \x\dots,
\end{eqnarray*}
which is precisely $Z[j]$.

The proof is rather spectacular; but only a simple counting principle,
well known in combinatorics~\cite[Section~5.1]{Stanley2}, is involved
here. The same principle plays a role in the cluster expansion of the
$\Sf$-matrix~\cite[Ch.~4]{RedBook}: the sum of operators associated
with Wick diagrams is equal to the normally ordered exponential of the
sum of operators associated with the connected Wick diagrams.

The most important application in QFT concerns the
\textit{renormalization group}; but this would take us too far afield.
See our comments in Section~12.
\end{xmpl}

\begin{xmpl}
In~\eqref{eq:appalling} we might need to solve for the $\sigma$'s
instead of the~$b$'s. Notice that to study diffeomorphisms as formal
power series of the type $f(t)=t+\sum_{n\geq2}a_n(f){t^n/n!}$ it is
not indispensable to use the coordinate functions~$a_n$. Instead, one
can use for example the new set of coordinates $d_n:=a_{n+1}/(n+1)!$,
for $n\ge1$. Consider then
$$
1 + \sum_{n=1}^\infty d_n
:= \exp\biggl(\sum_{k=1}^\infty\om_n\biggr) \sepword{or}
\sum_{k=1}^\infty\om_n = \log\biggl(1 + \sum_{n=1}^\infty d_n\biggr),
$$
to be solved for the $\om_n$. Since $\log(1+t)=\sum_{k\ge1}(-1)^{k-1}
(k-1)!t^k/k!$, from the very \bfb\ formula~\eqref{eq:wonder-series} it
follows that
$$
\om_n = \frac{1}{n!}\sum_{k=1}^n
(-1)^{k-1}(k-1)!B_{n,k}(d_1,2d_2,\dots,(n+1-k)!d_{n+1-k}) =:
S_n(\row d1n).
$$
The polynomials $S_n$ are the so-called Schur polynomials; they are
well known from the theory of symmetric functions. The first four are:
\begin{align*}
S_1(d_1) &= d_1;  \quad
S_2(d_1,d_2) = d_2 - \frac{d_1^2}{2};  \quad
S_3(d_1,d_2,d_3) = d_3 - d_1 d_2 + \frac{d_1^3}{3};
\\
S_4(d_1,d_2,d_3,d_4)
&= d_4 - \frac{d_2^2}{2} - d_1d_3 + d_1^2 d_2 - \frac{d_1^4}{4}.
\end{align*}
In view of~\eqref{eq:appalling}, the inverse map expressing the $d_n$
in terms of
the~$\om_n$ is given by
\begin{align*}
d_n &=
\frac{1}{n!}\sum_{k=1}^nB_{n,k}(\om_1,2\om_2,\dots,(n+1-k)!\om_{n+1-k})
\\
&= \sum_{c_1+2c_2+\cdots+nc_n=n}\,
\frac{\om_1^{c_1}\cdots\om_n^{c_n}}{c_1!\cdots c_n!}
=: AS_n(\om_1,\dots,\om_n);
\end{align*}
in the second equality the formula~\eqref{eq:wonder-formula} has been
used.

Now, we can construct an isomorphism between~$\Hl$ and~$B(V)$, where
$V$ is the vector space with a denumerable basis $Y_1,Y_2,\dots$, by
the algebraic correspondence
$$
l_i \mapsto S_i(Y_1,\dots,Y_i),
$$
with inverse
$$
Y_i \mapsto AS_i(l_1,\dots,l_i).
$$
An easy computation~\cite{Foissy} allows one to verify that
$$
\Dl_{B(V)}[AS_i(Y_1,\dots,Y_i)] =
\biggl[\sum_{j=0}^n AS_j\ox AS_{n-j}\biggr](Y_1,\dots,Y_i).
$$
This is enough to conclude as well that the elements
$$
S_i(l_1,\dots,l_i)
$$
are all primitive, and constitute a basis for the space of primitive
elements of~$\Hl$, which is thus primitively generated. We shall come
back to the isomorphism between $\Hl$ and~$B(V)$ in Section~14, where
it plays an important role.
\end{xmpl}

\begin{xmpl}
Among the techniques for solving differential equations of the type
$$
\frac{dY}{dt} = A(t)Y(t), \qquad Y(t) = 1,
$$
with the variable $Y$ being an operator valued function, there is the
Magnus technique~\cite{MagnusIndeed}, where the solution is sought in
the form $\exp\bigl(\sum_{k=1}^\infty\Omega_k(t)\bigr)$ ---and an
explicit form for the $\Omega_k$ is given--- and the (Feynman--Dyson)
time-ordered exponential, which is in fact an iterative solution of
the form
$$
1 + \sum_{l=1}^\infty P_l,
$$
where the $P_l$ are time-ordered products. Relations between both
types of solutions have been rather painfully obtained in applied
science papers~\cite{Burumbum,Salzman}. Here they are transparent; of
course, it does not reduce to $\Omega_k = S_k(\row P1k)$, because
now the $P_l$ need not commute. It is however possible to introduce
symmetrized Schur polynomials like:
\begin{align*}
{\tilde S}_3(d_1,d_2,d_3) &= d_3 - \thalf(d_1d_2+d_2d_1) +
\frac{d_1^3}{3};
\\
{\tilde S}_4(d_1,d_2,d_3,d_4) &= d_4 - \thalf d_2^2 -
\thalf(d_1d_3+d_3d_1) + \frac{d_1^2d_2+d_1d_2d_1+d_2d^2_1}{3} -
\frac{d_1^4}{4},
\end{align*}
and so on; and then certainly
$$
\Omega_k = {\tilde S}_k(\row P1k).
$$
More details on this will be given in~\cite{Kalliope}.
\end{xmpl}

\begin{xmpl}
Formulae for reversion of formal power series, going back to
Lagrange~\cite{LaMemoire}, have enchanted generations of
mathematicians and physicists, and there is an immense literature on
them. It is very easy to prove that an element
$f(x)=ax+bx^2+cx^3+\ldots$ of $x\C[[x]]$ like the ones we have used to
study diffeomorphisms has a reciprocal or compositional inverse
$f^{(-1)}(x)$ such that $f(f^{(-1)}(x))=x$ and $f^{(-1)}(f(x))=x$ if
and only if $a\ne0$, in which case the reciprocal is unique; and any
left or right compositional inverse must coincide with it. For
computing it, many methods are available. One finds in the handbook of
mathematical functions~\cite{Bluebible} the recipe: given
$$
y = ax + bx^2 + cx^3 + dx^4 + ex^5 + \dots
$$
then
$$
x = Ay + By^2 + Cy^3 + Dx^4 + Ey^5 + \dots
$$
where
\begin{align}
aA &= 1,
\notag \\
a^3B &= -b
\notag \\
a^{5}C &= 2b^2-ac
\notag \\
a^{7}D &= 5abc-a^2d-5b^3
\notag \\
a^{9}E &= 6a^2bd+3a^2c^2+14b^4-a^3e-21ab^2c
\notag \\
\dots &
\label{eq:old-favourite}
\end{align}

Let us translate the search for $f^{(-1)}$ into
algebraic-combinatorial terms; everything that follows should be
pretty obvious. As $\F\simeq\Rr^\cop(G)$, at least `morally'
speaking, one expects to get back to the group $G$ (or rather
$G^\opp$) by means of the Tannaka--Kre\u{\i}n paradigm. Since
$\R$ is commutative, the set $\Hom_\alg(\F,\R)$ of all algebra
morphisms is a group under convolution. Now, the action of $f\in
\Hom_\alg(\F,\R)$ is determined by its values on the $a_n$. The
map
$$
f \mapsto f(t) =\sum_{n=1}^\infty f_n\frac{t^n}{n!},
$$
where $f_n := \dst{f}{a_n}$, establishes a bijection from
$\Hom_\alg(\F,\R)$ onto the set of formal (exponential) power
series over the reals such that $f_0 = 0$ and $f_1 =1$. We know that
such series form a group under the operation of functional
composition.

This correspondence is an \textit{anti-isomorphism} of groups: there
is really nothing to prove, but we may go through the motions again.
Indeed, let $f,g \in \Hom_\alg(\F,\R)$; then
$$
f*g(a_n) = m(f \ox g)\Dl a_n = \sum_{k=1}^n g_k
B_{n,k}(1,f_2,\dots,f_n),
$$
where we took in consideration that $f_1=\dst{f}{1}=1$. This is the
same as the $n$-th coefficient of~$h(t)=g(f(t))$.

In other words, $G=\Hom_\alg(\F^\cop,\R)$. By our discussion
in Section~5, the antipode of $\F^\cop$ is $S^{-1}=S$. Therefore,
making allowance for our prior choice of $a=1$ ---an almost trivial
matter on which we shall reflect later--- and the use of ordinary
instead of exponential series, whenever $f$ and $g$ are formal
exponential power series with $f_0 = 0$ and $f_1 = 1$ verifying
$f\circ g(x)=x$ or $g\circ f(x)=x$, the formulae
\eqref{eq:old-favourite} correspond to
$$
a_n(g) = Sa_n(f).
$$
Indeed $Sa_2=-a_2$ gives $B=-b;\,Sa_3=-a_3 + 3a_2^2$ gives $3!C = -3!c
+ 3(2!b)^2$, that is $C=2b^2-c;\,Sa_4=-a_4 - 15a_2^3 + 10a_2a_3$ gives
$4!D=-4!d + 10(2!b)(3!c) - 15(2!b)^3$, that is $D=5bc-d-5b^3$; and so
on.

Lagrange reversion is usually proved by Cauchy's theorem in the
context of analytic functions, where it is known as the
Lagrange--B\"urmann formula; or by matrix algebraic methods. Its
derivation by Hopf algebraic methods in~\cite{Precursors} is
particularly elegant, but it is worthwhile to note that the equivalent
of formula~\eqref{eq:reverted} that they use was known prior to that
derivation.
\end{xmpl}

\section{From Fa\`a di Bruno to Connes--Moscovici}

The reader familiar with the ``noncommutative geometry"
Connes--Moscovici algebras will have noticed the similitude between
the commutative Hopf subalgebra of~$\H_\CM(1)$ ---the simplest of
those--- and the \bfb\ algebra $\F$. In fact, they are one and the
same. To see this, remember that to study diffeomorphisms as formal
power series it is not necessary to use the coordinates $a_n$. Without
losing information, a description of $\F$ can be done in terms of the
new set of coordinates
$$
\dl_n(f) := \bigl[\log f'(t)\bigr]^{(n)}(0), \quad n\ge1.
$$
Consider then
$$
h(t) := \sum_{n\geq1}\dl_n(f){t^n/n!} = \log f'(t) = \log\bigl(1 +
\sum_{n\ge1}a_{n+1}(f)t^n/n!\bigr).
$$
{}From the formula~\eqref{eq:wonder-series} it follows that
\begin{equation}
\dl_n = \sum_{k=1}^n (-1)^{k-1}(k-1)!B_{n,k}(a_2,\dots,a_{n+2-k})
=: L_n(\row a2{n+1}).
\label{eq:CMgenerators}%
\end{equation}
The polynomials $L_n$ (closely related to the Schur polynomials) are
called \textit{logarithmic polynomials} in combinatorics: they give
the successive derivatives of (exponential) series of the form
$\log\bigl(f(t)\bigr)$.

On the other hand, from the series expression of~$\exp(h(t)) = f'(t)$
we see that
\begin{equation}
a_{n+1} = \sum_{k=1}^n B_{n,k}(\dl_1, \dots, \dl_{n+1-k})
=: Y_n(\row {\dl}1n),
\label{eq:x-to-deltas}%
\end{equation}
and we have inverted~\eqref{eq:CMgenerators}. Recall that
$$
Y_n(\row {\dl}1n) =
\sum_{\la\in\Pi_n} \dl_1^{\la_1} \cdots \dl_n^{\la_n},
$$
where as usual $\la_j$ is the number of blocks of size $j$ in a
partition $\la$ of~$[n]$.

Now, from~\eqref{eq:CMgenerators} we get $\dl_1 = a_2, \dl_2
= a_3 - a_2^2, \dl_3 = a_4 - 3a_2 a_3 + 2a_2^3$, and $\dl_4 = a_5 -
3a_3^2 - 4a_2a_4 + 12a_2^2a_3 - 6a_2^4$, and since the coproduct is an
algebra morphism, by use of~\eqref{eq:deltaex} we obtain the coproduct
in the Connes--Moscovici coordinates.  For instance, for the first few
generators,
\begin{align*}
\Dl\dl_1 &= \dl_1 \ox 1 + 1 \ox \dl_1,  \\
\Dl\dl_2 &= \dl_2 \ox 1 + 1 \ox \dl_2 + \dl_1 \ox \dl_1,  \\
\Dl\dl_3 &= \dl_3 \ox 1 + 1 \ox \dl_3 + 3\dl_1 \ox \dl_2
+ (\dl_1^2 + \dl_2) \ox \dl_1 , \\
\Dl\dl_4 &= \dl_4 \ox 1 + 1 \ox \dl_4 + 6\dl_1 \ox \dl_3
+ (7\dl_1^2 + 4\dl_2) \ox \dl_2 +
(3\dl_1 \dl_2 + \dl_1^3 + \dl_3) \ox \dl_1.
\end{align*}
In conclusion, the commutative subalgebra of the Connes--Moscovici
Hopf algebra is but an avatar of the \bfb\ algebra: algebras that
differ only by the basis presentation we of course do not distinguish.

\smallskip

The task is now to assemble the whole of~$\H_\CM(1)$ from~$\F$.
For that, it is required to look closer at the structure of $G$, and
just a bit more of Hopf algebra theory.

A \textit{matched pair} of groups is by definition a group~$G$ with
two subgroups~$G_1,G_2$ such that $G=G_2G_1, G_1\cap G_2=1$. The
affine subgroup $G_2$ of transformations in $\Diff^+(\R)$ given by
$t\to\b+\a t$, with $\b\in\R,\a\in\R^+$ and the group $G_1$ of
transformations tangent to the identity, considered in Section~8, give
such a bijection onto $G$.

Since matched pairs are conspicuous by their absence in most
textbooks, we discuss the situation in some detail. The definition
implies that there is a natural action of~$G$ (and in particular
of~$G_2$) on the \textit{homogeneous space} $G_1\simeq G_2\backslash
G$, and vice versa of~$G$ (thus of~$G_1$) on~$G_2\simeq G/G_1$. Both
actions are given by composition. In detail: given $\psi\in G$, we
decompose it as $\psi = kf$, with $k \in G_2$, $f\in G_1$, and
$$
\b = \psi(0), \quad \a = \psi'(0), \quad f(t) =
\frac{\psi(t)-\psi(0)}{\psi'(0)}.
$$
The right action of~$\psi\in G$ on~$f\in G_1$ is given by
$$
[f\rt\psi](t)
= \frac{f(\psi(t))-f(\psi(0))}{\psi'(0)f'(\psi(0))};
$$
note $[f\rt\psi](0) = 0,[f\rt\psi]'(0) = 1$; and in
particular
\begin{equation}
[f\rt(\b,\a)](t) = \frac{f(\a t+\b)-f(\b)}{\a f'(\b)}.
\label{eq:first-act}
\end{equation}
The left action of~$G$ on~$G_2$ is seen from consideration
of~$\psi(\b+\a t)$, and we obtain in particular for $\psi=f\in G_1$:
\begin{equation}
f\lt(\b,\a) = (f(\b),\a f'(\b)).
\label{eq:second-act}
\end{equation}
Note that $(k_1f_1)(k_2f_2) = k_1(f_1\lt k_2)(f_1\rt k_2)f_2$ and
$(kf)^{-1} = (f^{-1}\lt k^{-1})(f^{-1}\rt k^{-1})$, as general
equalities of the theory of matched pairs.

We ought to translate the previous considerations into Hopf-algebraic
terms. In what follows we consider bialgebras acting on algebras from
the left. On~$\C$ they are assumed to act by $h\.1=\eta(h)\,1$.

\begin{defn}
\label{df:H-module algebra}
A (left, Hopf) \textit{$H$-module algebra} $A$ is an algebra which is a
left module for the algebra~$H$ such that the defining maps $u:\C\to
A$ and $m: A\ox A\to A$ intertwine the action of~$H$:
\begin{equation}
u(h\.1) = h\.u(1); \quad m(h_\ox\.(a\ox b)) = h\.m(a\ox b)
\label{eq:Hopf-Leib}%
\end{equation}
that is, $h\.1_A=\eta(h)\,1_A$ and $h\.(ab) = \sum
(h_{(1)}\.a)(h_{(2)}\.b)$ whenever $a,b\in A$ and $h\in H$.
\end{defn}

The first condition is actually redundant~\cite{Miriam}. It is easy to
see that the definition corresponds to the usual notions of groups
acting by automorphisms, or Lie algebras acting by derivations. If $h$
is a \textit{primitive} element of~$H$, then~\eqref{eq:Hopf-Leib}
entails that $h\.1_A = 0$ and that $h\.(ab) = (h\.a)b + a(h\.b)$: as
already observed, primitive elements act by derivations.
Therefore~\eqref{eq:Hopf-Leib} may be regarded as a generalized
Leibniz rule.

(Analogously one defines $H$-module coalgebras, $H$-module bialgebras
and so on; we have no use for them in this article.)

When there is an algebra covariant under a group or Lie algebra, one
constructs in a standard way the semidirect product algebra. In the
same spirit~\cite{Maje}:

\begin{defn}
\label{df:smashproduct}
Let~$H$ be a Hopf algebra and $A$ a (left) Hopf $H$-module algebra.
The (untwisted) \textit{smash product} or \textit{crossed product}
algebra, denoted~$A\# H$ or~$A\rtimes H$, is the vector space $A\ox H$
endowed with unit~$1\ox1$ and the product
$$
(a \ox h)(b \ox k) := \sum a (h_{(1)}\.b) \ox h_{(2)} k.
$$
\end{defn}

We verify associativity of the construction:
\begin{align*}
(a \ox h)\bigl((b \ox k)(c \ox l)\bigr)
&= (a \ox h)(b(k_{(1)}\.c)\ox k_{(2)} l)
\\
&= a\Bigl(h_{(1)}\.\bigl(b(k_{(1)}\.c)\bigr)\Bigr)\ox h_{(2)}k_{(2)} l
\\
&= a(h_{(1)}\.b)(h_{(2)}k_{(1)}\.c) \ox h_{(3)}k_{(2)} l
\\
&= (a(h_{(1)}\.b)\ox h_{(2)}k)(c\ox l)
\\
&= \bigl((a \ox h)(b \ox k)\bigr)(c \ox l).
\end{align*}

To alleviate the notation, we can identify $a\equiv a\ox1$ and $h\equiv
1\ox h$.  Then $ah = (a\ox1)(1\ox h)$, whereas
$$
ha = (1\ox h)(a\ox1) = h_{(1)}\.a\ox h_{(2)} = (h_{(1)}\.a)h_{(2)}.
$$
Both $A$ and $H$ are subalgebras of~$A\rtimes H$. A very simple
example is given by $H=U(\g)$, where~$\g$ is the 1-dimensional Lie
algebra acting as $\frac{d}{dx}$ on $A=\C[x]$. Then $A\#H$ is the Weyl
algebra. In our case $H$ will be the enveloping algebra $\U(\g_2)$ of
the affine group Lie algebra $\g_2$; and the left module algebra is
none other than $\F$.

We exhibit then the action of~$\U(\g_2)$ on~$\F$, and check that the
\bfb\ algebra is a Hopf $\U(\g_2)$-module algebra. Consider the action
on $G_1$ given by~\eqref{eq:first-act}. We proceed step by step. With
$\b=0$, one gathers
$$
a_n([f\rt(0,\a)]) := [f\rt(0,\a)]^{(n)}(0) =
\a^{n-1}a_n(f),
$$
for $n\ge2$.  At the infinitesimal level, with $(0,\a) =: \exp sY$, we
conclude the existence of an action:
\begin{equation}
Y\.a_n = (n-1)a_n.
\label{eq:no-alforjas}
\end{equation}
Similarly
\begin{equation}
Y\.\dl_n = n\dl_n.
\label{eq:menos-alforjas}
\end{equation}
Note that $Y\.$ is the grading operator for~$\F$. At this stage, one
could construct already the crossed product of~$\F$ by the grading.

We proceed to the other generator. With $\a=1$,
equation~\eqref{eq:first-act} now gives
$$
[f\rt(\b,1)]^{(n)}(0) =
\biggl(\frac{f(t+\b)-f(\b)}{f'(\b)}\biggr)^{(n)}(0).
$$
At the infinitesimal level, with $(\b,1) =: \exp\b X$, we conclude:
\begin{equation}
X\.a_n = a_{n+1} - a_2a_n.
\label{eq:neo(old)-generation}
\end{equation}
At this point, if not before, one sees the wisdom of Connes and
Moscovici's logarithmic coordinates, as then~\eqref{eq:first-act}
mutates into
$$
\log[f\rt(\b,\a)]'(t) = \log f'(\b+\a t),
$$
up to a constant; from which one infers the simpler action
\begin{equation}
X\.\dl_n = \dl_{n+1}.
\label{eq:old-generation}
\end{equation}

Both $X$ and $Y$ act by derivation; thus $\F$ is a $\U(\g_2)$-module
algebra. Here the relations $Za = Z\.a + aZ$ hold, for $Z = X$ or~$Y$.
Note that $\F$ is \textit{not} a $\U(\g_2)$-module bialgebra. By
inspection of~\cite{ConnesMHopf}, and summarizing so far, we conclude
that the algebra $\H_\CM(1)$ is the smash product $\F\rtimes\U(\g_2)$
of the enveloping algebra~$\U(\g_2)$ of the affine Lie algebra by the
\bfb\ Hopf algebra; it is generated just by $Y,X,a_2=\dl_1$.

We can as well regard the algebra $\H_\CM(1)$ as the enveloping
algebra of the extension $\L'$ of $\g_2$ by the abelian Lie algebra
$\L$ spanned by the $a_n$, which is a derivation $\g_2$-module.
Recall that an extension of this type is an exact sequence
$$
0 \to \L \xrightarrow{i} \L' \xrightarrow{h} \g_2 \to 0,
$$
where $[x,i(a)]:=i(h(x)\.a)$, for $x\in\L',a\in\L$; this lifts to an
exact sequence of enveloping algebras, in our case
$$
0 \to \F \to \H_\CM(1) \to \U(\g_2) \to 0.
$$
It is permissible to write $[X,\dl_n]$ for $X\.\dl_n$, and so on; but
we do not use that notation.

\marker
The burning question now is, what is the ``good'' coproduct on the
smash product algebra $\F\#\U(\g_2)$? But before tackling that, we
pause to redeliver~\eqref{eq:no-alforjas} from Connes and
Moscovici's~\eqref{eq:menos-alforjas}
and~\eqref{eq:neo(old)-generation} from \eqref{eq:old-generation} by
the combinatorial argument. Since $X,Y$ are derivations
\begin{align}
Y\.(\dl_1^{c_1} \cdots \dl_n^{c_n})
&= (1c_1 + 2c_2 + \cdots + nc_n) \dl_1^{c_1} \cdots
\dl_n^{c_n},
\label{eq:Ycomm}%
\\
X\.(\dl_1^{c_1} \cdots \dl_n^{c_n})
&= \sum_{j=1}^n c_j \dl_1^{c_1} \cdots \dl_j^{c_j-1}
\dl_{j+1}^{c_{j+1}+1}  \cdots \dl_n^{c_n}.
\label{eq:Xcomm}%
\end{align}
Thus, using~\eqref{eq:x-to-deltas} and~\eqref{eq:Ycomm}
\begin{align*}
Y\.a_{n+1}
&= Y\.Y_n(\row {\dl}1n)
= \sum_{c\in\Pi_n}
Y\.(\dl_1^{c_1}  \cdots \dl_n^{c_n})
\\
&= \sum_{c\in\Pi_n} (1c_1 + 2c_2 + \cdots + nc_n) \, \dl_1^{c_1}
\cdots \dl_n^{c_n}
\\
&= n Y_n(\row {\dl}1n) = na_{n+1}.
\end{align*}
We have verified~\eqref{eq:no-alforjas}. Similarly,~\eqref{eq:Xcomm}
entails
$$
X\.a_{n+1}=  \sum_{c\in\Pi_n}
X\. (\dl_1^{c_1}  \cdots \dl_n^{c_n})
= \sum_{c \in\Pi_n} \sum_{j=1}^n c_j \,
\dl_1^{c_1} \cdots \dl_j^{c_j-1} \dl_{j+1}^{c_{j+1}+1}
\cdots \dl_n^{c_n}.
$$
Notice that
$$
1c_1 + 2c_2 + \cdots  + j(c_j -1) + (j+1)(c_{j+1}+1)
+(j+2)c_{j+2} + \cdots + nc_n = n+1;
$$
therefore we must think of partitions of $[n+1]$.  Now, by deleting
the number $n+1$ such partitions give a partition of $[n]$.
Furthermore, if $c$ is a partition of $[n]$, then the partitions of
$[n+1]$ that give rise to~$c$ by dropping $n+1$, are obtained by
either adding the singleton $\{n+1\}$ or by inserting $n+1$ in one
of the blocks of $c$.  Conversely, all partitions of $[n+1]$ are
obtained from one of $[n]$ by either of these two procedures.
Moreover, if $c$ is of class $1^{c_1}2^{c_2} \cdots n^{c_n}$, and
$n+1$ is included in a block of size $j$ of $c$ we obtain a
partition of $[n+1]$ of class $1^{c_1}2^{c_2} \cdots j^{c_j-1}
(j+1)^{c_{j+1} +1} \cdots n^{c_n}$, and there are as many of these
partitions as blocks of size $j$; namely $c_j$.  Thus
\begin{align*}
a_{n+2} &= Y_{n+1}(\row{\dl}1{n+1})
= \sum_{e\in\Pi_{n + 1}} \dl_1^{e_1} \cdots \dl_{n+1}^{e_{n+1}}
\\
&= \sum_{c\in\Pi_n} \sum_{j=1}^n c_j
\dl_1^{c_1} \cdots \dl_j^{c_j-1} \dl_{j+1}^{c_{j+1}+1}
\cdots \dl_n^{c_n}
+ \dl_1 \sum_{c\in\Pi_n} \dl_1^{c_1} \cdots \dl_n^{c_n}
\\
&= X\.a_{n+1} + \dl_1 \, a_{n+1},
\end{align*}
which is precisely~\eqref{eq:neo(old)-generation}.

We return to the search for a \textit{compatible} coalgebra structure
on~$\F\rtimes\U(\g_2)$. Given a Hopf algebra~$H$, one considers
on~$\C$ the coaction given by $\ga(1)=u(1)$.

\begin{defn}
\label{df:smashcoproduct}
A (right) Hopf $H$-\textit{comodule coalgebra}~$C$ is a coalgebra
which is a right comodule for the coalgebra~$H$ such that the counit
map and the coproduct on~$C$ intertwine the coaction of~$H$:
$$
(\eta\ox\id)\ga = \ga\eta; \quad (\Dl\ox\id)\ga = \ga_\ox\Dl.
$$
In this context, the \textit{smash coproduct} or crossed coproduct
coalgebra $H\ltimes C$ is defined as the vector space $H \ox C$
endowed with the counit~$\eta_H\ox\eta_C$ and the coproduct
$$
\Dl(a \ox c)
= \sum a_{(1)} \ox c_{(\uno)}^{(1)} \ox a_{(2)}c_{(1)}^{(2)} \ox c_{(2)}.
$$
\end{defn}

Now comes ``the revenge of the \bfb\ coalgebra'', since for the
identification of the Connes--Moscovici Hopf algebra we need $H=\F$ to
coact on the coalgebra~$C=\U(\g_2)$. One ought to be careful here, as
we are about to use the coalgebra structure of~$\F$ for the first
time, and actually $\F\not\simeq\Rr(G_1)$, but
$\F\simeq\Rr^\cop(G_1)$!

To see why $\F$ naturally coacts on~$\U(\g_2)$, note that, to
implement the action~\eqref{eq:second-act} on~$G_2$ of the
diffeomorphisms tangent to the identity, and reasoning as above, we
would have an action of ${\F'}^\opp$ on a suitable algebra
$C(G_2)$ of functions on~$G_2$. This we can choose to regard as a
coaction of~$\F^\cop$ on~$C(G_2)$~\cite[Section~1.2]{Calypso}, or
better on its dual coalgebra~$U(\g_2)$. At the infinitesimal level, if
$(\b,\a) =\exp(\b X)\exp(\log\a Y)$ we see that $f\.Y=Y$ for all $f$,
but
$$
f\.X = \frac{d}{d\b}\biggr|_{\a=1,\b=0}\, [f\rt(\a,\b)]
= X + f''(0)Y = X + a_2(f)Y.
$$
The astuteness of Connes and Moscovici's definition of
$\H_\CM(1)$ is precisely that this information on the structure
of the diffeomorphism group is made patent. To the last expressions does
correspond the coaction $\ga: U(\g_2)\to U(\g_2)\ox\F^\cop$ given
by $\ga(Y) = Y\ox1, \ga(X) = X\ox1 + Y\ox a_2$.

The smash coproduct structure~$\F^\cop\ltimes\U(\g_2)$
is given by
\begin{align}
\Dl^\cop X
&:= \Dl(1\ox X) = 1\ox1\ox1\ox X + 1\ox X\ox1\ox1 + 1\ox Y\ox a_2\ox1
\notag \\
&= 1\ox X + X\ox 1 + Y\ox a_2,
\label{eq:almost-there}
\end{align}
the other pertinent coproducts staying unchanged. It is time to return
to the use of~$\F$, instead of~$\Rr(G_1)$, and to account for that it is
enough to turn~\eqref{eq:almost-there} round; therefore we finally have
$$
\Dl X = X\ox 1 + 1\ox X + a_2\ox Y.
$$
In conclusion, we have proved the following result.

\begin{thm}
\label{pr:CMequalsFdB}
The Connes--Moscovici bialgebra is the crossed product algebra and
coalgebra of the enveloping algebra of the Lie algebra of the affine
group and the \bbfb\ bialgebra.
\end{thm}

This conclusion is restated in~\cite{KhalkhaliR}. The construction is
similar to Majid's bicrossedproduct bialgebra, although with important
differences of detail. In~\cite{Maje} and~\cite{Majeverde} the
canonical link to the matched pair of groups situation is also
highlighted; and one might borrow the notation $\F\bowtie\U(\g_2)$
for~$\H_\CM(1)$. Furthermore, $\F\bowtie\U(\g_2)$ is automatically a
Hopf algebra; in fact
$$
0 = m(\id \ox S)\Dl X = m(X \ox 1 + 1 \ox SX + a_2 \ox SY)
=  X + SX - a_2\,Y
$$
entails that the only nontrivial formula for the good antipode is
given by $SX = -X + a_2Y$. The same result is obtained from the
formula $S(aZ) = \sum SZ^{(1)}S(aZ^{(2)})$ in~\cite[Ch.~6]{Maje}, with
$a=1,Z=X$. Note that $S$ is of infinite order, as
$$
S^{2n}X = X + na_2 \sepword{and} S^{2n+1}X = -X + a_2Y - na_2.
$$
The natural pairing between $\H_\CM(1)$ and $C(G_2)\rtimes G_1$
given by
$$
\dst{aZ_{k}}{gf} = g(k)a(f),
$$
where $g$ is a function on~$G_2$ and $Z_k$ is the element of~$\g_2$
that gives $k\in G_2$ by exponentiation, is also transparent.

Similarly, $\H_\CM(n)\simeq\F(n)\bowtie(\R^n\rtimes\gl(n;\R))$,
with $\F(n)$ the $n$-coloured \bfb\ Hopf algebra (on which something
will be said later).

\smallskip

Given the role accorded in this work to the \bfb\ algebras, a
reference to the somewhat checkered past of the \bbfb\ formula might
not be out of place. Consult the excellent article by
Johnson~\cite{DoctorJohnson}, where, besides a nice derivation of the
formula very much in the spirit of our Section~7, it is recalled that
there is no proof for~\eqref{eq:faa-diB} in~\cite{FaadiBruno}, and
explained that formulae equivalent to it were present in the
mathematical literature from the beginning of the ninetenth century.
Also the further historical investigation~\cite{Craik} deserves a
look.

\newpage

\section*{Part III: Hopf Algebras of Graphs and Distributive
Lattices}
\addcontentsline{toc}{section}{\protect\numberline{III}\enspace
HOPF ALGEBRAS of GRAPHS and DISTRIBUTIVE LATTICES}

\section{Hopf algebras of Feynman graphs}

We start this part by an invitation, too: some readers will want to
see a worked out example from perturbative quantum field theory, where
the combinatorics of renormalization leads to a finite `renormalized'
graph by means of local counterterms, from a given Feynman graph. The
analytical complications militate against developing an example that
is really challenging from the combinatorial viewpoint; but this
cannot be helped. Needless to say, the experts can skip the following
discussion.

To make matters as simple as possible, our example will be taken from
the (Ginzburg--Landau) $\vf^4_4$ scalar model in Euclidean field
theory. We work in the dimensional regularization scheme in momentum
space~\cite{Nobel99,ConoSur} and display the counterterms \textit{ab
initio}. The free energy functional is given by
$$
E_{\mathrm{free}}[\vf] =
\int d^Dx\,\biggl[\frac{1}{2}\vf\square\vf + \frac{1}{2}m^2\vf^2\biggr].
$$
Here $\square$ denotes the $D$-dimensional Laplacian. The interaction
part is extended by counter\-terms, of the form
\begin{equation}
E_{\mathrm{int}}[\vf] =
\int d^Dx\,\biggl[\frac{\mu^\eps\tilde{g}}{4!}\vf^4 +
c_g\frac{\mu^\eps\tilde{g}}{4!}\vf^4 + \frac{c_\vf}{2}\vf\square\vf +
\frac{c_{m^2}}{2}m^2\vf^2\biggr].
\label{eq:traicion}
\end{equation}
Denote $\eps:=4-D$. The definition of the original vertex includes the
mass parameter $\mu$, introduced to make $\tilde{g}$ dimensionless:
$$
g = \tilde{g}\mu^\eps = \cross .
$$
The counterterms produce additional vertices in the diagrammatic
expansion. In particular:
$$
c_g\,\tilde{g}\mu^\eps = \crossfatpt .
$$
Dimensional analysis indicates that all the counterterms are
dimensionless; therefore they can only depend on $\eps,\tilde{g}$ or
combinations like $m^2/\mu^2,k^2/\mu^2$. It turns out that the last
one appears just in intermediate stages as $\log(k^2/\mu^2)$ and that
these nonlocal terms cancel in the final expressions ---this is the
key to the whole affair. Thus $c_{m^2},c_{\vf},c_{g}$ depend only on
$\eps,\tilde{g},m^2/\mu^2$ (and the dependence on $m^2/\mu^2$ can be
made to disappear).

The quantities $\vf,m,\tilde{g}$ in the previous displays are the
\textit{renormalized} field, mass and coupling constant. The original
form of the theory is recovered by multiplicative renormalization:
$$
\quad Z_\vf = 1 + c_\vf, \quad  Z_{m^2} = 1 + c_{m^2}, \quad
Z_g = 1 + c_g.
$$
The total energy functional becomes
$$
E[\vf] = \int d^Dx\, \biggl[ \frac{1}{2} Z_\vf \,\vf\square\vf +
\frac{1}{2} Z_{m^2} \,m^2\vf^2 +
\frac{\mu^\eps\tilde g}{4!} Z_g \,\vf^4 \biggr],
$$
with the $\eps$-dependent coefficients. Introduction of the bare field
and the bare mass and coupling:
$$
\vf_B = \sqrt{Z_\vf}\,\vf, \quad
m_B^2 = \frac{Z_{m^2}}{Z_\vf}, \quad
\tilde{g}_B = \frac{Z_{g}}{Z_\vf^2} \,\mu^\eps \tilde{g},
$$
brings the energy functional to the standard form
$$
E[\vf_B] = \int d^Dx\,\biggl[ \frac{1}{2}\, \vf_B\square\vf_B +
\frac{1}{2}\, m_B^2\vf_B^2 +
\frac{\tilde{g}_B}{4!}\, \vf_B^4\biggr].
$$
The bare quantities here are functions of the renormalized quantities
$\tilde{g}$, $m$, the mass scale $\mu$, and~$\eps$.

The Feynman rules for the model are next recalled; instead of using
directly the modified Feynman rules including the counterterms, we
will `discover' the latter in the process of renormalization. We plan to
concentrate on the proper (1PI) vertex function
$\Ga^{(4)}(k_1,\dots,k_4)$, which is defined only for
$k_1 +\cdots+ k_4 = 0$ and is represented by \textit{amputated}
diagrams. This means that we only need:
\begin{itemize}
\item{} A propagator factor $\dfrac{1}{p_j^2+m^2}$, with
$j \in \set{1,\dots,I=2p-2}$, where $p$ is the approximation order
(number of vertices), for each internal line. Each internal momentum
$p_j$ is expressed in terms of the loop momenta and the four external
momenta.

\item{} An integration $(2\pi)^{-D}\int\,d^4l_m$ over each
(independent) loop momentum with the index
$m\in\set{1,\dots,L=I-p+1=p-1}$.

\item{} The factors $\tilde{g}\mu^\eps$, one for each vertex.

\item{} The weight factor of the graph.
\end{itemize}

We also briefly recall the counting of ultraviolet divergences.
According to the rules, a Feynman integral $I_\Ga$ with $p$ vertices
and four external lines contains $L=p-1$ loop integrations and thus in
the numerator of the integrand a power $D(p-1)$ of the momentum
appears. Each of the internal lines contributes a propagator. Thus
there are altogether
$$
\omega(\Ga) := D(p-1) - 2(2p-2) = \eps(1-p)
$$
powers of momentum in such a Feynman integral. This power
$\omega(\Ga)$ is the \textit{superficial degree of divergence} of
$\Ga$. As $\eps \downarrow 0$, all four-point proper graphs
are superficially logarithmically \textit{divergent}. A graph is said
to have \textit{subdivergences} if it contains a superficially
divergent subdiagram $\ga$, that is to say, with $\omega(\ga) \geq 0$ in
that limit. The only possibly divergent subintegrations are those of
the two- and four-point subdiagrams. Up to one loop, we have
$$
\Ga^{(4)}(k_1,\dots,k_4) =  \cross + \frac{3}{2} \,\fish +
\text{counterterm} + O(g^3).
$$
Actually `$\frac{3}{2}\fish$' stands for three integrals of the
same form, in terms of the `Mandelstam' variables $s = (k_1+k_2)^2$,
$t = (k_1+k_3)^2$ and $u = (k_1+k_4)^2$. In the next to leading order:
$$
\Ga^{(4)}(k_1,\dots,k_4) =  \cross + \frac{3}{2}\,\fish +
3\,\winecup + \frac{3}{4}\,\bikini + \text{counterterms} + O(g^4).
$$
A similar comment applies in relation to the $s,t,u$ variables. (We
ignore a fish-cum-tadpole graph that would easily be taken into
account anyway; only at three-loop order would we have to handle in
parallel the renormalization of two-point diagrams in earnest.)

We compute the fish graph $\fish$~:
$$
I_{\mathrm{fish}}(k) = \tilde{g}^2 \mu^{2\eps}
\int \frac{d^Dp}{(2\pi)^D}\,\frac{1}{p^2+m^2}\,\frac{1}{(p+k)^2+m^2},
$$
where $k = k_1 + k_2$, say. Using Feynman's formula
$$
\frac{1}{A^aB^b} = \frac{\Ga(a+b)}{\Ga(a)\Ga(b)}
\int_0^1 dx\, \frac{x^{a-1}(1-x)^{b-1}}{[Ax + B(1-x)]^{a+b}},
$$
we obtain
\begin{align*}
I_{\mathrm{fish}}(k)
&= g^2 \int_0^1 dx \int \frac{d^Dp}{(2\pi)^D}\,
\frac{1}{\{(p^2 + m^2)(1-x) + [(p + k)^2 + m^2]x\}^2}
\\
&= g^2 \int_0^1 dx \int \frac{d^Dp}{(2\pi)^D}\,
\frac{1}{(p^2 + 2pkx + k^2x + m^2)^2}
\\
&= \frac{g^2\,\Ga(2-D/2)}{(4\pi)^{D/2}}
\int_0^1 dx\, \frac{1}{(sx(1-x) + m^2)^{2-D/2}}
\\
&= \frac{\tilde g^2 \mu^\eps \Ga(\eps/2)}{(4\pi)^{2}}
\int_0^1 dx\, \biggl[\frac{4\pi\mu^2}{(sx(1-x)+m^2)}\biggr]^{\eps/2}.
\end{align*}
Expanding now partially in powers of $\eps$, this yields
\begin{equation}
I_{\mathrm{fish}}(s) =
\tilde{g} \mu^\eps \frac{\tilde{g}}{(4\pi)^{2}}
\biggl[\frac{2}{\eps} + \psi(1) +
\int_0^1 dx\, \log\frac{4\pi\mu^2}{sx(1-x)+m^2}\, + O(\eps)\biggr].
\label{eq:fishandchips}
\end{equation}
Here $\psi(z) := \Ga'(z)/\Ga(z)$ is the digamma function.  Some
comments are already in order.  The divergence happens for~$D$ an even
integer greater than or equal to 4 (so it is clearly a logarithmic
one).  The remaining integral is finite as long as $m^2 \neq 0$.  In
equation~\eqref{eq:fishandchips} we have separated
$\tilde{g}\mu^\eps$, that will become the coupling constant, and is
not expanded in powers of~$\eps$; only the expression multiplying it
contributes to the renormalization constant $Z_g$ with a pole term
independent of the free mass scale $\mu$.  That parameter appears only
in the finite part, and the arbitrariness of its choice exhibits a
degree of ambiguity in the regularization procedure.

Now comes the renormalization prescription. Suppose we wanted the
value of $\Ga^{(4)}(k_i)$ at $k_i = 0$ to be (finite and) equal, at
least at the present order, to the renormalized coupling constant
\begin{equation}
\Ga^{(4)}(0) = g.
\label{eq:ren-recipe-bis}
\end{equation}
This would lead us to take
$$
Z_g = 1 + \tfrac{3}{2}\, \tilde{g} \,I_{\mathrm{fish}}(0),
$$
that is, a counterterm given by
$$
c_g(\eps,\tilde{g},m^2/\mu^2) =
-\frac{\tilde{g}}{(4\pi)^{2}}\biggl[\frac{6}{\eps} + 3\psi(1) -
3\log(4\pi m^2/\mu^2)\biggr].
$$
However, other choices like $c_g = -6\tilde{g}/(4\pi)^{2}\eps$, not
`soaking up' the finite terms, would be admissible.  We do not bother
to write the (finite, if a bit involved) result for $\Ga^{(4)}(s,t,u)$
in any of those cases.

\medskip

Let us tackle now the two-loop diagrams of the four-point function.
The bikini graph $\bikini$ is uncomplicated:
$$
I_{\mathrm{bikini}}(k) =
-g^3 \int \frac{d^Dp}{(2\pi)^D}\, \frac{1}{((p-k)^2+m^2)(p^2+m^2)}
\int \frac{d^Dq}{(2\pi)^D}\, \frac{1}{((q-k)^2+m^2)(q^2+m^2)},
$$
a product of two independent integrals. Here $k$ again denotes any one
of $k_1 + k_2$, $k_1 + k_3$, or $k_1 + k_4$. We obtain
\begin{align*}
I_{\mathrm{bikini}}
&= -\tilde{g} \mu^\eps \frac{\tilde{g}^2}{(4\pi)^{4}}\,
\biggl[ \frac{2}{\eps} + \psi(1) +
\int_0^1 dx\,\log\frac{4\pi\mu^2}{k^2x(1-x)+m^2}\, + O(\eps) \biggr]^2
\\
&= -\tilde{g} \mu^\eps \frac{\tilde{g}^2}{(4\pi)^{4}}\,
\biggl[ \frac{4}{\eps^2} + \frac{4}{\eps} \,\psi(1) +
\frac{4}{\eps} \int_0^1 dx\, \log\frac{4\pi\mu^2}{k^2x(1-x)+m^2}\, +
O(\eps^0) \biggr].
\end{align*}
For the first time, we face a combinatorial problem. It would
\textit{not} do to square and subtract (one third of) the previously
obtained vertex counterterm, as this procedure gives rise to a
nonlocal divergence of the form $(\log k^2)/\eps$. The solution is
easy enough: we take into account the counterterms corresponding to
the subdivergences by substituting $\bigl(I_{\mathrm{fish}}(k) -
I_{\mathrm{fish}}(0)\bigr)^2$ for $I_{\mathrm{bikini}}(k)$. This
happens to accord with the renormalization prescription and both the
double and the single pole then cancel out. In other words,
renormalization `factorizes', and the \textit{three} counterterms that
come from taking the functional~\eqref{eq:traicion} seriously do
appear.

\medskip

Things are slightly more complicated for the wine-cup graph
$\!\!\winecup$, both analytically and combinatorially. We find the
integral
$$
-g^3 \int \frac{d^Dp}{(2\pi)^D}\, \frac{d^Dq}{(2\pi)^D}\,
\frac{1}{((k_1+k_2-p)^2+m^2)(p^2+m^2)}\,
\frac{1}{(q^2+m^2)((p-q+k_3)^2+m^2)},
$$
or
$$
I_{\mathrm{winecup}}(k) = -g^3 \int \frac{d^Dp}{(2\pi)^D}\,
\frac{1}{((k_1+k_2-p)^2+m^2)(p^2+m^2)}\, I_{\mathrm{fish}}(p + k_3).
$$
(We have taken the loop momentum of the integral on the loop with two
sides as~$q$, and $p$ for the loop momentum of the integral on the
loop with three sides, and have used $k_1 + k_2 = -k_3 - k_4$.)
Therefore,
$$
I_{\mathrm{winecup}}(k) =
\frac{\tilde{g}^3\mu^{2\eps}\,\Ga(\eps/2)}{(4\pi)^2} \int_0^1 dx
\int \frac{d^Dp}{(2\pi)^D}\,
\frac{\biggl[\dfrac{4\pi\mu^2}{(k_3+p)^2x(1-x)+m^2)}\biggr]^{\eps/2}}
{((k_1+k_2-p)^2 + m^2)(p^2 + m^2)}.
$$
Combining the denominators in the usual way, after integrating with
respect to $p$, we obtain
\begin{align*}
& I_{\mathrm{winecup}}(k) =
\frac{\tilde{g}^3 \mu^\eps (4\pi\mu^2)^\eps \Ga(\eps)}{(4\pi)^4}
\int_0^1 dx\, [x(1-x)]^{-\eps/2} \int_0^1 dy\, y(1-y)^{\eps/2-1}
\int_0^1 dz
\\
&\x \biggl[yz(1-yz) (k_1+k_2)^2 + y(1-y) k_3^2 - 2yz(1-y) k_3(k_1+k_2)
+ m^2 \biggl( y + \frac{1-y}{x(1-x)} \biggr) \biggr]^{-\eps}.
\end{align*}
There is a pole term coming from the endpoint singularity in the
integral at $y = 1$. Around this point the square bracket can be
expanded:
$$
\biggl[ \cdots \biggr] = \biggl[1 - \eps \log\biggl( z(1-z) +
\frac{m^2}{s} \biggr) + O(\eps^2) \biggr].
$$
We are left with
\begin{align}
\frac{\tilde{g}^3\mu^\eps(4\pi\mu^2)^\eps}{(4\pi)^4}\,
\frac{1}{\eps}\, (1 + \eps\psi(1))
& \int_0^1 dx\, [x(1-x)]^{-\eps/2} \int_0^1 dy\, y(1-y)^{\eps/2-1}
\nonumber \\
&\x \int_0^1 dz\, \biggl[1 - \eps \log\biggl( z(1-z) +
\frac{m^2}{s} \biggr) + O(\eps^2) \biggr].
\label{eq:abrir-el-melon}
\end{align}
Now,
$$
\int_0^1 dx\,[x(1-x)]^{-\eps/2} = \frac{\Ga^2(1-\eps/2)}{\Ga(2-\eps)}.
$$
Similarly for the integral with respect to~$y$. We reckon that the
integral in~\eqref{eq:abrir-el-melon} is
$$
\frac{\Ga^2(1-\eps/2)\Ga(\eps)}{\Ga(2-\eps)\Ga(2+\eps/2)}\,
\biggl[ 1 - \eps \int^1_0 dx\, \log\biggl( x(1-x) + \frac{m^2}{s}
\biggr) \biggr].
$$
After some work,
$$
\frac{\Ga^2(1-\eps/2)\Ga(\eps)}{\Ga(2-\eps)\Ga(2+\eps/2)} =
\frac{2}{\eps}\, \frac{1}{(1-\eps)(1+\eps/2)}\, [1 + O(\eps)],
$$
and in conclusion
$$
I_{\mathrm{winecup}}(s) =
\tilde{g} \mu^\eps \frac{\tilde{g}^2}{(4\pi)^4}\, \frac{2}{\eps^2}\,
\biggl[1 + \frac{\eps(1 + 2\psi(1))}{2} - \eps \int_0^1 dx\,
\log\biggl( \frac{sx(1-x)+m^2}{4\pi\mu^2} \biggr) \biggr] + O(\eps^0).
$$
There is no way to contemplate the renormalization of this without
getting rid of the dreaded nonlocal divergence. This can be done
precisely by taking account of the subdivergence in the wine-cup
diagram, that is, subtracting the term (one half of)
$I_{\mathrm{fish}}(s)I_{\mathrm{fish}}(0)$. The nonlocal divergence in
this expression is seen to exactly cancel the nonlocal divergence in
the previous display!

The result of such a subtraction is still divergent, but it contains
only terms independent of momentum, that now can be reabsorbed in the
subtraction of the overall divergence.

Although the combinatorial resources required so far are trivial, it
is pretty clear that at a higher approximation order every detail of
renormalization computations becomes nightmarish.  It makes sense,
then, also from the viewpoint of general renormalizability theorems,
to clarify the combinatorial aspect as much as possible independently
of the analytical ones.  This was accomplished to a large extent by
the precursors, culminating in the Zimmermann forest
formula~\cite{Zim}.  A step further was taken when Kreimer recognized
the Hopf algebra structure lurking behind it.  This is what eventually
allows us to extract the general combinatorial meaning of the forest
formula in this paper.

Thus motivated, we turn to the Connes--Kreimer paradigm.

\marker
Bialgebras of Feynman graphs, encoding the combinatorics of
renormalization, were introduced in~\cite{ConnesKrRHI}.  The precise
definition we use in this paper was first given in~\cite{Etoile}.  It
is appropriate for massless scalar models in configuration space; in
the examples we shall always envisage the $\vf^4_4$ model. 
Nevertheless, similar constructions hold in any given quantum field
theory, such as the massive $\vf^3_6$ model considered
in~\cite{ConnesKrRHI}.  Just like in the original paper by Connes and
Kreimer and its follow-ups, we avoid cumbersome formalism by relying
on the pictorial intuition provided by the diagrams themselves.  The
reader is by now supposed thoroughly familiar with the concept of
(superficial) degree of divergence.

We recall that a \textit{graph} or diagram $\Ga$ of the theory is
specified by a set $V(\Ga)$ of \textit{vertices} and a set $L(\Ga)$ of
\textit{lines} (propagators) among them; \textit{external} lines are
attached to only one vertex each, \textit{internal} lines to two.
Diagrams with no external lines (`vacuum diagrams') will not be taken
into account; in $\vf^4_4$ theory only graphs with an even number
of external lines are to be found. We are also excluding diagrams with
a single vertex, and thus it is desirable to banish tadpole diagrams,
in which a line connects a vertex to itself, as well.

Given a graph $\Ga$, a \textit{subdiagram} $\ga$ of $\Ga$ is specified
by a subset of at least two elements of~$V(\Ga)$ and a subset of the
lines that join these vertices in~$\Ga$. By exception, the empty
subset~$\emptyset$ will be admitted as a subdiagram of~$\Ga$. As well
as~$\Ga$ itself. Clearly, the external lines for a subdiagram~$\ga$
include not only a subset of original incident lines, but also some
internal lines of~$\Ga$ not included in~$\ga$.

The connected pieces of $\Ga$ are the maximal connected subdiagrams. A
diagram is called \textit{proper} (or 1PI) when the number of its
connected pieces would not increase on the removal of a single
internal line; otherwise it is \textit{improper}. An improper graph is
the union of proper components plus subdiagrams containing a single
line. A diagram is called~1VI when the number of its connected pieces
can increase upon the removal of a single vertex.

A \textit{subgraph} of a proper graph is a subdiagram that contains
all the elements of $L(\Ga)$ joining its vertices in the whole graph;
as such, it is determined solely by the vertices. A subgraph of an
improper graph $\Ga$, distinct from $\Ga$ itself, is a proper
subdiagram each of whose components is a subgraph with respect to the
proper components of~$\Ga$.

\begin{figure}[htb]
\centering
\begin{picture}(120,20)
\put(-5,20){\line(1,0){130}}
\put(-5,-20){\line(1,0){130}}
\qbezier(30,20)(-5,0)(30,-20)
\qbezier(30,20)(65,0)(30,-20)
\qbezier(90,20)(55,0)(90,-20)
\qbezier(90,20)(125,0)(90,-20)
\put(9,-25){\dashbox{1}(42,50)}
\put(69,-25){\dashbox{1}(42,50)}
\end{picture}
\vspace{2pc}
\caption{The ``roll'' in $\vf^4_4$ theory}
\label{fig:roll}
\end{figure}
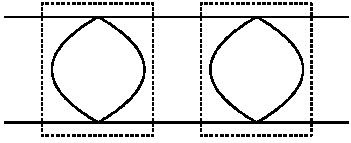

We write $\ga \subseteq \Ga$ if and only if $\ga$ is a subgraph of
$\Ga$ as defined (not just a subdiagram), although practically
everything we have to say would work the same with a less restrictive
definition.  (For renormalization in configuration space, it is more
convenient to deal with subgraphs than with more general subdiagrams. 
Moreover, Zimmermann showed long ago that only subtractions
corresponding to subgraphs that are renormalization parts need to be
used~\cite{ZimVW} in renormalization, and this dispenses us from
dealing with a more general definition.)  When a subdiagram contains
several connected pieces, each one of them being a subgraph, we still
call it a subgraph.  For example, Figure~\ref{fig:roll} illustrates
the case of a subgraph of the $\vf^4_4$ model, made out of two
connected pieces, which in spite of containing all the vertices, does
not coincide with the whole graph.

Two subgraphs $\ga_1,\ga_2$ of $\Ga$ are said to be
\textit{nonoverlapping} when $\ga_1 \cap \ga_2 = \emptyset$ (that is,
$\ga_1$ and~$\ga_2$ have no common vertices) or $\ga_1 \subseteq
\ga_2$ or $\ga_2 \subseteq \ga_1$; otherwise they are overlapping.
Given $\ga \subseteq \Ga$, the quotient graph or cograph $\Ga/\ga$
(reduced graph in Zimmermann's~\cite{Zim} parlance) is defined by
shrinking each \textit{connected component} of~$\ga$ in $\Ga$ to a
point, that is to say, each piece of $\ga$ (bereft of its external
lines) is considered as a vertex of $\Ga$, and all the lines in $\Ga$
not belonging to $\ga$ belong to $\Ga/\ga$. This is modified in the
obvious way when $\ga$ represents a propagator correction (i.e., has
two external lines). A nonempty $\Ga/\ga$ will be proper iff $\Ga$ is
proper. The graphs $\Ga$ and $\Ga/\ga$ have the same external
structure, that is, structure of external lines. It is useful to think
of the external structure as a kind of colour: although subgraphs of a
given graph $\Ga$ may have colours different from~$\Ga$, cographs
cannot.

\begin{figure}[ht]
\begin{center}
\vspace{2pc}
\parbox{5pc}{
\begin{picture}(50,18)
\put(-20,5){$\Delta$}
\put(-10,5){$\Biggl($}
\put(25,10){\circle{40}}
\put(25,-10){\circle{20}}
\qbezier(25,30)(35,10)(50,20)
\qbezier(25,30)(15,10)(0,20)
\put(52,5){$\Biggr)$}
\end{picture}
}\qquad
would contain the term
\qquad
\parbox{7pc}{
\begin{picture}(80,10)
\put(0,5){\line(-1,2){5}}
\put(0,5){\line(-1,-2){5}}
\put(10,5){\circle{20}}
\put(30.2,5){\circle{20}}
\put(40.2,5){\line(1,2){5}}
\put(40.2,5){\line(1,-2){5}}
\put(52,2){$\otimes$}
\put(85,5){\circle{30}}
\put(85,-10){\circle{15}}
\put(85,20){\line(3,2){10}}
\put(85,20){\line(-3,2){10}}
\put(82,18){$\bullet$}
\end{picture}
}
\end{center}
\vspace{1pc}
\caption{Cograph which is a tadpole part}
\label{fig:tadpole-corr}
\vspace{1pc}
\end{figure}
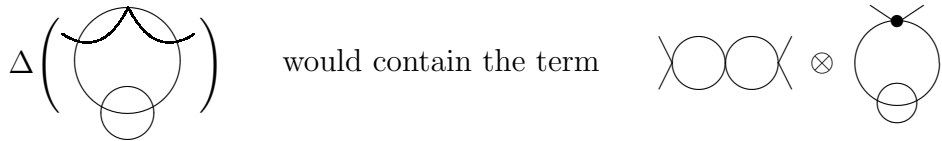

\marker
Now, a bialgebra $\H$ is defined as the polynomial algebra generated
by the empty set $\emptyset$ and the connected Feynman graphs that are
(superficially) divergent and/or have (superficially) divergent
subgraphs (renormalization parts in Zimmermann's parlance), with set
union as the product operation. Hence $\emptyset$ is the unit element
$1\in\H$. The counit is given by $\eta(\Ga) := 0$ on any generator,
except $\eta(\emptyset) = 1$. The really telling operation is the
coproduct $\Dl: \H \to \H \ox \H$; as it is to be a homomorphism of
the algebra structure, we need only define it on connected diagrams.
By definition, the (reduced) coproduct of $\Ga$ is given by
\begin{equation}
\Dl'\Ga\, :=
\sum_{\emptyset \varsubsetneq \ga \varsubsetneq \Ga} \ga \ox \Ga/\ga.
\label{eq:Feynman-coprod}%
\end{equation}
The sum is over all divergent, proper, not necessarily connected
subgraphs of $\Ga$ (not including $\emptyset$ and $\Ga$) such that
\textit{each piece} is divergent, and such that $\Ga/\ga$ is not a
tadpole part. We put $\Ga/\Ga=1$. When appropriate, the sum runs also
over different types of local counterterms associated
to~$\ga$~\cite{ConnesKrRHI}; this is not needed in our model example,
because it corresponds to a massless theory, where propagators carry
only one type of counterterms. It is then suggestive that the concept
of primitive element for~$\Dl$ coincides with that of primitive
diagram in~QFT. We show in Figure~\ref{fig:tadpole-corr} how
appearance of tadpole parts in $\Ga/\ga$ can happen. The cograph
corresponding to the ``bikini'' subgraph in the upper part of the
graph in Figure~\ref{fig:tadpole-corr} is a tadpole correction. It
gives rise to a one-vertex reducible subgraph that can be
suppressed~\cite{Pastores}.

For the proof of the bialgebra properties of $\H$, we refer
to~\cite{Polaris}; for graphical examples of coproducts,
see~\cite{Etoile}. From now on, we restrict consideration to the
nontrivial subbialgebra of proper and superficially divergent graphs
---the situation considered in~\cite{ConnesKrRHI}--- that we still
refer to as~$\H$. Actually $\H$ is a graded bialgebra. Obvious grading
operators are available: if $\#(\Ga)$ denotes the number of vertices
in $\Ga$ (i.e., the coupling order), then we define the degree of a
generator (connected element) $\Ga$ as $\nu(\Ga) := \#(\Ga) - 1$; the
degree of a product is the sum of the degrees of the factors. This
grading is compatible with the coproduct, and clearly scalars are the
only degree~0 elements. Other gradings are by the number $I(\Ga)$ of
internal lines in $\Ga$ and by loop number $l(\Ga):=I(\Ga)-\nu(\Ga)$.
For the $\vf^4_4$ model, $l(\Ga) = \nu(\Ga) + 1$ for two-point
graphs and $l(\Ga) = \nu(\Ga)$ for four-point graphs.

By our remarks in Section~5, $\H$ is a Hopf algebra, and we have at
least two formulae for the antipode: $S_G$ and $S_B$. Nevertheless, in
this context it is preferable to write the antipode in a more
pictorial language, taking advantage of geometrical intuition.

\begin{defn}
\label{df:chain}
A \textit{chain} $\Cc$ of a proper, connected graph $\Ga$ is a
sequence $\ga_1 \varsubsetneq\ga_2\varsubsetneq \cdots \varsubsetneq
\ga_k$ of proper, divergent, \textit{not necessarily connected}
subgraphs of $\Ga$, not including $\emptyset$ and $\Ga$. We denote by
$\Ch(\Ga)$ the set of chains of $\Ga$. The \textit{length} of a chain
$\Cc$ is the number of `links' $l(\Cc) = k + 1$, and we write
$\Om(\Cc) := \ga_1\,(\ga_2/\ga_1) \cdots (\ga_k/\ga_{k-1})
\,(\Ga/\ga_k)$. We allow also the empty set to be a chain.
\end{defn}

With this notation we can define the antipode as follows:
\begin{equation}
S_\DS(\Ga) := \sum_{\Cc\in\Ch(\Ga)} (-1)^{l(\Cc)} \,\Om(\Cc).
\label{eq:ds-antip}%
\end{equation}

This definition corresponds, on the one hand, to the correct version
of the Dyson--Salam formula for renormalization. On the other hand,
formula~\eqref{eq:ds-antip} looks similar to the explicit expression
for the antipode given by Schmitt for his incidence Hopf
algebras~\cite{Schmitt1}; we shall come back to that.

We show that $S_\DS$ is nothing but a reformulation of $S_G$; in other
words:

\begin{thm}
\label{pr:Dyson-equals-chains}
The expansion of $S_\DS$ coincides identically with the
expansion of~$S_G$.
\end{thm}

\begin{proof}
First, given a proper, connected graph $\Ga$, we rewrite
$$
S_\DS(\Ga) := \sum_k (-1)^{k+1} \sum_{\Cc\in\Ch_k(\Ga)} \Om(\Cc),
$$
where $\Ch_k(\Ga)$ denote the set of chains of length~$k+1$. Thus, it
is enough to prove that
$$
(-1)^{k+1} \sum_{\Cc\in\Ch_k(\Ga)} \Om(\Cc) =
(u \eta -\id)^{*(k+1)}(\Ga).
$$
For this purpose, first notice that
\begin{equation}
\sum_{\emptyset \varsubsetneq \ga_1 \varsubsetneq \ga_2 \cdots
\varsubsetneq \ga_k}
\ga_1 \ox \ga_2/\ga_1 \oxyox \ga_k/\ga_{k-1} \ox \Ga/\ga_k
= \Dl'_k \cdots \Dl'_2 \Dl'\Ga.
\label{eq:foldup}%
\end{equation}
Indeed, by definition of the coproduct the statement is true for
$k = 1$. Moreover, if \eqref{eq:foldup} holds for $k-1$, then
\begin{align*}
\sum_{\emptyset \varsubsetneq \ga_1 \varsubsetneq \ga_2 \cdots
\varsubsetneq \ga_k \varsubsetneq \Ga}
&\ga_1 \ox \ga_2/\ga_1 \oxyox \ga_k/\ga_{k-1} \ox \Ga/\ga_k
\\
&= \sum_{\emptyset \varsubsetneq \ga_2 \varsubsetneq \ga_3
\cdots \varsubsetneq \ga_k  \varsubsetneq \Ga}
\Dl'_k(\ga_2 \ox \ga_3/\ga_2 \oxyox \ga_k/\ga_{k-1} \ox
\Ga/\ga_k)
\\
&= \Dl'_k \biggl( \sum_{\emptyset \varsubsetneq \ga_1 \varsubsetneq
\ga_2 \cdots\varsubsetneq \ga_{k-1}  \varsubsetneq \Ga}
\ga_1 \ox \ga_2/\ga_1 \oxyox \ga_{k-1}/\ga_{k-2} \ox\Ga/\ga_{k-1}
\biggr)
\\
&= \Dl'_k \Dl'_{k-1} \cdots \Dl'_2 \Dl'\Ga.
\end{align*}
Thus, by \eqref{eq:quasicojo-formula}
\begin{align*}
(-1)^{k+1} \sum_{\Cc\in\Ch_k(\Ga)} \Om(\Cc)
&= (-1)^{k+1} \sum_{\emptyset \varsubsetneq \ga_1 \varsubsetneq \ga_2
\cdots \varsubsetneq \ga_k  \varsubsetneq \Ga}
\ga_1 \, \bigl(\ga_2/\ga_1\bigr) \cdots
\bigl( \ga_k/\ga_{k-1} \bigr) \, \bigl(\Ga/\ga_k \bigr)
\\
&= (-1)^{k+1} m\,m_2 \cdots m_k\,\Dl'_k \cdots \Dl'_2 \Dl' (\Ga)
= (u \eta -\id){*(k+1)}(\Ga).
\hspace{1.8em} \qed
\end{align*}
\hideqed
\end{proof}

The proof shows how the chains are generated from the coproduct.

For illustration, now that we are at that, we
record~\eqref{eq:Bogol-recur} in the language of the algebra of
graphs:
$$
S_B\Ga := - \Ga - \sum_{\emptyset \varsubsetneq \ga \varsubsetneq \Ga}
(S_B\ga)\, \Ga/\ga.
$$

Let us illustrate as well the construction of the graded dual Hopf
algebra $H'$, for Hopf algebras of Feynman graphs. Each connected
element $\Ga$ gives a derivation or element $Z_\Ga \: \H \to \C$ of
the Lie algebra of infinitesimal characters on~$H$, defined by
\begin{align*}
\dst{Z_\Ga}{\Ga_1\cdots \Ga_k} &= 0
\quad\text{unless $k=1$ and $\Ga_1 = \Ga$};
\\
\dst{Z_\Ga}{\Ga} &= 1.
\end{align*}
Also, $\dst{Z_\Ga}{1} = 0$ since $Z_\Ga\in\Der_\eta\H$. Clearly any
derivation~$\dl$ vanishes on the ideal generated by products of two or
more connected elements. Therefore, derivations are determined by
their values on the subspace spanned by single graphs, and reduce to
linear forms on this subspace.

\section{Breaking the chains: the formula of Zimmermann}

\begin{defn}
\label{df:forest}
A (normal) \textit{forest} $\F$ of a proper, connected graph $\Ga$ is
a set of proper, divergent and connected subgraphs, not including
$\emptyset$ and $\Ga$, such that any pair of elements are
nonoverlapping. $F(\Ga)$ denotes the set of forests of $\Ga$. The
\textit{density} of a forest $\F$ is the number $d(\F) = |\F| + 1$,
where~$|\F|$ is the number of elements of $\F$. We allow also the
empty set to be a forest. Given $\ga\in\F\cup\{\Ga\}$ we say that
$\ga'$ is a \textit{predecessor} of~$\ga$ in~$\F$ if
$\ga'\varsubsetneq \ga$ and there is no element $\ga''$ in $\F$ such
that $\ga'\varsubsetneq \ga'' \varsubsetneq \ga$; that is to say,
$\ga$ covers $\ga'$. Let
$$
\Th(\F) := \prod_{\ga \in \F\cup\{\Ga\}}  \ga / \tilde\ga,
$$
where $\tilde\ga$ denote the disjoint union of all predecessors of
$\ga$ in~$\F$. When $\ga$ is minimal, $\tilde\ga = \emptyset$, and
$\ga/\tilde\ga = \ga$.
\end{defn}

Notice that if a forest $\F$ is a chain, then $\Th(\F) =\Om(\F)$, and
conversely if a chain $\Cc$ is a forest, $\Om(\Cc) = \Th(\Cc)$.
Obviously not every forest is a chain; but also not every chain is a
forest, because product subgraphs can occur in chains and cannot in
forests.

Zimmermann's version for the antipode is given by
\begin{equation}
S_Z(\Ga) := \sum_{\F \in F(\Ga)} (-1)^{d(\F)} \Th(\F).
\label{eq:Zim-glory}
\end{equation}
We assert that $S_Z$ provides yet another formula for the antipode of
$\H$. A standard proof of this, except for minor combinatorial
details, is just the mechanical one of checking that a given formula
yields a convolution inverse for $\id$~\cite{Hektor}. It is not
entirely uninformative; but gives little insight into why there should
be no cancellations in the actual computation; and cancellations
in~\eqref{eq:Zim-glory} there are not, since in order to have
$\Th(\F)=\Th(\F')$ it is required that $d(\F)=d(\F')$! Instead of
repeating such a proof, and with a view to generalizations, here we
give a longer, but elementary and much more instructive treatment that
(we hope) clarifies how chains of different length give cancelling
contributions.

\begin{figure}[ht]
\begin{center}
\vspace{1pc}
\begin{picture}(0,50)
\put(0,0){\line(3,1){72}}
\put(0,0){\line(-3,1){72}}
\put(0,0){\line(1,2){20}}
\put(0,0){\line(-1,2){20}}
\put(20,40){\line(2,-1){52}}
\put(-20,40){\line(-2,-1){52}}
\qbezier(-20,40)(0,60)(20,40)
\qbezier(-20,40)(0,20)(20,40)
\end{picture}
\end{center}
\caption{Diagram $\Ga$ without extra cancellations in $S_Z(\Ga)$ with
respect to $S_{DS}(\Ga)$}
\label{fig:diagrama-prieto}
\vspace{1pc}
\end{figure}
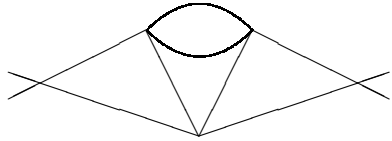

Of course the assertion means that in general there are fewer forests
than chains. Experience with Feynman graphs indicates that overlapping
tends to reduce the cancellation phenomenon. There are ``very
overlapping'' diagrams like the one in the~$\vf^4_4$ model
pictured in Figure~\ref{fig:diagrama-prieto}, for which the sets of
chains and forests coincide.

Despite examples like this one, it is apparent that Zimmermann's
formula is, from the combinatorial viewpoint, altogether subtler than
those of Bogoliubov or Dyson and Salam. We plunge now into showing how
all the cancellations implicit in the Dyson--Salam approach for graphs
are taken into account and \textit{suppressed}. To each chain $\Cc$ of
$\Ga$ we associate the forest $\F_\Cc$ consisting of all the connected
components of the different elements of~$\Cc$.

\begin{lema}
\label{pr:OmequalTh}%
Let $\Cc$ be a chain; then $\Om(\Cc) = \Th(\F_\Cc)$.
\end{lema}

\begin{proof}
Clearly the statement is true for chains with only one element. Assume
the result holds for chains in $\bigcup_{i=2}^{k-1} \Ch_i(\Ga)$ where
$\Ga$ is an arbitrary graph in~$\H$, and let $\Cc := \ga_1 \subset
\ga_2\subset \cdots \subset \ga_k$ be a chain in $\Ch_k(\Ga)$. Then
$\D :=\ga_1 \subset \ga_2\subset \cdots \subset \ga_{k-1}$ belongs to
$\Ch_{k-1}(\ga_k)$, and if $\la_1, \dots , \la_s$ are the connected
components of $\ga_k$, then $\F_\Cc = \F_\D \cup
\{\la_1,\dots,\la_s\}$. Now, clearly $\la_1, \dots , \la_s$ are the
maximal elements of $\F_\Cc$, therefore, in $\F_\Cc$, $\Ga/\ga_k =
\Ga/\la_1 \cdots \la_s = \Ga/ \tilde\Ga$, thus by the induction
hypothesis
$$
\Om(\Cc) = \Om(\D) \bigl(\Ga/\ga_k \bigr)
= \Th(\F_\D) \bigl(\Ga/\ga_k \bigr)
= \prod_{\ga\in \F_\Cc \bigcup \{\Ga\}} \ga/\tilde \ga
= \Th(\F_\Cc).
\eqno\qed
$$
\hideqed
\end{proof}

We can as well associate chains to forests in such a way that, if
$\Cc$ has been associated to $\F$, then $\F_\Cc=\F$. If a forest~$\F$
is a chain, then $\F$ itself is the only chain so associated to $\F$,
and clearly $d(\F) = \ell(\F)$. Each forest $\F$ has associated at
least one chain. Indeed, let $L_1 := \{\la^1_1,\dots,\la^1_{n_1}\}$ be
the list of maximal elements of $\F$, let $L_2 :=
\{\la^2_1,\dots,\la^2_{n_2}\}$ be the set of predecessors in $\F$ of
elements in $L_1$, let $L_3$ be the collection of predecessors in~$\F$
of elements in~$L_2$, and so on. If the listing of the elements
of~$\F$ is completed in the $k$-th iteration, we let $\Cc_\F$ be
the chain
$$
\Cc_\F := \la^k_1 \cdots \la^k_{n_k} \subset \cdots \subset
\la^2_1\cdots \la^2_{n_2} \subset \la^1_1 \cdots \la^1_{n_1};
$$
which by definition is associated to $\F$. We denote by $\Ch(\F)$ the
set of chains (of different length in general) associated to a given
forest $\F$ of $\Ga$.

So far, we find
$$
S_{DS}(\Ga) = \sum_{\Cc\in \Ch(\Ga)} (-1)^{\ell(\Cc)} \Om(\Cc)
= \sum_{\Cc\in \Ch(\Ga)} (-1)^{\ell(\Cc)} \Th(\F_\Cc)
= \sum_{\F\in F(\Ga)}
\biggl( \sum_{\Cc\in \Ch(\F)} (-1)^{\ell(\Cc)} \biggr) \Th(\F).
$$

\begin{thm}
\label{pr:Zimmerman}
For each proper, connected graph $\Ga$, $S_{DS}(\Ga) = S_Z(\Ga)$.
\end{thm}

\begin{proof}
To conclude that $S_{DS}(\Ga) = S_Z(\Ga)$ it remains to prove that for
each forest $\F$, $\sum_{\Cc\in \Ch(\F)} (-1)^{\ell(\Cc)} =
(-1)^{d(\F)}$. For this the idea is to enumerate the chains associated
to a forest with $k+1$ elements from a list of the chains associated
to the forest obtained by deleting one element. Thus we proceed by
induction on the density of the forests. By the previous remarks the
statement holds if $\F$ is a chain, in particular if $\F$ has only one
element. Suppose that the claim holds whenever $\F$ has at most $k$
elements, and let $\G$ be a forest with $k+1$ elements, say $\G = \F
\cup \{\ga\}$. Let $\Cc_1, \dots, \Cc_r$ be the collection of all
chains in $\Ch(\F)$. If $\Cc_i$ is the chain $\ga^i_1\subset \ga^i_2
\subset \cdots \subset \ga^i_{k_i}$, let $\ga^i_{t_i}$ be the first
(from the left) term of the chain $\Cc_i$ such that a connected
component of it covers $\ga$. If no term satisfies that condition, set
$t_i = k_i + 1$. On the other hand, let $\ga^i_{s_i}$ be the last
(from the right) link of the chain with (one or several) connected
components contained in~$\ga$, not appearing in a previous link of the
chain. If no $\ga^i_j$ has connected components contained in $\ga$,
take $s_i=0$. We shall denote by~$\widehat\ga^i_j$ the element of $\H$
obtained from $\ga^i_j$ by replacing the product of all its connected
components contained in~$\ga$ by~$\ga$ itself. Then we can construct
the following $2(t_i-s_i)-1$ chains in~$\Ch(\G)$:
\begin{align*}
\Cc^0_i &:=
\ga^i_1\subset \cdots \subset \ga^i_{s_i} \subset
\widehat\ga^i_{s_i} \subset \widehat\ga^i_{s_i+1}
\subset \cdots \subset
\widehat\ga^i_{t_i-1} \subset \ga^i_{t_i}
\subset \cdots \subset \ga^i_{k_i}
\\
\Cc^1_i &:=
\ga^i_1\subset \cdots \subset \ga^i_{s_i} \subset
\widehat\ga^i_{s_i+1} \subset \widehat\ga^i_{s_i+2}
\subset \cdots \subset
\widehat\ga^i_{t_i-1} \subset \ga^i_{t_i}
\subset \cdots \subset \ga^i_{k_i}
\\
\Cc^2_i &:=
\ga^i_1\subset \cdots \subset \ga^i_{s_i+1} \subset
\widehat\ga^i_{s_i+1} \subset \widehat\ga^i_{s_i+2}
\subset \cdots \subset
\widehat\ga^i_{t_i-1} \subset \ga^i_{t_i}
\subset \cdots \subset \ga^i_{k_i}
\\
&\hskip 4cm   \vdots
\\
\Cc^{2j-1}_i &:=
\ga^i_1\subset \cdots \subset \ga^i_{s_i+j-1} \subset
\widehat\ga^i_{s_i+j} \subset \cdots \subset \widehat\ga^i_{t_i-1}
\subset \ga^i_{t_i} \subset \cdots \subset \ga^i_{k_i}
\\
\Cc^{2j}_i &:=
\ga^i_1\subset \cdots \subset \ga^i_{s_i+j} \subset
\widehat\ga^i_{s_i+j} \subset \cdots \subset \widehat\ga^i_{t_i-1}
\subset \ga^i_{t_i} \subset \cdots \subset \ga^i_{k_i} \\
&\hskip 4cm   \vdots
\\
\Cc^{2(t_i-s_i)-3}_i &:=
\ga^i_1\subset \cdots \subset \ga^i_{t_i-2}
\subset \widehat\ga^i_{t_i-1} \subset \ga^i_{t_i}
\subset \cdots \subset \ga^i_{k_i}
\\
\Cc^{2(t_i-s_i)-2}_i &:=
\ga^i_1\subset \cdots \subset \ga^i_{t_i-1}
\subset \widehat\ga^i_{t_i-1} \subset \ga^i_{t_i}
\subset \cdots \subset \ga^i_{k_i}\,.
\end{align*}
Notice that $\ell(\Cc^{2u}_i) = \ell(\Cc_i) +1$ for $u= 0,1,\dots,
(t_i-s_i)-1$, whereas $\ell(\Cc^{2v -1}_i) = \ell(\Cc_i)$ for $v=
1,2,\dots, (t_i-s_i)-1$.  Since every chain in $\Ch(\G)$  is equal to
$\Cc^j_i$ for some pair $(i,j)$, it follows, by the induction
hypothesis, that
\begin{align*}
\sum_{\Cc\in \Ch(\G)} (-1)^{\ell(\Cc)}
&= \sum_{i=1}^r \sum_{j=0}^{2(t_i-s_i)-2} (-1)^{\ell(\Cc^j_i)}
= \sum_{i=1}^r (-1)^{\ell(\Cc^0_i)} \\
&= - \sum_{i=1}^r (-1)^{\ell(\Cc_i)}
= - (-1)^{d(\F)} = (-1)^{d(\G)}.
\tag*\qed
\end{align*}
\hideqed
\end{proof}

\section{Incidence Hopf algebras}

The kinship between the Connes--Moscovici algebra, the Kreimer Hopf
algebra~\cite{KreimerOriginal} of renor\-malization and
Connes--Kreimer algebras of rooted trees and Feynman graphs is by now
well known; it was the discovery of this kinship what gave the current
impulse to the subject. Yet the ``classical'' \bfb\ algebra appears to
be of the same kind, and it is known to fit in the framework of
incidence bialgebras of lattices and posets, that goes back to the
pioneering work by Rota, together with Joni, Doubilet and
Stanley~\cite{JoniR,Doub}. Rota introduced bialgebras in
combinatorics, with coproducts becoming natural tools to systematize
decompositions of posets and other combinatorial objects. This has
been developed by Schmitt~\cite{Schmitt1,Schmitt2}. The lecture notes
by D\"ur~\cite{Duroconellos} also touch upon the subject.

We want to explore the relations between Connes--Kreimer--Moscovici
algebras and the lore of incidence Hopf algebras, eventually
culminating in the importation of Zimmermann's formula into the latter
realm. Before that, we will obtain the \bfb\ algebra $\F$ (yes, the
symbol $\F$ is overworked in this paper) anew as the incidence Hopf
algebra on the family of partitions of finite sets.

\medskip

A family $\Pc$ of finite intervals (that is, finite partially ordered
sets, or \textit{posets} for short, of the form
$\{x,z\} := \set{y : x\le y \le z}$) is called (interval)
\textit{closed} if it contains all the subintervals of its elements;
that is, if $x\le y\in P\in\Pc$, then $\{x,y\}\in\Pc$. We will write
$x<y$ whenever $x\le y$ and $x\ne y$. Analogously to Section~11, we
will say that $y$ covers~$x$ whenever $x<y$ and there is no~$z$ such
that $x<z<y$. If $P=\{x,y\}\in\Pc$, denote $0_P := x$ and $1_P := y$.

An \textit{order-compatible equivalence} relation on a closed family
$\Pc$ is an equivalence relation~$\sim$ such that whenever the
intervals $P,Q$ are equivalent in $\Pc$, then there exists a bijection
$\phi:P\to Q$ such that $\{0_P,x\}\sim\{0_Q,\phi(x)\}$ and
$\{x,1_P\}\sim\{\phi(x),1_Q\}$ for all $x\in P$. Denote by~$[P]$
the equivalence class of $P$ in the quotient family $\Pc_\sim$; such
equivalence classes are called \textit{types}.

\begin{defn}
\label{df:inccoal}
Let $C(\Pc_\sim)$ be
the vector space generated by the types. It is a coalgebra under the
maps $\Dl: C(\Pc_\sim)\to C(\Pc_\sim)\ox C(\Pc_\sim)$ and $\eta:
C(\Pc_\sim)\to\C$ defined by
\begin{equation}
\Dl[P] = \sum_{x\in P}[\{0_P,x\}]\ox[\{x,1_P\}],
\label{eq:incidencedelta}
\end{equation}
and
$$
\eta[P] = 1 \sepword{if} |P| = 1,  \qquad
\eta[P] = 0 \text{ otherwise}.
$$
Since $\sim$ is order-compatible, $\Dl$ is well defined. The
verifications are elementary. We call~$C(\Pc_\sim)$ the
\textit{incidence coalgebra} of $\Pc$ modulo $\sim$.
\end{defn}

It is also easily seen that all incidence coalgebras are quotients of
the ``full'' incidence coalgebra, associated to the trivial
(isomorphism) equivalence relation, by the coideals spanned by
elements $P-Q$, when~$P\sim Q$.

The family $\Pc$ is \textit{hereditary} if it is closed as well under
Cartesian products: $P \x Q \in \Pc$ for every pair $P,Q$ in $\Pc$;
the partial order $(x_1,y_1)\le(x_2,y_2)$ iff $x_1\le x_2$ in~$P$ and
$y_1\le y_2$ in~$Q$ is understood. A \textit{chain} in $\Pc$ is an
interval $\Cc$ such that $x,y\in\Cc$ implies either~$x\le y$ or~$y\le
x$. A chain~$\Cc$ is said to have length $\ell(\Cc)= n$ if
$|\Cc|=n+1$. Clearly a chain of length~$n$ can be written as
$x_0<x_1<\cdots<x_n$. The set of chains that are subintervals of an
interval $P$, such that $x_0=0_P$ and $x_n=1_P$, is denoted
by~$\Ch(P)$. We write $\Om(\Cc)$ for the type of the Cartesian product
$\{x_0,x_1\}\x\{x_1,x_2\}\x \cdots\x\{x_{n-1},x_n\}$. A chain is
saturated if no further `links' can be found between its elements;
that is, for $1\le i\le n,\,x_{i-1}\le y \le x_i$ implies $y=x_{i-1}$
or $y=x_i$.

Incidence coalgebras are filtered by length in a natural way. An
interval $P$ is said to have \textit{length} $n$ if the longest chain
in~$P$ has length $\ell(P)=n$. If $P\sim Q$, then $l(P)=l(Q)$, so
length is well defined on the set of types; also clearly $\Ch(P)\sim
\Ch(Q)$ if~$P\sim Q$. Let $C_n$ be the vector subspace
of~$C(\Pc_\sim)$ generated by types of length $n$ or less; then
$C_0\subseteq C_1\subseteq\ldots$ is a filtering of~$C(\Pc_\sim)$.
Indeed, for any interval $P$ and $x$ in $P$, the union of a chain of
$\{0_P,x\}$ and a chain of $\{x,1_P\}$ gives a chain of~$P$, hence
$l([\{0_P,x\}])+l([\{x,1_P\}])\le l(P)$, and so $\Dl C_n \subseteq
\sum_{k=0}^nC_k\ox C_{n-k}$. An interval is \textit{graded} when the
length of all saturated chains between $0_P$ and $1_P$ is the same.
Whenever all elements of the family~$\Pc$ are graded, $C(\Pc_\sim)$ is
a graded coalgebra.

\medskip

Suppose furthermore that there is an order-compatible equivalence
relation $\sim$ on a hereditary family $\Pc$ satisfying:
\begin{enumerate}
\item[(i)] if $P \sim Q$, then $P \x R \sim Q \x R$ and
$R \x P \sim R \x Q$ for all $R \in \Pc$.
\item[(ii)] if $Q$ has just one element, then
$P \x Q \sim P \sim Q \x P$.
\end{enumerate}

In particular, the second of these conditions imply that all one-point
intervals are of the same type (hereinafter denoted by~1). If they both
hold, then $C(\Pc_\sim)$ turns out to be a connected bialgebra, with
the product induced by the Cartesian product of intervals. In this
context, types that are not (nontrivial) Cartesian products are called
\textit{indecomposables}. The set of indecomposable types is
denoted~$\Pc_\sim^\circ$.

\begin{thm}
\label{pr:incidence-areHopf}
Let us rebaptize $C(\Pc_\sim)$ as $H(\Pc_\sim)$. In view of our
remarks in Section~5 it is in fact a Hopf algebra, called an
\textit{incidence Hopf algebra}; and a formula for the antipode is
\begin{equation}
S_\DS[P] = \sum_{\Cc \in \Ch([P])}(-1)^{l(\Cc)}\Om(\Cc).
\label{eq:manes-de-DS}
\end{equation}
\end{thm}

\begin{proof}
We follow our argument of~\cite{Hektor} for the Hopf algebras of
Feynman graphs, which translates almost verbatim into general incidence
algebras theory; whereby it is directly checked that $S_\DS$ is an
inverse of $\id$ under convolution.  By definition
\begin{align*}
(S_\DS * \id) [P]
&= \sum_{x\in P} S_\DS([\{0_P,x\}]) \, [\{x,1_P\}]
= S_\DS([P])
+ \sum_{0_P\le x<1_P} S_\DS([\{0_P,x\}])\, [\{x,1_P\}]
\\
&= S_\DS([P]) + \sum_{0_P\le x<1_P}\,
\sum_{\D \in \Ch(\{0_P,x\})} (-1)^{l(\D)} \Om(\D) \, [\{x,1_P\}].
\end{align*}
Now, if $\D \in \Ch(\{0_P,x\})$, say $\D = x_0<x_1<\dots<x_n =x$, then
$\Cc = x_0<x_1<\dots<x_n<x_{n+1}=1_P \in \Ch(P)$. Moreover,
\begin{equation}
\Om(\Cc) = \Om(\D) \, [\{x,1_P\}], \sepword{and}\quad
l(\Cc) = l(\D) + 1.
\label{eq:chain-length}%
\end{equation}
Reciprocally, given a chain $\Cc = x_0<x_1<\dots<x_{n+1} \in \Ch(P)$,
then $\D = x_0<x_1<\dots<x_n =: x \in \Ch(\{0_P,x\})$, and
\eqref{eq:chain-length} holds. Therefore
$$
S_\DS * \id ([P])
= S_\DS([P]) - \sum_{\Cc \in \Ch(P)} (-1)^{l(\Cc)} \Om(\Cc)
= 0 = u \eta([P]);
$$
in other words, $S_\DS$ is a left inverse for $\id$, and therefore
it is an antipode.
\end{proof}

It is still possible to have a Hopf algebra, albeit not a connected
one, by weakening condition~(ii) above: one requires that a neutral
element~1 exist in~$\Pc$ such that $P\sim P\x1\sim1\x P$ for all
$P\in\Pc$, and that all $P\in\Pc$ with $|P|=1$ be invertible.

By now the reader should not be surprised to find that the expansions
for $S_\DS$ and $S_G$ are identical: one formally can reproduce
exactly the argument of Theorem~\ref{pr:Dyson-equals-chains} to obtain
$$
m^k {\Dl'}^k([P]) = \sum_{0_P=x_0<x_1<\cdots<x_{k+1}=1_P} \
\prod_{i=1}^{k+1} [\{x_{i-1}, x_i\}]
= \sum_{\Cc\in\Ch_k(P)} \Om(\Cc).
$$
In summary, the (correct version of) the Dyson--Salam scheme for
renormalization is analogous to the explicit expression for the
antipode~\eqref{eq:manes-de-DS}, typical of incidence Hopf algebras.

\medskip

The dual convolution algebra $H^*(\Pc_\sim)$ of~$H(\Pc_\sim)$ is
called the incidence algebra of~$\Pc$ modulo $\sim$ (unfortunately the
standard terminology is a bit confusing). It can be identified to the
set of maps from $\Pc_\sim$ to $\C$; and the group of multiplicative
functions $\Hom_\alg(H(\Pc_\sim),\C)$ can be identified with the set
of functions whose domain is the set of indecomposable types.
Explicitly,
$$
\dst{f*g}{[P]} = \sum_{x\in
P}\dst{f}{[\{0_P,x\}]}\,\dst{g}{[\{x,1_P\}]}.
$$
In this context, the classical theory of the \textit{M\"obius
function} is reformulated as follows. Incidence Hopf algebras come to
the world with a distinguished character: the \textit{zeta function}
$\zeta$ of~$\Pc$, given by $\dst{\zeta}{[P]}=1$ for all types; it is
patently multiplicative. Its convolution inverse $\mu$ is the
\textit{M\"obius character} or M\"obius function of
posets~\cite{RotaMoebius}. Therefore, $\mu = \zeta S$, and different
formulae for the antipode $S$ provide different ways to compute $\mu$.
For instance,~\eqref{eq:manes-de-DS} gives
$$
\mu([P]) = -1 + c_2 - c_3 +\dots,
$$
where $c_n$ is the number of chains of length~$n$. Rota defined
in~\cite{RotaMoebius} the \textit{Euler characteristic} of $P$ as
$E([P]) := 1 + \mu([P])$, and related it with the classical Euler
characteristic in a suitable homology theory associated to~$P$.

As for the examples: if one considers the family $\Pc$ of finite
Boolean algebras and let $\sim$ be isomorphism, it is seen that $\Pc$
is hereditary; and if we denote by $Y$ the isomorphism class of the
poset of subsets of a one-element set, the associated incidence Hopf
algebra is the binomial Hopf algebra with coproduct~\eqref{eq:binom}.
Also, if we take for~$\Pc$ the family of linearly ordered sets and
$\sim$ the isomorphism relation again, and we let $Y_n$ denote the
type of the linearly ordered set of length~$n$, we recover the ladder
Hopf algebra~$\Hl$, introduced after Theorem~\ref{pr:the-other-half},
as an incidence Hopf algebra.

\marker
The similarity of the previous setup to the theory of algebras of
Feynman graphs is striking. A dictionary for translating the family of
Feynman graphs, reduced by isomorphism classes, into the language of
incidence algebras would then seem to be quickly established,
purporting to identify $\le$ with $\subseteq$ and to
show~\eqref{eq:Feynman-coprod} as an avatar
of~\eqref{eq:incidencedelta}. In order to underline the parallelism,
one could try to depart somewhat from the usual notation: indeed,
writing $\emptyset,\Ga$ for any $0_P,1_P$ respectively; $x$ for
$\{0_P,x\}$ and $\Ga/x$ for $\{x,1_P\}$, would make the notations
completely parallel. It is also clear that \textit{connected} in the
sense of graph types should translate into \textit{indecomposable} in
the incidence algebra framework.

However, the direct attempt to identify Connes--Kreimer algebras and
the Rota incidence algebras does \textit{not} work: $\H$ represents a
particular class of incidence algebras, and it is important to
recognize the traits that characterize it inside that framework: one
can be badly misled by formal analogies.  We show how this can happen
by means of an example (see~\ref{xm:legs} further down) inside the
commutative Hopf algebra of rooted trees.

An (undecorated) \textit{rooted tree} is a connected poset in which
each element (vertex) is covered by at most another element: this
selects a unique distinguished uncovered element, called the root. An
\textit{admissible cut} $c$ of a rooted tree~$T$ is a subset of its
lines such that the path from the root to any other vertex includes at
most one line of~$c$. Deleting the cut branches produces several
subtrees; the component containing the original root (the trunk) is
denoted $R_c(T)$. The remaining branches also form rooted trees, where
in each case the new root is the vertex immediately below the deleted
line; $P_c(T)$ denotes the juxtaposition of these pruned branches.

\begin{defn}
\label{df:CKtrees}
The (Connes--Kreimer) Hopf algebra of \textit{rooted trees} $H_R$ is
the commutative algebra generated by symbols $T$, one for each
isomorphism class of rooted trees, plus a unit~$1$ corresponding to
the empty tree; the product of trees is written as the juxtaposition
of their symbols. A product of trees is called a (rooted)
\textit{wood} (we can hardly call it a forest). The counit $\eta: H_R
\to \R$ is the linear map defined by~$\eta(1) := 1_\R$ and
$\eta(T_1T_2\cdots T_n) = 0$ if $\row T1n$ are trees.
In~\cite{ConnesKrTrees} the coproduct on~$H_R$ is defined as a map
$\Dl \: H_R \to H_R \ox H_R$ on the generators (extending it as an
algebra homomorphism) as follows:
\begin{equation}
\Dl 1 := 1 \ox 1; \quad
\Dl T := T \ox 1 + 1 \ox T + \sum_{c\in C(T)} P_c(T) \ox R_c(T).
\label{eq:coprod-tree}
\end{equation}
Here $C(T)$ is the list of admissible cuts.
\end{defn}

It should be clear that the ladder Hopf algebra $\Hl$ above is a
(cocommutative) Hopf subalgebra of~$H_R$: the type of the linearly
ordered set of length~$n$ is identified to the ``stick'' with~$n$
vertices.

Let the \textit{natural growth} $N\: H_R\to H_R$ be the unique
derivation defined, for each tree $T$ with vertices $\row v1n$, by
$$
N(T) := T_1 + T_2 +\cdots + T_n,
$$
where each $T_j$ is obtained from $T$ by sprouting one new leaf from
each vertex~$v_j$. In particular, let $t_1,t_2$ respectively denote
the trees with one and two vertices, $t_{31},t_{32}$ the two trees
with three vertices, where the root has respectively fertility one and
two; for trees with four vertices, we shall denote by $t_{41}$ the
rooted tree where all vertices have fertility~$1$ (a stick), by
$t_{42}$ and $t_{43}$ the four-vertex trees with $2$ or~$3$ outgoing
lines from the root (respectively a hook and a claw), and by $t_{44}$
the tree whose root has fertility~$1$ and whose only vertex with
length~$1$ has fertility~$2$ (a biped); this being the notation
of~\cite[Ch.~14]{Polaris}. We obtain
\begin{align*}
N(t_1) &= t_2,
\\
N^2(t_1) &= N(t_2) = t_{31} + t_{32},
\\
N^3(t_1) &= N(t_{31} + t_{32}) = t_{41} + 3 t_{42} + t_{43} + t_{44}.
\end{align*}
{}From~\eqref{eq:coprod-tree},
\begin{align*}
\Dl N(t_1) &= N(t_1) \ox 1 + 1 \ox N(t_1) + t_1 \ox t_1,
\\
\Dl N^2(t_1) &= N^2(t_1) \ox 1 + 1 \ox N^2(t_1) + (t_1^2+t_2) \ox t_1
		 + 3t_1 \ox t_2,
\\
\Dl N^3(t_1) &= N^3(t_1) \ox 1 + 1 \ox N^3(t_1) + N^2(t_1) \ox t_1
		   + 4 t_2 \ox t_2
\\
&\qquad + 6 t_1 \ox N^2(t_1) + 7 t_1^2 \ox t_2 + 3 t_1t_2 \ox t_1
		   + t_1^3 \ox t_1;
\end{align*}
this the reader will find familiar, as these are completely analogous
to the expressions for the coproduct of the $\dl_2,\dl_3,\dl_4$ in the
\bfb\ algebra. This is no accident. The number of times that the
tree~$T$ with~$n$ vertices appears in~$N^{n-1}(t_1)$ is known, and the
reader can consult for instance~\cite{ConnesKrTrees}
or~\cite[Section~14.1]{Polaris} for a proof that indeed with the
identifications $t_1\equiv\dl_1,N^{n-1}(t_1)\equiv\dl_n$, the \bfb\
algebra $\F$ is a \textit{Hopf subalgebra} of~$H_R$. By the way, the
identification of the natural growth subalgebra in $H_R$ to $\F$ is
not unexpected, in that there is a relation between differentials and
rooted trees, that goes back to Cayley~\cite{Cayley}, giving precisely
that correspondence; see~\cite{BrouderRK}, and
also~\cite[Section~14.2]{Polaris} and the more
recent~\cite{GirelliKMSecond}.

Hopf algebras of rooted trees have extremely important normative
properties, into which unfortunately we cannot go here. We refer the
reader to~\cite{MaestrosSutiles} and to~\cite{Turaev}. For the
relation between the algebra of rooted trees and the shuffle algebra,
consult~\cite{Murua}.

\begin{xmpl}
\label{xm:legs}
Consider now, for instance, the tree $t_{32}$. With this definition,
its coproduct is
$$
\Dl t_{32} = t_{32} \ox 1 + 1 \ox t_{32} + 2t_1 \ox t_2
+ t_1^2 \ox t_1.
$$
This is shown in Figure~\ref{fig:cut-rule}. Now, we can make every
tree into an interval $\{0_T,1_T\}$ by collating all its leaves to a
notional element $0_T$. But even so, if we think of a tree as just
such a poset and identify $\{0_T,v\}$ with the tree generated by a cut
just above~$v$, and omit the reasonable doubt about the meaning of
$\{v,1_T\}$, this definition would \textit{not} agree
with~\eqref{eq:incidencedelta}: the last term certainly does not
appear.
\end{xmpl}

\begin{figure}[htb]
\centering
\newcommand{\punto}{\;\begin{picture}(2,2)(0,-3)
\put(0,0){\circle{4}}
\end{picture}\;}
\newcommand{\match}{\;\begin{picture}(5,10)(0,2)
\put(0,10){\circle{4}}
\put(0,10){\line(0,-1){10}}
\put(0,0){\circle*{3}}
\end{picture}\;}
\newcommand{\legs}{\;\begin{picture}(20,10)(0,2)
\put(10,10){\circle{4}}
\put(10,10){\line(-1,-1){10}}
\put(0,0){\circle*{3}}
\put(10,10){\line(1,-1){10}}
\put(20,0){\circle*{3}}
\end{picture}\;}
$$
\Dl\Bigl( \legs \Bigl)
= \legs \ox 1 + 1 \ox \legs + 2\,\punto \ox \match
		 + \punto\punto \ox \punto
$$
\caption{Subinterval rule $\neq$ cut rule}
\label{fig:cut-rule}
\end{figure}

The point here is that the interval $\{0_T,1_T\}$ does not contain all
of its convex subposets. What ``convex'' means is recalled in
Section~13. Once this is clarified, the proof of Zimmermann's formula
in the previous section will be seen to adapt to a large class of
incidence Hopf algebras. To do this in some generality, and to dispel
the misunderstandings we alluded to above, a new bevy of concepts in
incidence Hopf algebra theory is required. Before turning to
that, let us obtain $\F$ anew, as advertised, in terms of the theory
of this section.

\subsection{The \bbfb\ algebra as an incidence bialgebra}

The Fa\`a di Bruno algebra, it could be suspected by now, is an
incidence bialgebra, corresponding to the case where the intervals
belong to the family of posets that are partitions of finite sets;
this was realized in~\cite{Doub}.  We start by ordering the partitions
of a finite set~$S$.  One says that $\{\row A1n\} = \pi \le \tau =
\{\row B1m\}$, or that $\pi$ refines $\tau$, if each $A_i$ is
contained in some $B_j$.  The set of partitions $\Pi(S)$ of $S$ is an
interval, where the biggest element is the partition with just one
block, and the smallest the partition whose blocks are all singletons.

\begin{prop}
\label{pr:subintinfaa}
The subinterval $\{\pi,\tau\}$ of $\Pi(S)$ is isomorphic to the poset
$\Pi_1^{\la_1} \x \cdots \x \Pi_k^{\la_k}\cdots$, where $\la_j$ is
the number of blocks of $\tau$ that are the union of exactly $j$
blocks of $\pi$ (so $\sum\la_j=|\tau|$ and $\sum j\la_j=|\pi|$).
\end{prop}

\begin{proof}
Let $B^1_1, \dots, B^1_{\la_1}$ be the blocks of $\pi$ that are also
blocks of $\tau$, and for each integer $k$ let
$\{B_1^{k1},\dots,B_1^{kk}\}, \dots,
\{B_{\la_k}^{k1},\dots,B_{\la_k}^{kk}\}$ be
the collections of $k$ blocks of $\pi$ that produce a block of $\tau$.
Clearly
\begin{equation}
\Pi(\{B^1_1\} \x \cdots \x \{B_1^{21},B_1^{22}\} \x \cdots \x
\{B_{\la_k}^{k1},\dots,B_{\la_k}^{kk}\} \x \cdots) \cong
\Pi_1^{\la_1} \x \cdots \x \Pi_k^{\la_k}\cdots.
\label{eq:star}
\end{equation}
If $\sigma \in \{\pi,\tau\}$, then $\sigma$ is obtained by dividing
some of the blocks of $\tau$ into pieces that are unions of blocks of
$\pi$, which amounts to taking a partition of some of the products on
the right hand side of~\eqref{eq:star}.
\end{proof}

It is natural therefore to assign to each interval $\{\pi,\tau\}$ the
sequence $\la = (1^{\la_1},\cdots,k^{\la_k},\cdots)$ ---only finitely
many components are nonzero--- and to declare two intervals in the
family $\Pc = \set{\Pi(S) : S\;\hbox{is a finite set}}$ to be
equivalent when the corresponding vectors are equal; there are of
course an infinite number of intervals in each equivalence class. It
is immediate that $\Pc$, with this equivalence relation $\sim$, is
interval closed and hereditary. Equivalently, one can think of the
poset of finite partitions of a countable set (such that only one
block is infinite) ordered by refinement. Furthermore, from the
correspondences
$$
[\{\pi,\tau\}] \longleftrightarrow \la \longleftrightarrow
x_1^{\la_1} x_2^{\la_2} \cdots x_k^{\la_k} \cdots,
$$
it is also natural to declare $\widetilde\Pc$ isomorphic, as an
algebra, to the algebra of polynomials of infinitely many variables
$\C[\row x1k, \dots]$.

To describe the associated coproduct explicitly it is enough to find
what its action is on the indecomposable types~$[\Pi_n]$. If
$[\{\pi,\tau\}] = [\Pi_n]$ ---which corresponds to the vector $(0,
\dots, 0,1,0, \dots)$ with the 1 in the $n$-th place, or to the
polynomial $x_n$--- then $\pi$ has $n$ blocks and $\tau$ just one.
Moreover, if $\sigma \in \{\pi,\tau\}$, then $[\{\sigma,\tau\}]$
corresponds to $x_k$ for some $k \le n$, whereas $[\{\pi,\sigma\}]$
goes with a vector $\a$ satisfying $\a_1 + \cdots + \a_n = k$, and
$\a_1 + 2\a_2 + \cdots + n\a_n = n$. We conclude
from~\eqref{eq:incidencedelta} that
$$
\Dl x_n = \sum_{k=1}^n \sum_{\a}\binom{n}{\a;k}
x_1^{\a_1} x_2^{\a_2} \cdots x_n^{\a_n} \ox x_k,
$$
and we have recovered~\eqref{eq:peace-and-strength}, with the
identification $a_n\equiv [\Pi_n]$.  Now, there is the proviso
$[\Pi_1]=1$, implicit in the main construction of Section~7.
In consequence, $\F=H(\set{\Pi(S):S\;\hbox{is a finite set}}_\sim)$.

\marker
Several final remarks on \bfb\ algebras are in order.

A variant of the construction of $\F$ allows one to obtain an enlarged
\bfb\ algebra, that we shall denote by $\F_{\mathrm{enl}}$. For that
we adjoin to $\C[\row x2k, \dots]$ the grouplike elements $x_1$ and
$x_1^{-1}$. Obviously the coproduct
formula~\eqref{eq:peace-and-strength} holds and the antipode
generalizes as follows:
$$
S_{\mathrm{enl}} x_n
= x_1^{-n}\sum_{k=1}^{n-1}(-1)^k\frac{(n-1+k)!}{(n-1)!}
B_{n-1,k}\Bigl(\frac{x_2x_1^{-1}}{2},\frac{x_3x_1^{-1}}{3},
\dots\Bigr),
$$
yielding (together with the discussion in that section)
formula~\eqref{eq:old-favourite} of Section~8.  This can be shown to
arise from a variation of the chain formula to take into account the
one-point segments.

\medskip

The following is taken from~\cite{Precursors}. A (multi)coloured set
is a finite set $X$ with a map $\theta:X\to \{1,\dots,N\}$. The
\textit{colour} of $x\in X$ is the value of~$\theta(x)$. If
$X_r:=\set{x\in X:\theta(x)=r}$, the size $|X|$ of~$X$ is the row
vector $(|X_1|,\dots,|X_N|)$. A \textit{coloured partition} of a
coloured set~$X$ is a partition of~$X$ whose set of blocks is also
coloured, with the condition $\theta(\{x\})=\theta(x)$ for singletons.
In what follows, take $N=2$ for simplicity of notation. Let
$|X|=(n_1,n_2)$. Coloured partitions form a poset $\Pi_{n_1,n_2}$,
with $\pi\le\rho$ if $\pi\le\rho$ as partitions and, for each block
$B$ of~$\pi$ which is also a block of~$\rho$,
$\theta_\pi(B)=\theta_\rho(B)$. We banish from the family of posets
one of the maximal elements $1_X$, so there are the two families
$\Pi_{n_1,n_2}^r$ with $r=1$ or 2 according to which maximal element
is kept. On the hereditary classes of segments of such coloured
partition posets under the relation of colour-isomorphism it is
possible to define a Hopf algebra structure, called the coloured \bfb\
Hopf algebra $\F(N)$. There is a corresponding anti-isomorphism
between the groups (by composition) of complex formal series in two
variables of the form
\begin{equation}
f^r(x_1,x_2) = x_r +
\sum_{n_1+n_2>1}f^r_{n_1n_2}\frac{x_1^{n_1}x_2^{n_2}}{n_1!n_2!}
\label{eq:ren-group}
\end{equation}
and the group of multiplicative functions $f$ on coloured \bfb\ Hopf
algebras, given by $f^r_{n_1n_2}:=\dst{f}{[\Pi_{n_1n_2}^r]}$.  The
antipode on $\F(N)$ provides Lagrange reversion in several
variables; we refrain from going into that, but refer the reader
to~\cite{LessOldHenrici} for a classical kind of proof; and
to~\cite{Precursors,SierraNevada} in the spirit of this review.

For the coloured \bfb\ algebras, $G(\F(N)^\circ)$ reproduces the
opposite $G^{\rm opp}_N$ of the formal diffeomorphism group in~$N$
variables; for undecorated trees, the dual group was identified by
Brouder as the \textit{Butcher group} of Runge-Kutta
methods~\cite{BrouderRK}; series of trees are composed by appropriate
grafting. The dual group $G(\H^\circ)$ of $\H$ for a quantum field
theory has been termed \textit{diffeographism group} by Connes and
Kreimer. The name is entirely appropriate, as one should regard also
the apparatus of Feynman graphs as a (very sophisticated)
approximation machinery for the computation of (quantum corrections
to) the coupling constants of a physical theory. In view of the ``main
theorem of renormalization''~\cite{OldRaymond}, these are typically
given by series like~\eqref{eq:ren-group}. In the outstanding paper a
Hopf algebra homomorphism $\F(N)\to\H$ (with the coupling constants as
colours) and its dual group morphism $G(\H^\circ)\to G_N^\opp$ are
exhibited. This second morphism is first constructed at the level of
the infinitesimal characters, and then lifted to the group; the
transpose map from $\F(N)$ to~$\H$ coincides with the one obtained
when calculating the coupling constants in terms of Feynman graphs.
For simple theories like massless $\vf^4_4$, only the simple \bfb\
algebra and $G_1^\opp$ are involved. The previous considerations
authorize us to regard the diffeographism group as the `general
renormalization group'. Unfortunately, this is the place where we
stop; but we can refer on this point, besides~\cite{ConnesKrRHII},
to~\cite{AlainGalois,GirelliKMFirst,AlainMatilde}, in a different vein
to~\cite{BergCartier}\footnote{This paper came out at an early stage
of the subject, with the excellent idea of developing its Lie
algebraic aspect in the context of triangular matrix representations;
but its treatment of factorization for the Bogoliubov recursive
procedure of renormalization is inconsistent}; and to the excellent
review~\cite{Puydedome}.

\medskip

Schmitt~\cite{Schmitt2} defines a \textit{uniform family} as a
hereditary family $\Pc$ of graded posets together with a relation
$\sim$, giving rise to an incidence Hopf algebra, such that, moreover:
\begin{itemize}
\item
The monoid $\Pc$ is commutative.
\item
If $[P]$ is indecomposable, then $\{y,1_P\}$ for $y<1_P$ is
indecomposable.
\item
For all $n\ge1$, there is exactly one type in $\Pc_\sim^\circ$ of
degree $n$.
\end{itemize}

Obviously the families of posets giving rise, respectively, to the
ladder Hopf algebra $\Hl$ and to the \bfb\ algebra $\F$, are uniform. 
This allows nice determinant formulae for the
antipode~\cite{Schmitt2,DoctorJohnson}, that need not concern us now. 
We may define a quasi-uniform family by dropping the last condition;
the coloured posets giving rise to the~$\F(N)$ do correspond to
quasi-uniform families.  The algebras $H_R$ and $\H$ (even more so
modulo the considerations in the next Section) do correspond to
quasi-uniform families as well.

\section{Distributive lattices and the general Zimmermann formula}

This section is partly motivated by the closing remarks
of~\cite{Schmitt2}. Our goal is to employ the distributive lattice of
order ideals associated with a general partially ordered set in order
to interpret the combinatorics of cuts in the algebra of rooted trees,
and further, to resolve the combinatorics of overlapping divergences.
We need some definitions. We work with finite posets. An
\textit{antichain} in a poset~$P$ is a subset of~$P$ such that any two
of its elements are incomparable in~$P$. An \textit{order ideal} is a
subset~$I$ that includes all the descendants of its elements, that is,
$x\in I$ whenever $x$ is smaller than some $y\in I$. We include the
empty set among the ideals of~$P$. The \textit{principal ideal}
generated by $y\in P$ is the ideal $\La_y := \set{x\in P :\, x \leq
y}$. Let $A$ be an antichain; the order ideal \textit{generated by}
$A$ is the subset of $P$ of those elements smaller than some $y\in A$.
A subposet $K$ is \textit{convex} when it contains the intervals
associated to any pair of its elements; in particular an interval is
convex. Differences of order ideals are also convex.

\begin{defn}
\label{df:lattice}
A \textit{lattice} is a particular class of poset, in which every pair
of elements~$s,t$ has a greatest lower bound $s\w t$ (``meet'') and a
least upper bound $s\vee t$ (``join''). That is, there exist
respectively an element $s\w t$ satisfying $s\w t\le s$, $s\w t\le t$
such that any other with the same property is smaller, and one element
$s\vee t$ satisfying $s\vee t\ge s$, $s\vee t\ge t$, and any other
element with the same property is greater.  A lattice~$L$ is
\textit{distributive} if for any $s,t,u$ in $L$
$$
s\w(t\vee u) = (s\w t)\vee(s\w u) \sepword{or equivalently}
s\vee(t\w u) = (s\vee t)\w(s\vee u).
$$
\end{defn}

Distributive lattices are always intervals. Sublattices of distributive
lattices are distributive. The collection of all subsets of any set is
a lattice, with meet $s\w t := s\cap t$ and join $s\vee t := s\cup t$.
In this example, these are the usual distributive laws between unions
and intersections. The sets of (finite) partitions of finite or
countable sets are lattices; but already $\Pi_3$ is not distributive.

The main example of distributive lattices is the poset $J_P$ of ideals
of any finite poset $P$ (ordered by inclusion). In fact, it is the
only example. An important theorem by
Birkhoff~\cite[Section~3.4]{Stanley1} states that for every finite
distributive lattice~$L$ there exists a unique (up to isomorphism)
poset $P$, such that $L \cong J_P$. For instance, this correspondence
sends an antichain~$A$ with $n$ elements into the set of its subsets,
ordered by inclusion, that is, the Boolean lattice of rank~$n$. In
general, the set~$P$ can be taken as the subset of \textit{join
irreducible} elements of~$L$, that is those elements $s$ that can not
be written as $s = t\vee u$ with $t<s$ and $u<s$. An order ideal
of~$P$ is join irreducible in~$J_P$ iff it is a principal order ideal.
Thus there is a one-to-one correspondence $\Lambda_y\leftrightarrow y$
between the join irreducibles of~$J_P$ and the elements of~$P$.

A poset $P$ is \textit{connected} if it cannot be written as the
disjoint union of two nontrivial posets; in other words, given any two
elements $x,y\in P$, one can find a sequence of elements
$x_{0}=x,x_1,\dots,x_{n-1},x_n=y$ such that any two successive
elements of the sequence are related by~$\le$ or~$\ge$. Now, to the
family of all finite posets, or to any subfamily~$\Pc$ of it closed
under the formation of disjoint unions and containing all the convex
subsets of its elements, we can associate the new family of posets
$\J_\Pc := \set{J_P \: P\in \Pc}$. An interval $\{I,I'\}$ in $J_P$, is
isomorphic to $J_{I'\setminus I}$, where $I'\setminus I$ is regarded
as a subposet of $P$. It is easy to see that distributive lattices are
always graded, with the degree or `rank' of~$J_P$ being precisely the
cardinality $|P|$ of~$P$. Consider the isomorphism equivalence
relation both in~$\Pc$ and~$\J_\Pc$. Since $J_{P\cup Q}\cong J_P\x
J_Q$, the set of indecomposable types $\J_\Pc^\circ$ of $\J_\Pc$ is
precisely the set of isomorphism classes of connected posets of $\Pc$.
Furthermore, by~\eqref{eq:incidencedelta} the coproduct in the
incidence Hopf algebra $H({\J_\Pc}_\sim)$ is given by
$$
\Dl[J_P] = \sum_{I\in J_P}[0_{J_P},I\}]\ox[\{I,1_{J_P}\}]
= \sum_{I\in J_P}[\{\emptyset,I\}]\ox[\{I,P\}]
= \sum_{I\in J_P} [J_I] \ox [J_{P\setminus I}].
$$
(Note that since both $I$ and $P\setminus I$ are convex their types
belong to $\Pc_\sim$.)

Motivated by all this, we introduce a new commutative Hopf algebra
structure in $\Pc_\sim$, defined by the product
\begin{equation}
PQ = QP := P\cup Q
\label{eq:revelation}
\end{equation}
and the coproduct
\begin{equation}
\Dl[P] := \sum_{I\in J_P} [I] \ox [P\setminus I],
\label{eq:second-revelation}
\end{equation}
with the obvious unit and counit. By construction, this Hopf algebra
is isomorphic to $H({\J_\Pc}_\sim)$. With some abuse of notation, in
the remainder of this section we call $H(\Pc_\sim)$ the Hopf algebra
given by~\eqref{eq:revelation} and \eqref{eq:second-revelation}. In
this setting a chain $\Cc$ of $P$ is defined by a chain of $J_P$, that
is, a sequence of order ideals $\emptyset=I_0\varsubsetneq I_1
\varsubsetneq \cdots \varsubsetneq I_n=P$, and
$$
\Om(\Cc) = [\{I_0,I_1\}\{I_1,I_2\}\cdots\{I_{n-1},I_n\}]
=[J_{I_1\setminus I_0}J_{I_2\setminus I_1}\cdots
J_{I_n\setminus I_{n-1}}].
$$
When $\Cc$ is regarded as a subposet of $P$ we write
$$
\Om(\Cc)
= [I_1\setminus I_0][I_2\setminus I_1]\cdots[I_n\setminus I_{n-1}].
$$
By Proposition~\ref{pr:incidence-areHopf} the antipode in $H(\Pc_\sim)$
is given by
$$
S_\DS[P] = \sum_{\Cc \in \Ch([P])}(-1)^{l(\Cc)}\Om(\Cc),
$$
where $\Ch([P])$ denotes the set of chain of $[P]$.

In complete analogy to the Hopf algebra of Feynman graphs we consider
the following.

\begin{defn}
\label{df:generalforest}
A \textit{forest} $\F$ of $P$ is a collection of connected subposets
of $P$ such that if $I_1$ and $I_2$ are in $\F$, then either
$I_1\cap I_2= \emptyset$, or $I_1\subset I_2$ or $I_2\subset I_1$. If
$I\in \F$, then $I'$ is a predecessor of $I$ in $\F$ if there is no
$I''\in \F$ such that $I'\varsubsetneq I''\varsubsetneq I$, and we
denote by~$\tilde I$ the disjoint union of all predecessors of $I$
in~$\F$. As in Section~11, we define
$$
\Th(\F) = \prod_{I\in\F\cup\{P\}} [I\setminus\tilde I].
$$
\end{defn}

We come to the main result: a general formula of the Zimmermann type
for the antipode of the incidence Hopf algebras~$H(\Pc_\sim)$.

\begin{thm}
\label{pr:generalZimmerman}
Let $\Pc$ be a family of posets that contains all the convex subsets
of its elements and is closed under the formation of disjoint unions,
and consider on $\Pc$ the isomorphism equivalence relation.  For any
$[P]\in H(\Pc_\sim)$, let $F([P])$ be the set of all forests of~$[P]$. 
Then
$$
S([P])= S_Z([P]):= \sum_{\F \in F([P])} (-1)^{d(\F)} \Th([P]).
$$
\end{thm}

\begin{proof}
Our proof of Theorem~\ref{pr:Zimmerman}, and its preliminaries, just
go through with no more than notational change, and there is no point
in repeating them.
\end{proof}

We turn to the examples. In order to obtain~\eqref{eq:coprod-tree}
from the theory of incidence Hopf algebras, one can proceed as
follows: consider the family $W$ of posets constituted by all rooted
woods; consider $\J_W := \set{J_P : P\in W}$ modulo isomorphism. Then
$\J_W$ is hereditary and we proceed with the construction of the
incidence algebra associated to it; the indecomposable types are
precisely the trees, and $H(W_\sim)\simeq H({\J_W}_\sim)$ now denotes
the polynomial algebra on them with the coproduct of above. This is
seen to reproduce the good coproduct~\eqref{eq:coprod-tree}. For
instance, in Example~\ref{xm:legs}, as well as the trivial ideals
for~$t_{32}$, there are the ideals given by each leaf, and both leaves
together. In conclusion, $H(W_\sim)\simeq H_R$. Clearly $H_R$ is a
quasi-uniform incidence algebra. Note that if a tree~$t$ is a stick
with $n$ vertices, then $J_t$ is a stick with $n+1$ vertices.

\begin{figure}[htb]
\centering
\begin{picture}(35,60)(10,0)
\put(30,60){\circle{4}}
\put(34,61){$a$}
\put(30,60){\line(0,-1){20}}
\put(30,40){\circle*{3}}
\put(34,41){$b$}
\put(30,40){\line(0,-1){20}}
\put(30,20){\circle*{3}}
\put(34,21){$c$}
\put(60,40){\vector(1,0){30}}
\end{picture}
\qquad
\begin{picture}(40,80)(-30,-10)
\put(20,60){\circle*{3}}
\put(24,61){$a$}
\put(20,40){\line(0,1){20}}
\put(20,40){\circle*{3}}
\put(24,41){$b$}
\put(20,20){\line(0,1){20}}
\put(20,20){\circle*{3}}
\put(24,21){$c$}
\put(20,0){\line(0,1){20}}
\put(20,0){\circle*{3}}
\put(24,1){$\scriptstyle\emptyset$}
\end{picture}
\caption{The distributive lattice for the tree $t_{31}$}
\label{fig:stick-lat}
\end{figure}
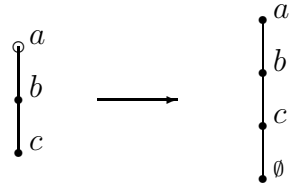

In this way we are able to complement the graphic description of
rooted trees by the Hasse diagrams of their distributive lattices of
ideals (the Hasse diagram of a finite poset~$P$ is drawn by
representing the elements of~$P$ by vertices and the cover relations
by edges~\cite{RusoCombinado}). We illustrate with the accompanying
figures the correspondence of (incidence Hopf algebras of)
distributive lattices to (the Hopf algebra of) rooted trees, up to
four vertices.

\begin{figure}[htb]
\centering
\begin{picture}(40,20)(10,0)
\put(20,40){\circle{4}}
\put(24,41){$a$}
\put(20,40){\line(1,-1){20}}
\put(40,20){\circle*{3}}
\put(45,18){$c$}
\put(20,40){\line(-1,-1){20}}
\put(0,20){\circle*{3}}
\put(-9,18){$b$}
\put(0,20){\line(0,-1){20}}
\put(0,0){\circle*{3}}
\put(-9,-2){$d$}
\put(65,20){\vector(1,0){30}}
\end{picture}
\qquad
\begin{picture}(80,80)(-50,0)
\put(20,68){\circle*{3}}
\put(24,69){$a$}
\put(20,40){\line(0,1){28}}
\put(20,40){\circle*{3}}
\put(25,40){$bc$}
\put(0,20){\line(1,1){20}}
\put(0,20){\circle*{3}}
\put(-9,18){$b$}
\put(0,20){\line(1,-1){40}}
\put(20,0){\circle*{3}}
\put(10,-3){$d$}
\put(20,40){\line(1,-1){40}}
\put(40,20){\circle*{3}}
\put(45,20){$cd$}
\put(20,0){\line(1,1){20}}
\put(40,-20){\line(1,1){20}}
\put(40,-20){\circle*{3}}
\put(45,-25){$\scriptstyle\emptyset$}
\put(60,0){\circle*{3}}
\put(65,0){$c$}
\end{picture}
\vspace{18pt}
\caption{The distributive lattice for the tree $t_{42}$}
\label{fig:crook-lat}
\end{figure}
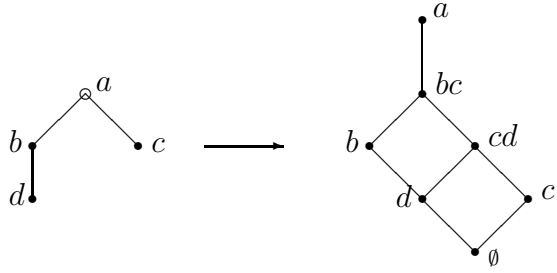

\begin{figure}[htb]
\centering
\begin{picture}(40,20)(10,0)
\put(20,20){\circle{4}}
\put(25,21){$a$}
\put(20,20){\line(-1,-1){20}}
\put(0,0){\circle*{3}}
\put(-9,-2){$b$}
\put(20,20){\line(0,-1){20}}
\put(20,0){\circle*{3}}
\put(11,-2){$c$}
\put(20,20){\line(1,-1){20}}
\put(40,0){\circle*{3}}
\put(45,-2){$d$}
\put(65,10){\vector(1,0){30}}
\end{picture}
\qquad
\begin{picture}(80,80)(-50,0)
\put(20,20){\line(0,1){40}}
\put(20,60){\circle*{3}}
\put(24,61){$a$}
\put(20,40){\circle*{3}}
\put(25,41){$bcd$}
\put(0,20){\line(1,1){20}}
\put(20,40){\line(1,-1){20}}
\put(20,20){\circle*{3}}
\put(0,0){\line(1,1){20}}
\put(20,20){\circle*{3}}
\put(23,19){$bd$}
\put(20,20){\line(1,-1){20}}
\put(40,0){\circle*{3}}
\put(45,-2){$d$}
\put(40,0){\line(0,1){20}}
\put(40,20){\circle*{3}}
\put(45,19){$cd$}
\put(0,0){\line(1,-1){20}}
\put(0,0){\circle*{3}}
\put(-10,-2){$b$}
\put(0,0){\line(0,1){20}}
\put(0,20){\circle*{3}}
\put(-13,19){$bc$}
\put(0,20){\line(1,-1){20}}
\put(20,0){\circle*{3}}
\put(24,-2){$c$}
\put(20,0){\line(1,1){20}}
\put(20,-20){\line(0,1){20}}
\put(20,-20){\line(1,1){20}}
\put(20,-20){\circle*{3}}
\put(25,-25){$\scriptstyle\emptyset$}
\end{picture}
\vspace{18pt}
\caption{The distributive lattice for the tree $t_{43}$}
\label{fig:claw-lat}
\end{figure}
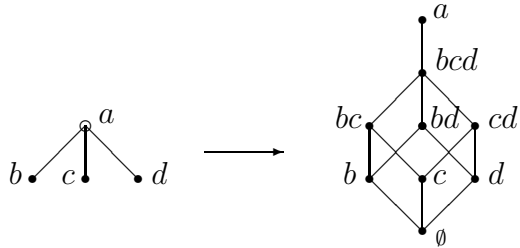

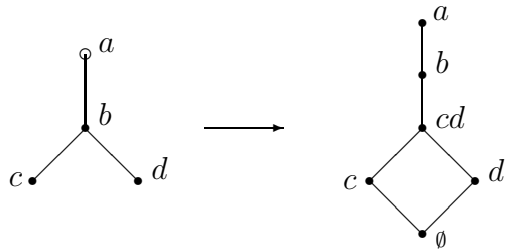
\begin{figure}[htb]
\centering
\begin{picture}(40,20)(10,0)
\put(20,48){\circle{4}}
\put(25,49){$a$}
\put(20,48){\line(0,-1){28}}
\put(20,20){\circle*{3}}
\put(25,21){$b$}
\put(20,20){\line(-1,-1){20}}
\put(0,0){\circle*{3}}
\put(-9,-2){$c$}
\put(20,20){\line(1,-1){20}}
\put(40,0){\circle*{3}}
\put(45,1){$d$}
\put(65,20){\vector(1,0){30}}
\end{picture}
\qquad
\begin{picture}(80,80)(-50,0)
\put(20,20){\line(0,1){40}}
\put(20,60){\circle*{3}}
\put(24,61){$a$}
\put(20,40){\circle*{3}}
\put(25,41){$b$}
\put(0,0){\line(1,1){20}}
\put(20,20){\circle*{3}}
\put(25,20){$cd$}
\put(20,20){\line(1,-1){20}}
\put(40,0){\circle*{3}}
\put(45,0){$d$}
\put(0,0){\line(1,-1){20}}
\put(0,0){\circle*{3}}
\put(-10,-3){$c$}
\put(20,-20){\line(1,1){20}}
\put(20,-20){\circle*{3}}
\put(25,-25){$\scriptstyle\emptyset$}
\end{picture}
\vspace{18pt}
\caption{The distributive lattice for the tree $t_{44}$}
\label{fig:biped-lat}
\end{figure}


With a bit of practice, one is able to read quickly the lattice paths;
for instance, in figure~6, one sees that there are 12 chains and 8
forests associated to the tree~$t_{42}$. There is no particular
advantage in computing the antipode in this way; however on other
matters the distributive lattice viewpoint is very helpful. For
bialgebras of Feynman graphs, things are just a tad more difficult. We
also illustrate with a few figures the correspondence of incidence
Hopf algebras to some Feynman graphs for the four-point function
in~$\vf_4^4$ theory, up to three loops. The reader will have noticed
that Boolean lattices (and thus antichains) must be ascribed to
articulate (one-vertex reducible) diagrams in~$\vf^4_4$ theory. This
is in consonance with the fact, noted in Section~11, that
renormalization factorizes for such graphs. The case of the cat's-eye
diagram illustrates the possibility, for overlapping divergencies, of
chooing a subposet of~$\J_{\Pc}$, for $\Pc$ the divergence poset. Note
the grading by length of Feynman graphs (that for connected graphs
coincides as a filtering with the filtering by depth).

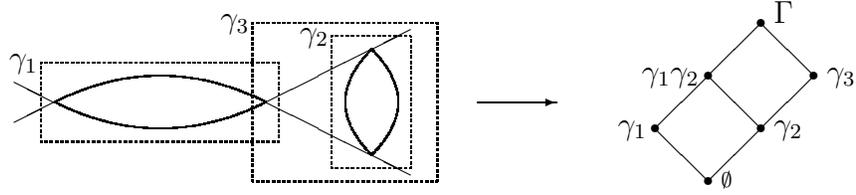
\begin{figure}[htb]
\centering
\begin{picture}(180,20)(0,-10)
\qbezier(0,0)(40,20)(80,0)
\qbezier(0,0)(40,-20)(80,0)
\put(0,0){\line(-2,1){15}}
\put(0,0){\line(-2,-1){15}}
\put(80,0){\line(2,1){55}}
\put(80,0){\line(2,-1){55}}
\qbezier(110,0)(110,10)(120,20)
\qbezier(110,0)(110,-10)(120,-20)
\qbezier(130,0)(130,10)(120,20)
\qbezier(130,0)(130,-10)(120,-20)
\put(-5,-15){\dashbox{1}(90,30)}
\put(-17,13){$\ga_1$}
\put(105,-25){\dashbox{1}(30,50)}
\put(93,23){$\ga_2$}
\put(75,-30){\dashbox{1}(70,60)}
\put(63,28){$\ga_3$}
\put(160,0){\vector(1,0){30}}
\end{picture}
\qquad
\begin{picture}(80,40)(-20,0)
\put(0,0){\circle*{3}}
\put(-14,-2){$\ga_1$}
\put(0,0){\line(1,1){40}}
\put(20,20){\circle*{3}}
\put(-5,18){$\ga_1\ga_2$}
\put(40,40){\circle*{3}}
\put(45,40){$\Ga$}
\put(0,0){\line(1,-1){20}}
\put(20,-20){\circle*{3}}
\put(25,-22){$\scriptstyle\emptyset$}
\put(20,20){\line(1,-1){20}}
\put(40,0){\circle*{3}}
\put(45,-2){$\ga_2$}
\put(40,40){\line(1,-1){20}}
\put(60,20){\circle*{3}}
\put(65,18){$\ga_3$}
\put(20,-20){\line(1,1){40}}
\end{picture}
\vspace{2pc}
\caption{The distributive lattice for the feeding-shark diagram}
\label{fig:shark-lat}
\end{figure}


\begin{figure}[htb]
\centering
\begin{picture}(140,20)(10,0)
\put(20,10){\qbezier(0,0)(40,40)(80,0)}
\put(20,10){\qbezier(0,0)(40,-40)(80,0)}
\put(100,10){\line(1,1){20}}
\put(100,10){\line(1,-1){20}}
\put(20,10){\line(-1,1){20}}
\put(20,10){\line(-1,-1){20}}
\qbezier(50,10)(50,20)(60,30)
\qbezier(50,10)(50,0)(60,-10)
\qbezier(70,10)(70,20)(60,30)
\qbezier(70,10)(70,0)(60,-10)
\put(45,-15){\dashbox{1}(30,50)}
\put(54,8){$\ga_1$}
\put(10,-25){\dashbox{1}(70,65)}
\put(12,28){$\ga_2$}
\put(40,-20){\dashbox{1}(70,65)}
\put(98,28){$\ga_3$}
\put(140,10){\vector(1,0){30}}
\end{picture}
\qquad
\begin{picture}(80,40)(-40,0)
\put(20,40){\circle*{3}}
\put(25,40){$\Ga$}
\put(0,20){\line(1,1){20}}
\put(0,20){\circle*{3}}
\put(-14,18){$\ga_2$}
\put(0,20){\line(1,-1){20}}
\put(20,0){\circle*{3}}
\put(7,-5){$\ga_1$}
\put(20,40){\line(1,-1){20}}
\put(40,20){\circle*{3}}
\put(45,18){$\ga_3$}
\put(20,0){\line(1,1){20}}
\put(20,0){\line(0,-1){28}}
\put(20,-28){\circle*{3}}
\put(24,-32){$\scriptstyle\emptyset$}
\end{picture}
\vspace{2pc}
\caption{The distributive lattice for the cat's-eye diagram}
\label{fig:gato-lat}
\end{figure}
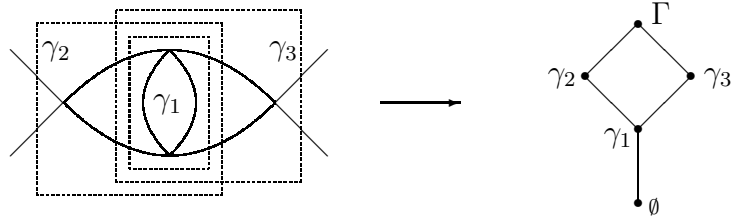

In conclusion for this part, we have understood the question on the
relation between the Hopf algebras of renormalization of Connes and
Kreimer and the incidence Hopf algebras of Rota, Schmitt and others,
in terms of the incidence Hopf algebra on the family of (sub)poset
ideals $\J_\Pc$, where $\Pc$ denotes the family of posets made of
Feynman graphs and subgraphs pertaining to a given model theory.

\marker
If in the antipode formulae for connected elements $T$ of $H_R$ one
formally substitutes~1 for all trees, one obtains zero whenever $T\ne
t_1$.  Concretely, the zeta function on trees is the character taking
the value 1 on every tree (i.e., the `geometric series' element of
$G(H_T^\circ)$, see~\cite{GirelliKMSecond}, and the M\"obius function
$\zeta^{*-1}$ sends~1 to~1, $t_1$ to~$-1$ and any other tree to zero. 
This reflects the well-known fact that the M\"obius function of any
distributive lattice $J$ is zero, except when~$J$ is the Boolean
algebra of rank~$n$, for which $\zeta^{*-1}(J) = (-1)^n$ ---whereby
on~$H(\Pc_\sim)$ the M\"obius function vanishes except for antichains;
for the latter, $\zeta^{*-1}(A)=(-1)^{|A|}$ holds.

\marker The formula of Zimmermann is also valid under more general
circumstances; for instance, for the \bfb\ algebra, although the
partition lattices are not distributive. What is needed is a good
description of the intervals in the algebra in terms of indecomposable
types. This occurs in the \bfb\ algebra: see
Proposition~\ref{pr:subintinfaa}. The expression of $S_Z$ then leads
to~\eqref{eq:reverted}; this was found in~\cite{Precursors}, as
repeatedly indicated. After the first version of our paper appeared,
in an interesting development, it has been found to work for the
\textit{free} \bfb\ algebra~\cite{AEP} as well. In our present
framework, it is enough to remark that, as discussed in Section~12,
the \bfb\ algebra $\F$ is a Hopf subalgebra of~$H_R$, to which the
distributive lattice paradigm applies.

\newpage

\section*{Part IV: The General Structure Theorems}
\addcontentsline{toc}{section}{\protect\numberline{IV}\enspace
THE GENERAL STRUCTURE THEOREMS}

\section{Structure of commutative Hopf algebras I}

The title of the section might have been: `why the boson Hopf algebra
is important'. Indeed $B(V)$, one of the simplest commutative and
cocommutative connected bialgebras, beyond its role in QFT will be
seen in this section to play a normative role in the theory.

We have invoked a couple of times the Milnor--Moore theorem. This is
the well known structure theorem for connected graded cocommutative
Hopf algebras, stating that such a Hopf algebra $H$ is necessarily
isomorphic, as a Hopf algebra, to~$\U(P(H))$. Remember
$P(\U(P(H)))=P(H)$. For algebras of finite type, from our remarks on
the filtering by depth at the end of Section~6, there is but a short
step to a proof for the Milnor--Moore theorem, and the reader is
invited to provide such proof. We refrain from giving any, as that is
found in several places, including the book~\cite{Polaris} by
ourselves; we remit as well to~\cite{Montgomery} and the original
paper~\cite{MilnorM}.

\begin{thm}
\label{pr:commutative-areBoson}
Let $H$ be a graded connected commutative skewgroup of finite type,
and let $V$ be a graded vector space such that $V\op H_+^2=H_+$. Then
there exists a graded algebra isomorphism between $B(V)$ and~$H$.
\end{thm}

\begin{proof}
One has $H\simeq H''\simeq\U(P(H'))'$ by the Milnor--Moore theorem. If
$\set{e_i}$ is a graded basis of~$P(H')$, then consider the vector
space~$V_0$ linearly generated by the dual basis (defined
by~$f_j(e_i)=\dl_{ij}$); note that the space algebraically generated
by $V_0$ is all of~$\U(P(H'))'$ ---this is essentially the
Poincar\'e--Birkhoff--Witt theorem~\cite{Savant}. Thus the map
$B\iota_{V_0}\:B(V_0)\to\U(P(H'))'$ lifting $V_0\ni f_j\to
f_j\in\U(P(H'))'$ is an algebra isomorphism.

One has $V_0\op H_+^2=H_+$. This is so in view of
$$
H = \U(P(H'))' = V_0\op P(H')^\perp = \C1\op V_0\op H_+^2.
$$
where~\eqref{eq:new-number} has been used in the last equality. Now,
there is nothing very particular about~$V_0$. Let $V$ another
supplementary graded vector subspace of~$H_+^2$ in~$H_+$, and denote
as usual by~$\C[V]$ the subalgebra of~$H$ generated by~$V$. To prove
that $\C[V]=H$ we (routinely) show by induction that
$\bigoplus_{k=0}^n H^{(n)}\subset\C[V]$. Clearly this is true for
$n=0$. Assume the claim is true for all $k<n$, and let $a$ be an
element of degree $n$, that can be taken homogeneous. One has $a = v +
b_1b_2$ with $v\in V$ and $b_i\in\bigoplus_{k=0}^{n-1}H^{(k)}$;
therefore $a\in\C[V]$, i.e., $V$ indeed generates~$H$. Moreover $V_0$
and $V$ are isomorphic graded vector spaces; therefore $B(V),B(V_0)$
and~$H$ are isomorphic graded algebras.

As a consequence, $H$ cannot have divisors of zero; also, the
filtering by depth in connected graded commutative Hopf algebras
(discussed extensively at the end of Section~6) has an associated
\textit{algebra} grading.
\end{proof}

It is useful to reflect more on Theorem~\ref{pr:commutative-areBoson}.
Borrowing the language of topology, we may say that $s\: H_+/H_+^2\to
H_+$ is a section of $\pi\:H_+\to H_+/H_+^2$ if $s$ is a right inverse
for~$\pi$. We have arrived at the conclusion that any section of~$\pi$
for a commutative connected graded bialgebra~$H$ induces an algebra
isomorphism between the free commutative algebra over $H_+/H_+^2$
and~$H$. This is Leray's theorem~\cite[Theorem~7.5]{MilnorM}, proved
here at least for~$H$ of finite type. (By the way, the theorem by
Milnor and Moore, together with the Poincar\'e--Birkhoff--Witt
theorem, incorporates the statement that a \textit{cocommutative}
connected graded bialgebra~$H$ is isomorphic as a coalgebra with the
cofree cocommutative coalgebra $Q_\cocom(P(H))$ over the space of
its primitive elements, with the isomorphism induced by any left
inverse for the inclusion $P(H) \hookto H$.) Note as well that
$H_+/H_+^2\simeq P(H')^*$, and as $P(H')$ is a Lie algebra,
$H_+/H_+^2$ inherits a natural Lie coalgebra structure, on which
unfortunately we cannot dwell.

Theorem~\ref{pr:commutative-areBoson} does not give all we would want,
as for instance we are still mostly in the dark about the primitive
elements of the Hopf algebra, that seem to play such an important role
in the Connes--Kreimer approach to renormalization theory. The Hopf
algebra of rooted trees contains many primitive elements, but, not being
cocommutative, cannot be primitively generated. Also Hopf algebras of
Feynman graphs are not primitively generated; and the \bfb\ algebra,
as we shall soon see, is very far from being primitively generated.

At least we know for sure, in view of~\eqref{eq:prim-indecomp}, that
primitive elements are indecomposable. The landscape thus emerging can
be summarized as follows (look back at Section~6, and, if you can
endure the category theory jargon, have a look at~\cite{BlauBuch},
too). Let~$H(n):= \C\bigl[\bigoplus_{l=1}^nH^{(l)}\bigr]$. Each~$H(n)$
is a graded subalgebra, and $H=\bigcup_{n=1}^\infty H(n)$. It is clear
that $H(n)^{(n+1)}= H^{(n+1)}\cap H_+^2$. Therefore, $P(H)\cap
H(n)^{(n+1)}=(0)$, and one can decompose~$H^{(n+1)}$ as
$$
H^{(n+1)} = H(n)^{(n+1)} \op P(H)^{(n+1)} \op W_H(n+1),
$$
for some suitable (nonuniquely defined) supplementary vector space
$W_H(n+1)$. Furthermore, since $H(n)$ is generated by $\sum_{l=1}^n
H^{(l)}$, and the compatibility of the coproduct with the grading
entails $\Dl H^{(n)} \subset H(n)\ox H(n)$ for each $n$, we see
that~$H(n)$ is an ascending sequence of Hopf algebras. Thus, in any
graded connected commutative bialgebra~$H$ there is a graded vector
subspace $W_H$, with $W_H^{(0)} = W_H^{(1)} = 0$, such that the unique
morphism of graded Hopf algebras $\nu : B(P(H))\ox B(W_H)\to H$,
extending the morphism
\begin{align}
(P(H) \ox 1) \op (1\ox W_H) &\to P(H) \op W_H \subseteq H
\nn \\
p \ox 1 + 1 \ox w &\mapsto p+w
\label{eq:adios-madrid}
\end{align}
is a graded algebra isomorphism. The quotient morphism $H\to H/H_+^2$
maps the graded subspace $P(H) + W_H$ isomorphically onto $H/H_+^2$.
Moreover, the following statements are equivalent: (1) $q_H: P(H)\to
H_+/H_+^2$ is an isomorphism; (2) $W_H = 0$; (3) $H$ is primitively
generated; (4) $\nu$ is an isomorphism of Hopf algebras.

Also, $B(P(H))$ is precisely the largest cocommutative Hopf subalgebra
(subcoalgebra, subbialgebra) of~$H$.

\marker To investigate the primitive elements (and primitivity degree)
in important commutative algebras, in the few years after the
introduction of the Connes--Kreimer paradigm, a few strategies, due to
Broadhurst and Kreimer, Foissy, and Chryssomalakos and coworkers, were
made available. The reader wishing to familiarize himself with the
structure of graded connected commutative Hopf algebras ought to
consult the original
papers~\cite{BK,godgift,Foissy,Chryssomalakosetal}. In the last of
these, so-called normal coordinate elements are introduced. We
indicate the origin of the `normal coordinate' terminology: the
$\psi$-coordinates on~$G(H^\circ)$ defined
by~$\psi_{\dl}(\exp\la\dl)=\la$ are elements of~$H$ that can be
interpreted geometrically as `normal coordinates'. We
commend~\cite{Chryssomalakosetal} for other geometrical insights, such
as the identification of primitive elements in~$H$ with closed left
invariant 1-forms on~$G(H^\circ)$.

However, the computational algorithm in~\cite{Chryssomalakosetal} for
`normal coordinate elements' ---consult for instance their
equation~(25)--- is painfully indirect. Our own strategy, inspired by
work of Reutenauer~\cite{Reutenauer}, Patras~\cite{Patras1,Patras2}
and Loday~\cite{LodayEM,LodayBook}, is to settle for the knowledge of
what in Section~5 we called quasiprimitive elements. They constitute a
privileged canonical supplementary space of~$H_+^2$ in~$H_+$.
Beautifully, and somewhat unexpectedly, it coincides with the space of
normal coordinate elements. This treatment grants us the best grip on
the reconstruction of~$H$; it is deeply related to cyclic homology,
but we will not go into that here.

As a rule, examples are more instructive than proofs. Therefore we
illustrate our purpose by giving first the `easy' example of the
ladder algebra~$H_\ell$. The finer structure theorem is proved in the
next section. The example of the \bfb\ algebra follows afterwards.

\begin{xmpl}
As a corollary of Leray's theorem, if $P(H)\to H_+/H_+^2$ is an
isomorphism, then $B\iota_{P(H)}$ is an isomorphism of Hopf algebras,
and $H$ is primitively generated.  In particular, there will be a
polynomial expression for the basis elements of~$H$ in terms of
primitive elements.  Now, in any Hopf algebra associated to a field
theory there exist ``ladder'' subalgebras of diagrams with only
completely nested subgraphs, and any of these is isomorphic to~$\Hl$. 
The structure of~$\Hl$ is completely understood in terms of the
isomorphism, already described in Example~8.3, between this bialgebra
with its filtering by depth (which in this case give rise to a Hopf
algebra grading), and a symmetric algebra.

As an incidence algebra, $\Hl$ comes from a uniform family; therefore
with the standard grading $\#$ there are exactly $p(n)$ linearly
independent elements of degree~$n$. Let us examine the first few
stages. For $n = 1$ there is a basis with only the stick with one
vertex~$\l_1$; for $n = 2$, there is a basis with~$\l_2$, the stick
with two vertices, and $\l_1^2$. Further bases are $\l_3$, $\l_1\l_2$,
$\l_1^3$ for $n = 3$; $\l_4$, $\l_1\l_3$, $\l_2^2$, $\l_1^2\l_2$,
$\l_1^4$ for $n = 4$; $\l_5$, $\l_1\l_4$, $\l_2\l_3$, $\l_1^2\l_3$,
$\l_2^2\l_1$, $\l_1^3\l_2$, $\l_1^5$ for $n = 5$.

On the other hand, there is an infinite number of primitives, given,
as we know, by the Schur polynomials in the sticks, one at each
$\#$-order; it must be so because there is one indecomposable at each
$\#$-order. The first few are:
\begin{align*}
p_1 &= \l_1; \quad
p_2 = \l_2 - \thalf \l_1^2; \quad
p_3 = \l_3 - \l_1\l_2 + \tthird \l_1^3;
\\
p_4 &= \l_4 - \l_1\l_3 - \thalf \l_2^2 + \l_1^2\l_2 - \tquarter\l_1^4;
\\
p_5 &= \l_5 - \l_1\l_4 - \l_2\l_3 + \l_1^2\l_3 + \l_1\l_2^2
- \l_1^3\l_2 + \tfifth \l_1^5.
\end{align*}

Following~\cite{BK}, we now introduce in the Hopf algebra~$\Hl$ the
dimension~$h_{n,k}$ of the space of homogeneous elements~$a$ such that
$\#(a)=n$ and $\dl(a)=k$; we know that if~$a$ is not a scalar, $1\le
k\le n$ holds. We claimed already $h_{n,1}=1$ for all~$n$. And
certainly $h_{n,n}\ge1$, as for instance $\l_n$ has depth~$n$ ---in
fact, $h_{n,n}=1$. To explore farther, in analogy to the definition of
the antipode as a geometric series, we introduce
$$
\log^*\id := \sum_{k\ge1}\frac{(-1)^{k-1}}{k}(\id-u\eta)^{*k}
= (\id-u\eta) - \frac{(\id-u\eta)^{*2}}{2}
+ \frac{(\id-u\eta)^{*3}}{3} - \cdots.
$$
which, by the argument used in Proposition~\ref{pr:geometricseries},
is a finite sum in $H_+$; and its convolution powers. We easily find
\begin{align}
\log^*\id\,\l_2 &= p_2; &
\log^*\id\,\l_1^2 &= 0.
\notag \\
\log^*\id\,\l_3 &= p_3; &
\log^*\id\,\l_1\l_2 &= \log^*\id\,\l_1^3 = 0;
\notag \\
\log^*\id\,\l_4 &= p_4; &
\log^*\id\,\l_1\l_3 &= \log^*\id\,\l_2^2 = \log^*\id\,\l_1^2\l_2
= \log^*\id\,\l_1^4 = 0;
\label{eq:first-dawn}
\end{align}
and so on, as $\log^*\id$ sends the indecomposables into the
primitives, and kills products
---see~Proposition~\ref{pr:Log-kills-products} below for details.  Now
$$
\frac{\log^{*2}\id}{2!}\,p_2  = 0; \quad
\frac{\log^{*2}\id}{2!}\,\l_1^2 = \l_1^2.
$$
This together with the first line in~\eqref{eq:first-dawn} means
$h_{2,1}=h_{2,2}=1$; it could not be otherwise.

Next,
$$
\frac{\log^{*2}\id}{2!}\,p_3 = 0; \quad
\frac{\log^{*2}\id}{2!}\,\l_1\l_2 =
\l_1\l_2 - \thalf\l_1^3 = \l_1p_2; \quad
\frac{\log^{*2}\id}{2!}\,\l_1^3 = 0.
$$
At order~3, finally:
$$
\frac{\log^{*3}\id}{3!}p_3 =  0; \quad
\frac{\log^{*3}\id}{3!}\l_1p_2 = 0; \quad
\frac{\log^{*3}\id}{3!}\l_1^3 = \l_1^3.
$$
Thus the `natural basis' for~$\Hl^{(3)}$ is
$$
(\l_3 - \l_1\l_2 + \tthird \l_1^3, \l_1\l_2 - \thalf\l_1^3, \l_1^3) =
(p_3, p_1p_2, p_1^3),
$$
with respective depths (primitivity degrees) $1,2,3$; thus
$h_{3,1}=h_{3,2}=h_{3,3}=1$.

Similarly, then
\begin{align*}
\frac{\log^{*2}\id}{2!}\,\l_1\l_3
&= \l_1\l_3 - \l_1^2\l_2 + \tthird \l_1^4 = \l_1p_3; \quad
\frac{\log^{*2}\id}{2!}\,\l_2^2
= \l_2^2 - \l_1^2\l_2 + \tquarter \l_1^4 = p_2^2;
\\
\frac{\log^{*2}\id}{2!}\,\l_1^2\l_2 &= 0; \quad
\frac{\log^{*2}\id}{2!}\,\l_1^4 = 0.
\end{align*}
Therefore there is a space of subprimitive elements of dimension~2
in~$\Hl^{(4)}$, that is $h_{4,2}=2$. Moreover,
$\frac{\log^{*3}\id}{3!}$ kills $p_4$, $\l_1p_3$, $p_2^2$ and
$\l_1^4$; also,
$$
\frac{\log^{*3}\id}{3!}\,\l_1^2\l_2 =
\l_1^2\l_2-\thalf\l_1^4 = \l_1^2p_2;
\sepword{finally}
\frac{\log^{*4}\id}{4!}\l_1^4 = \l_1^4.
$$
The natural basis for~$\Hl^{(4)}$ is:
$$
\bigl(\l_4 - \l_1\l_3 - \thalf \l_2^2 + \l_1^2\l_2 - \tquarter \l_1^4,
\l_1\l_3 - \l_1^2\l_2 + \tthird\l_1^4,
\l_2^2 - \l_1^2\l_2 + \tquarter \l_1^4,
\l_1^2\l_2 - \thalf\l_1^4, \l_1^4 \bigr) =
(p_4,p_1p_3,p_2^2,p_1^2p_2,p_1^4).
$$
Moreover $Sp_i=-p_i$ for all~$i$. The reader is invited to examine the
next homogeneous subspace. One sees that $h_{n,k}$ for~$\Hl$ is the
number of partitions of~$n$ into~$k$ integers; this was proved
in~\cite{BK} using a powerful technique involving Hilbert series.

In summary, the algebra of ladder graphs or sticks~$\Hl$ offers a case
in point for the simplest version of the result
in~\eqref{eq:adios-madrid}, for which $W_{\Hl}=0$.
\end{xmpl}

\section{Structure of commutative Hopf algebras II}

Consider, for $H$ a graded connected commutative Hopf algebra, a
suitable completion of the tensor product~$H \ox H'$, say~$H \barox
H'$. This is a unital algebra, with product $m\ox\Dl^t$ and unit~$1
\ox 1$. Now by Leray's theorem~\ref{pr:commutative-areBoson}, our $H$
is a boson algebra over a supplement~$V$ of~$H_+^2$ in~$H_+$; let $A$
index a basis for~$V$, let ${\tilde A}$ index the linear basis of $H$
consisting of monomials $X_u$ in elements of~$A$, and let~$Z_u$ denote
an element of the dual basis in~$H'$; then the product on~$H \barox
H'$ is given by the \textit{double series} product:
$$
\biggl(\,\sum_{u,v\in{\tilde A}}\a_{uv}X_u\ox Z_v\biggr)
\biggl(\,\sum_{w,t\in{\tilde A}}\b_{wt}X_w\ox Z_t\biggr)
:= \sum_{u,v,w,t\in{\tilde A}} \a_{uv} \b_{wt} \,X_u X_w\ox Z_v Z_t.
$$
This is done just like in~\cite{Reutenauer} for the
shuffle--deconcatenation Hopf algebras of words.  The linear embedding
$\End H \to H \barox H'$ given by
$$
f \mapsto \sum_{u\in{\tilde A}}f(X_u)\ox Z_u,
$$
is really a convolution algebra embedding
$$
(\End H,*) \to (H \barox H',m\ox\Dl^t).
$$
Indeed,
\begin{align}
\biggl(\,\sum_{u\in{\tilde A}} f(X_u) \ox Z_u\biggr)
\biggl(\,\sum_{v\in{\tilde A}} g(X_v) \ox Z_v\biggr)
&= \sum_{u,v\in{\tilde A}} f(X_u) g(X_v) \ox Z_u Z_v
\notag \\
&= \sum_{t\in{\tilde A}} \biggl(\,\sum_{u,v\in{\tilde A}}
f(X_u) g(X_v) \,\<Z_uZ_v,X_t> \biggr) \ox Z_t
\notag \\
&= \sum_{t\in{\tilde A}} \biggl(\,\sum_{u,v\in{\tilde A}}
f(X_u) g(X_v) \,\<Z_u\ox Z_v,\Dl X_t> \biggr) \ox Z_t
\notag \\
&= \sum_{t\in{\tilde A}} f*g(X_t) \ox Z_t.
\label{eq:conv-series}%
\end{align}

Notice that the identities $u\eta$ for convolution and~$\id$ for
composition in~$\End H$ correspond respectively to
$$
u\eta \mapsto 1\ox1;\qquad
\id \mapsto \sum_{u\in {\tilde A}}X_u\ox Z_u
$$
in the double series formalism.

On the other hand, let ${\tilde A}_+:={\tilde A}\setminus1$. Using the
same idea as in~\eqref{eq:conv-series}, we get
\begin{align*}
\log\biggl(\, \sum_{u\in{\tilde A}} X_u\ox Z_u \biggr)
&:= \sum_{k\ge1} \frac{(-1)^{k-1}}{k}
\biggl(\, \sum_{u\in{\tilde A}_+} X_u \ox Z_u \biggr)^k
\\
&= \sum_{k\ge1} \frac{(-1)^{k-1}}{k}
\sum_{u_1,\dots,u_k\in{\tilde A}_+}
X_{u_1} \cdots X_{u_k} \ox Z_{u_1} \cdots Z_{u_k}
\\
&= \sum_{w\in{\tilde A}} \sum_{k\ge1} \frac{(-1)^{k-1}}{k}
\sum_{u_1,\dots,u_k\in{\tilde A}_+}
\<Z_{u_1} \cdots Z_{u_k},X_w>\, X_{u_1} \cdots X_{u_k} \ox Z_w.
\end{align*}
Therefore
\begin{equation}
\log\biggl(\, \sum_{u\in{\tilde A}} X_u \ox Z_u \biggr)
= \sum_{w\in{\tilde A}} \pi_1(X_w) \ox Z_w,
\label{eq:basic-truth}
\end{equation}
where
$$
\pi_1(X_w) := \sum_{k\ge1} \frac{(-1)^{k-1}}{k}
\sum_{u_1,\dots,u_k\in{\tilde A}_+}
\<Z_{u_1} \cdots Z_{u_k},X_w>\, X_{u_1} \cdots X_{u_k}.
$$

Now, by definition
$$
\pi_1(X_w) = \log^*\id\,X_w.
$$
We moreover consider the endomorphisms
$$
\pi_n := \frac{\pi_1^{*n}}{n!},
$$
so that, by~\eqref{eq:conv-series}:
$$
\sum_{w\in A^*}\pi_n(X_w)\ox Z_w =
\frac{1}{n!}\biggl(\,\sum_{v\in A^*}\pi_1(X_v)\ox Z_v\biggr)^n.
$$
We may put $\pi_0 := u\eta$. Thus, if $a\in H$ is of order~$n$,
$\pi_m(a) = 0$ for $m > n$. Furthermore, for~$n > 0$,
\begin{equation}
\id^{*l}a = \exp^*(\log^*(\id^{*l}))a =
\sum_{m=1}^n\frac{(\log^*(\id^{*l}a))^m}{m!} =
\sum_{m=1}^nl^m\frac{(\log^*\id\,a)^m}{m!} = \sum_{m=1}^nl^m\pi_m(a).
\label{eq:greater-good}
\end{equation}
In particular $\id=\sum_{m\ge0}\pi_m$.

\begin{prop}
\label{pr:Trick}
For any integers $n$ and $k$,
\begin{equation}
\id^{*n}\,\id^{*k} = \id^{*nk} = \id^{*k}\,\id^{*n}.
\label{eq:burden of proof}
\end{equation}
\end{prop}

\begin{proof}
The assertion is certainly true for $k=1$ and all integers~$n$, and if
it is true for some $k$ and all integers~$n$, then taking into account
that $\id$ is an algebra homomorphism, \eqref{eq:algprop} and the
induction hypothesis give
$$
\id^{*n}\,\id^{*k+1}=\id^{*n}(\id^{*k}*\id)
=\id^{*nk}*\id^{*n}=\id^{*n(k+1)}.
$$
\end{proof}

Substituting the final expression of~\eqref{eq:greater-good}
in~\eqref{eq:burden of proof}, with very little work one
obtains~\cite[Theorem~2.9.c]{LodayEM}
\begin{equation}
\pi_m\pi_k = \dl_{mk}\,\pi_k.
\label{eq:make-believe}
\end{equation}

In other words the $\pi_k$ form a family of orthogonal projectors,
and therefore the space $H$ has the direct sum decomposition
\begin{equation}
H = \bigoplus_{n\ge0} H^{(n)} := \bigoplus_{n\ge0} \pi_n(H).
\label{eq:decomposition}
\end{equation}
Moreover, from~\eqref{eq:greater-good},
$$
\id^{*l} H^{(n)} = l^n H^{(n)},
$$
so the $H^{(n)}$ are the common eigenspaces of the operators
$\id^{*l}$ with eigenvalues $l^n$. Thus, the
decomposition~\eqref{eq:decomposition} turns $H$ into a graded
algebra. Indeed, if $a\in H^{(r)}$ and $b\in H^{(s)}$, then
$$
\id^{*l}(ab) = \id^{*l}a\,\id^{*l}b = l^{r+s}(ab),
$$
and therefore $m$ sends $H^{(r)}\ox H^{(s)}$ into~$H^{(r+s)}$.

If $H$ were cocommutative, nearly all the previous arguments in this
section would go through. For~\eqref{eq:burden of proof} one
uses~\eqref{eq:coalgprop} instead of~\eqref{eq:algprop}, and naturally
instead of an algebra grading one gets a coalgebra grading. In
particular, appealing to the identity~\eqref{eq:primitives}, and since
obviously $P(H)$ is contained in $\pi_1(H)$, we conclude that
$\pi_1(H)=P(H)$ in this case.

Assume again that $H$ is commutative, so $H'$ is cocommutative.
Our contention now is that the logarithm kills products.

\begin{prop}
\label{pr:Log-kills-products}
On a commutative, connected Hopf algebra~$H$, $\pi_1(H_+^2) = 0$
holds.
\end{prop}

\begin{proof}
Recall first that $H_+^2$ is orthogonal to~$P(H')$,
by~\eqref{eq:new-number}. Now, to avoid confusion, denote in this
proof by~$\pi'_1$ the first projector of $H'$. Since
$(\id-u\eta)_H=(\id-u\eta)^t_{H'}$, clearly $\pi_1$ is the transpose
of~$\pi'_1$, and then for an arbitrary~$u$
$$
\<\pi_1(X_iX_j),Z_u> = \<X_iX_j,\pi'_1(Z_u)> = 0,
$$
since $\pi'_1(Z_u)$ is primitive. Therefore $\pi_1(X_iX_j)$ must
vanish as claimed. (For the trained field theorist this was plausible,
as he knows that logarithms identify `connected' elements.)
\end{proof}

The grading associated to $\log^*\id$ does not, in general, coincide
with the previous grading. At any rate, because of homogeneity of the
convolution powers of~$\id$, we get an algebra bigrading of the Hopf
algebra, becoming a bialgebra bigrading when~$H$ is cocommutative as
well.

\marker
Now turn to the so-called normal coordinate elements (canonical
coordinates of the first kind would be a more appropriate name)
introduced in~\cite{Chryssomalakosetal} by the definition
$$
\sum_{u\in {\tilde A}}X_u\ox Z_u =:
\exp\biggl(\,\sum_{j\in A}\psi_j\ox Z_j\biggr).
$$
Note that the sum on the right hand side is only on $A$ (that is, only
on `letters', not on all `words'). From this, the authors immediately
conclude that any algebra basis element has a canonical decomposition
as follows:
\begin{align*}
X_j
&= \psi_j + \sum_{k\geq2} \frac{\<Z_{i_1}\cdots Z_{i_k},X_j>}{k!}
\,\psi_{i_1} \dots \psi_{i_k}
\\
&= \psi_j + \sum_{k\geq2}
\frac{\<Z_{i_1} \oxyox Z_{i_k}, \Dl^k X_j>}{k!}\,
\psi_{i_1} \dots\psi_{i_k}.
\end{align*}
The second form makes clear that only a finite number of terms
intervene, corresponding to sequences $J=(i_1,\dots,i_k)$ compatible
with $X_j$ in the sense that $\<Z_{i_1}\oxyox Z_{i_k},\Dl^k
X_j>\neq0.$ Even so, to extract $\psi_j$ from this change of infinite
basis is a painful business. However, looking back
at~\eqref{eq:basic-truth}, and exponentiating both sides, we see that
$$
\sum_{u\in {\tilde A}}X_u\ox Z_u =
\exp\biggl(\,\sum_{w\in {\tilde A}}\pi_1(X_w)\ox Z_w\biggr)
= \exp\biggl(\,\sum_{j\in A}\pi_1(X_j)\ox Z_j\biggr);
$$
since $\pi_1$ kills products, the sum of the right hand side extends
only over $A$. Therefore $\psi_j = \pi_1(X_j)$.

\smallskip

All the properties claimed for $H_R$ in~\cite{Chryssomalakosetal} are
seen as easy corollaries of the properties of the canonical projector
$\pi_1$. For instance, the diagonality of the antipode in the $\psi_j$
basis, or quasiprimitivity. Indeed, since $Sg = g^{-1}$ if $g$ is
grouplike, then
$$
\<g^{-1},\psi_j> = \<Sg,\psi_j> = \<g,S\psi_j>.
$$
But, if
$$
g = \exp\biggl(\,\sum_{k\in A}\a_kZ_k\biggr),
$$
then $\<g,\psi_j> = \a_j$ and $\<g^{-1},\psi_j> = -\a_j$, for
all $j\in A$. The conclusion that $S\psi_j = -\psi_j$ follows.

\smallskip

For applications in QFT, further investigation of (depth and)
quasiprimitivity in commutative algebras of the kind studied in this
review seems paramount.

\smallskip
\marker
Before reexamining our old friend $\F$, we make a skimpy remark on the
algebra~$H_R$ of rooted trees, much studied by several
authors~\cite{BK,godgift,Foissy,Chryssomalakosetal}. Even taken as a
proxy for the Hopf algebras of QFT, its complexity is staggering. The
number of rooted trees with~$n$ vertices is given by a famous sequence
$r=(r_1=1,r_2=1,2,4,9,20,48,115,\dots)$; then, with the standard
grading by number of vertices, clearly $\dim H^{(n)}=r_{n+1}$. There
are many primitives, beyond those contained in its Hopf
subalgebra~$\Hl$: for instance $h_{15,1}=30135$. We wish to remark
that the \bfb\ formulae are instrumental in finding $h_{n,k}$ in the
context of~$H_R$~\cite{Foissy}, again. The space~$W_{H_R}$ starts at
level~3, since Figure~\ref{fig:divide-and-rule} exhibits an
indecomposable nonprimitive (which has depth~2).

\begin{figure}[htb]
\centering
\newcommand{\punto}{\;\begin{picture}(2,2)(0,-3)
\put(0,0){\circle{4}}
\end{picture}\;}
\newcommand{\match}{\;\begin{picture}(5,10)(0,2)
\put(0,10){\circle{4}}
\put(0,10){\line(0,-1){10}}
\put(0,0){\circle*{3}}
\end{picture}\;}
\newcommand{\legs}{\;\begin{picture}(20,10)(0,2)
\put(10,10){\circle{4}}
\put(10,10){\line(-1,-1){10}}
\put(0,0){\circle*{3}}
\put(10,10){\line(1,-1){10}}
\put(20,0){\circle*{3}}
\end{picture}\;}
$$
\log^*\id\,\Bigl(\legs\Bigl)\, = \legs - \punto\match +
\tfrac{1}{6}\,\punto\punto\punto,
$$
\caption{Nonprimitive combination of nonproduct woods}
\label{fig:divide-and-rule}
\end{figure}

\begin{xmpl}
Matters for the \bbfb\ algebra are rather different from the ladder
algebra. Of course $\dim\F^{(n)} =p(n)$ still holds. Consider
$\F^{(2)}$, that is, the linear span of~$a_2^2=\dl_1^2$
and~$a_3=\dl_2+\dl_1^2$ ---remember that with our notation
$\#(a_n)=n-1$. Plainly, $a_2^2\in\F_+^2$. Now, although $a_3$ in the
indecomposable class is not primitive, it belongs to $\F_+^2\op
P(\F)^{(2)}$, since
$$
\log^*\id\,a_3 = a_3 - \tfrac{3}{2} a_2^2 = \dl_2 - \thalf\dl_1^2 =: p_2
$$
(the Schwarzian derivative) is primitive; the telltale sign was
$\tau\Dl a_3=\Dl a_3$.

Notice that no more primitives in $\F$, outside the space spanned
by~$p_1:=\dl_1$ and $p_2$, can possibly happen:
equation~\eqref{eq:new-number} tells us
$$
P(\F) = (\C1\op{\F'_+}^2)^\perp.
$$
But then from~\eqref{eq:formula-after-all} a dual basis of~$\F'$ is
made of products, except for two elements. Therefore $\dim P(\F)=2$.

Also, the projector $\log^*\id$ will produce primitives only on
indecomposable classes fulfilling $\Dl a=\tau\Dl a$; this cannot be
contrived here. We obtain
$$
\tilde{p}_3 := \log^*\id\,a_4 = a_4 - 5a_2a_3 + \tfrac{9}{2}a_2^3 =
\dl_3 - 2\dl_1\dl_2 + \thalf\dl_1^2,
$$
and $\log^*\id$ kills products as usual. We know that $\tilde{p}_3$ is
not primitive, rather it has depth~2, as it can easily be checked. It
still is quasiprimitive, in that
$$
m\Dl'\tilde{p}_3 = 0.
$$
The algebra grading respectively induced by the convolution powers of
the logarithm and by the depth filtering in this case do not coincide.
We have as well
$$
\frac{\log^{*2}\id}{2!}\,a_2a_3 = a_2a_3 - \tfrac{3}{2}a_2^3 = p_1p_2;
\quad \frac{\log^{*2}\id}{2!}\,a_4 = 5(a_2a_3 - \tfrac{3}{2}a_2^3);
\quad \frac{\log^{*2}\id}{2!}\,a_2^3 = 0.
$$
Finally
$$
\frac{\log^{*3}\id}{2!}\,a_2^3 = a_2^3 = p_1^3.
$$
A suitable basis for~$\F^{(3)}$ is then
$$
(a_4 - 5a_2a_3 + \tfrac{9}{2}a_2^3,a_2a_3 - \tfrac{3}{2}a_2^3,a_2^3) =
(\dl_3-2\dl_1\dl_2+\thalf\dl_1^2,\dl_1\dl_2-\thalf\dl_1^3,\dl_1^3) =
(\tilde{p}_3,p_1p_2,p_1^3),
$$
with respective depths (primitivity degrees) $2,2,3$.

In $\F^{(4)}$, we will have the linearly independent elements
$$
p_2^2,\ p_1\tilde{p}_3,\ p_1^2p_2,\ p_1^4,
$$
of respective depths $2,3,3,4$. We seek a suitable representative for
the missing indecomposable element. Proceeding systematically with the
projectors $\frac{\log^{*k}\id}{k!}$, we obtain
$$
\tilde{p}_4 := \log^*\id\,a_5 = a_5 - 15a_2a_4 - 5a_2^3 +
\tfrac{185}{6}a_2^2a_3 - 20a_2^4,
$$
next,
$$
\frac{\log^{*2}\id}{2!}\,a_2^3    = p_2^2;          \quad
\frac{\log^{*2}\id}{2!}\,a_2a_4   = p_1\tilde{p}_3; \quad
\frac{\log^{*3}\id}{3!}\,a_2^2a_3 = p_1^2p_2;       \quad
\frac{\log^{*4}\id}{4!}\,a_2^4    = p_1^4.
$$
A suitable basis for~$\F^{(4)}$ is then
$$
(\tilde{p}_4,\ p_2^2,\ p_1\tilde{p}_3,\ p_1^2p_2,\ p_1^4).
$$
Note that $S(p_1\tilde{p}_3)=p_1\tilde{p}_3,S(p_2^2)=p_2^2,S(p_1^2p_2)
=-p_1^2p_2,S(\tilde{p}_4)=-\tilde{p}_4$; and so in
general~$S\frac{\log^{*n}\id}{n!}a=(-)^n\frac{\log^{*n}\id}{n!}a$ (we
owe this remark to~F.~Patras). Going back to $\tilde{p}_4$, one can
check quasiprimitivity, and the fact it has depth~3. The pattern
repeats itself: at every order $\#$ in the original grading one
nonproduct generator of depth $(\# - 1)$ ---for which Broadhurst and
Kreimer give a recipe in~\cite{BK}--- is found.

Concerning $W_\F$: we know $a_3$ belongs to $\F(1)^{(2)}\op
P(\F)^{(2)}$, since $a_3-3/2a_2^2$ is primitive. Therefore, in this
case $W_\F^{(2)}$ is trivial. Now consider $\F^{(3)}$, the linear span
of $a_2^3,a_2a_3$ and~$a_4$. The element $a_4$ cannot be primitively
generated, and so there is a one-dimensional~$W_\F^{(3)}$. The same is
true for~$W_\F^{(n)}$, for every $n\ge3$.
\end{xmpl}

\medskip

To summarize the general situation: $\im\pi_1$ is the `good'
supplementary space for $H_+^2$ we were looking for. The elements~$a$
and~$\log^*\id\,a$ belong to the same class modulo $H_+^2$. The
logarithms of the nonproduct elements, all quasiprimitives, provide an
eminently suitable algebra basis for~$H$. We can think of~$H$, as a
graded algebra, as the polynomial algebra on the~$\log^*\id\,a$.
Moreover, the projectors behave well in regard to the action of the
antipode.

\section{Coda: on twisting and other matters}

We conclude provisionally these notes with pointers to two subjects of
current interest: Hopf algebras of quantum fields, and investigation
of `twisted antipodes'. Both are indebted to Rota's work. 

\smallskip

Consider now the theory of the free neutral scalar (for simplicity)
field $\vf(x)$. Then the space of quantum observables can be
identified to the Fock boson algebra on $V\,\equiv$ the space of
complex solutions of the Klein--Gordon (KG) equation
$$
(\square + m^2)v(x) :=
\frac{\del^2v(x)}{\del t^2} - \Dl v(x) + m^2v(x) = 0
$$
on Minkowski space $M_4$.  The algebra product is just the normal
product of fields.  Let $D$ be the (Jordan--Pauli) distribution
solving the Cauchy problem for the KG equation.  The Hilbert space of
states is in turn a Fock space, built on the space of real solutions
$V_\R$ made complex with the help of a complex structure, so that if
\begin{equation}
v_i(x) = \int D(x,y)h_i(y)\,d^4y
\label{eq:off-shell}
\end{equation}
for $i=1,2$ are real solutions of the KG equation, then
$$
\braket{v_1}{v_2} = \int h_1(x)D_+(x,y)h_2(y)\,d^4x\,d^4y,
$$
where $D_+$ is the standard Wightman function; the projection $h
\mapsto v$ to field equation solutions in~\eqref{eq:off-shell}
corresponds to dividing out the unrestricted fields by the ideal
generated by that equation, and will be implicitly used all the time. 
The quantum field is a $V$-valued distribution; it is defined by its
action~\cite{Daniel} by creating and annihilating a particle in the
distributional state $D(\cdot -x)$:
$$
\vf(x) = \frac{1}{\sqrt2}[a(D(\cdot -x)) + a^\7(D(\cdot -x))].
$$
This ensures that $(\square + m^2)\vf(h) :=
\vf((\square+m^2)h) = 0$, for any complex smearing function~$h$.
Note that
\begin{equation}
\<0, \vf(h_1)\vf(h_2)\,0> = \<v_1, v_2>.
\label{eq:first-VEV}
\end{equation}

On this concrete boson algebra we put the compatible cocommutative
coalgebra structure already described; the counit $\eta$ is at once
identified with the vacuum expectation value
$$
\eta(a) = \<0, a\,0>.
$$
It is instructive to consider the Wick monomials in the field operator
$\vf(x)$.  In regard to notation, to conform to usual practice, we
write $\wick:\vf(x_1) \cdots \vf(x_n):$ instead of
$\vf(x_1) \vee\cdots\vee \vf(x_n)$. The powers
$\vf(x)\vee\vf(x)$ will be denoted simply $\vf^2(x)$ in
place of the standard $\wick:\vf^2(x):$ (no other powers are
defined). For the purpose one posits
\begin{equation}
\vf^n(x) :=
\< \wick:\vf(x)\vf(x_2)\cdots\vf(x_n):,
\dl(x-x_2)\cdots\dl(x-x_n) >_{x_2\cdots x_n},
\label{eq:wick-weird}
\end{equation}
or, if one wishes:
$$
\vf^n(h) := \< \wick:\vf(x)\vf(x_2)\cdots\vf(x_n):,
h(x)\dl(x-x_2)\cdots\dl(x-x_n) >_{x,x_2\cdots x_n}.
$$
This is known to be well defined.  The diagonalization map for Wick
monomials can hardly be simpler: from~\eqref{eq:shuffles}
and~\eqref{eq:wick-weird} it is a couple of easy steps to obtain:
$$
\Dl\vf^n(x) =
\sum_{i=0}^n\binom{n}{i}\vf^{i}(x)\otimes\vf^{n-i}(x),
$$
i.e., the comultiplication is the binomial one (as befits
honest-to-God monomials).  One can alternatively use divided powers
$\vf^{(n)} := \vf^n/n!$.  More generally then:
$$
\Dl^l(\vf^{(n)}(x)) =
\sum\vf^{(m_1)}(x)\otimes\cdots\ox\vf^{(m_{l+1})}(x),
$$
with sum over all combinations of $l+1$ nonnegative integers $m_i$
such that $\sum_{i=1}^{l+1}m_i=n$.  Very easy is also to check,
from~\eqref{eq:wick-weird}, that
$$
\braket{\vf^n(x)}{\vf^m(y)} = \frac{\dl_{nm}}{n!}\,D^n_+(x,y).
$$
(Unlike other propagators, $D_{\pm}$ have powers of all orders.)

{}From this starting point, reference~\cite{Gang-of-Four} proceeds by
`twisting' the Hopf algebra structure of~$B(V)$ by suitable bilinear
forms $z$ on $V$.  The twisted product is given by
$$
a\8 b = \sum z(a_{(1)},b_{(1)}) \, a_{(2)} \vee b_{(2)}.
$$
Setting $z(1,1)=1$ is understood.  By cocommutativity, we equivalently
have
$$
a\8 b
= \sum a_{(1)}\vee b_{(1)}z(a_{(2)},b_{(2)})
= \sum a_{(1)}\vee b_{(2)}z(a_{(2)},b_{(1)})
= \sum a_{(2)} \vee b_{(1)} z(a_{(1)},b_{(2)}).
$$
Note that there is \textit{no} restriction on the degrees of $a,b\in
B(V)$.  The authors consider two different twisted products,
respectively corresponding to the operator and the time-ordered
products of elements of $B(V)$.  The associated bilinear forms are
$z(v_1,v_2)=\braket{v_1}{v_2}$ and the \textit{symmetric} pairing
$\roundbraket{\cdot}{\cdot}$, given by
$$
z(v_1,v_2) = \roundbraket{v_1}{v_2} = \roundbraket{v_2}{v_1}
= \braket{0}{\T[\vf(h_1)\vf(h_2)]\,0}
= \int h_1(x)D_F(x,y)h_2(y);
$$
we recall that the time-ordered product of free fields is defined by
$$
\T[\vf(x_1)\vf(x_2)] = \T[\vf(x_2)\vf(x_1)] =
\Th(t_1-t_2)\vf(x_1)\vf(x_2) + \Th(t_2-t_1)\vf(x_2)\vf(x_1),
$$
where $\Th$ is the Heaviside function (so, time increases from right
to left) and that
$$
\braket{0}{\T[\vf(x_1)\vf(x_2)]\,0} = D_F(x_1-x_2).
$$

The resulting algebra in the first case can be called the Weyl algebra,
since the canonical commutation relations
$$
\braket{0}{[\vf(h_1),\vf(h_2)]\,0}
= s(v_1,v_2)
$$
are satisfied; where the bilinear form $s$ is given by the integral on
the space $V_\R$ of solutions
$$
s(v_1,v_2)
:= \int_\Sigma (v_1\del_\mu v_2 - v_2\del_\mu v_1)\,d\sigma^\mu
= \int h_1(x)D(x,y)h_2(y)\,d^4x\,d^4y,
$$
that does not depend on $\Sigma$ itself, and defines a symplectic
form, which is complexified in the standard way. In the second case we
obtain a \textit{commutative} algebra. By use on it of the above
indicated comultiplication, recently Mestre and Oeckl have been able
to show the relations between the different $n$-point functions of
quantum field theory in a very economical manner~\cite{MexicanBiHat}.

\smallskip

An interesting property, generalizing~\eqref{eq:first-VEV}, is
$$
\eta(a\8 b) = z(a,b).
$$
In effect, by the defining property of $\eta$
\begin{align*}
\eta(a\8 b)
&= \sum \eta(a_{(1)} \vee b_{(1)})z(a_{(2)},b_{(2)})
= \sum \eta(a_{(1)})\eta(b_{(1)})z(a_{(2)},b_{(2)}) \\
&= z(\eta(a_{(1)})a_{(2)},\eta(b_{(1)})b_{(2)})
= z(a,b).
\end{align*}
The bilinear forms used are examples of~\textit{Laplace pairings} (a
concept originally introduced by Rota~\cite{GrosshansRS}); in turn
these are cocycles in a Hopf algebra cohomology.  More general
cocycles would seem to relate to interacting quantum fields, and to
the passage to noncommutative field theory~\cite{MexicanHat}.  It is
worth noting that the (much more difficult) Hopf algebra cohomology of
the noncocommutative Hopf algebras $H_R$ and $\H$ underlines Kreimer's
program in the direction of using Hopf algebraic techniques to
simplify the calculation of Feynman diagrams: see for
instance~\cite{DirkHCoh}.

In summary, the fundamentals of quantum field theory have been
co-algebrized. The price is worth paying for the complete automation
not only of the twisted product formulae, but of many indispensable
calculations in field theory that are not found, or only haphazardly,
in textbooks, and otherwise require a substantial amount of
combinatorics (in this respect, the splendid little book by
Caianello~\cite{OhSoleMio} still stands out). The Hopf algebra
approach frees us from the perennial, tiresome recourse of
decompositions in $\vee$-products of homogeneous elements of order~1
in every argument.

The benefits of this abstract framework were harvested
in~\cite{BrouderG-24,Gang-of-Four}, as their results translate into
strong versions of the Wick theorems of quantum field theory. As we
said in the introduction, however, the approach from quantum
theoretical first principles is still evolving (see~\cite{BrouderS} in
this connection), and the passage from the Hopf algebras that Brouder,
Oeckl and others have associated to quantum fields to the
renormalization Hopf algebra of Connes and Kreimer is perhaps in the
cards.

\marker
The reader should be aware that, with respect to the Hopf algebra
approach to renormalization, we have done less than to scratch the
surface. The heart of this approach is a multiplicative map~$f$ (the
``Feynman rule'') of~$\H$ into an algebra~$V$ of Feynman amplitudes:
for instance, in dimensional regularization the character takes values
in a ring of Laurent series in the regularization parameter~$\eps$. In
physics, the Feynman rules are essentially fixed by the interpretation
of the theory, and thus one tends to identify~$\Ga$ with~$f(\Ga)$.
Perhaps the main path-breaking insight of~\cite{KreimerOriginal} is
the introduction of the ``twisted antipode''. Let us then usher in the
other personages of this drama. There is a linear map $T: V \to V$,
which effects the subtraction of ultraviolet divergencies in each
renormalization scheme. The twisted (or ``renormalized'') antipode
$S_{T,f}$ is a map $\H\to V$ defined by~$S_{T,f}(\emptyset) =
1;S_{T,f}=T\circ f\circ S$ for primitive diagrams, and then
recursively:
$$
S_{T,f} \Ga = -[T\circ f]\Ga - T\biggl[
\sum_{\emptyset\varsubsetneq\ga\varsubsetneq\Ga}S_{T,f}(\ga)\,
f(\Ga/\ga)\biggr].
$$
In other words, $S_{T,f}$ is the map that produces the counterterms in
perturbative field theory.  The Hopf algebra approach works most
effectively because in many cases $S_{T,f}$ is multiplicative; for
that, it is not necessary for $T$ to be an endomorphism of the algebra
of amplitudes~$V$, but the following weaker
condition~\cite{RotaBaxter} is sufficient:
$$
T(hg) = T(T(h)g) + T(hT(g)) - T(h)T(g).
$$
This endows $V$ with the structure of a \textit{Rota--Baxter algebra}
of weight~1; it is fulfilled in the BPHZ formalism for massive
particles, and the dimensional regularization scheme with minimal
subtraction, for which the Connes--Kreimer paradigm is most cleanly
formulated. By the way, here is where~\cite{BergCartier} fails.
Finally, the renormalized amplitude $R_{T,f}$ is given by
$$
R_{T,f} := S_{T,f} * f.
$$
This map is also a homomorphism; compatibility with the coproduct
operation is given by its very definition as a convolution. The
(nonrecursive) forest formula for $R_{T,f}$ is precisely (the
complete) Zimmermann's formula~\cite{Zim} of quantum field theory.

A Rota--Baxter algebra of weight~$\theta$ is given by the
condition
$$
\theta T(hg) = T(T(h)g) + T(hT(g)) - T(h)T(g).
$$
Shuffle algebras and Rota--Baxter algebras of weight~0 are essentially
the same thing~\cite{Catecismo,Murua}. The theory of Rota--Baxter
algebras is examined in depth in~\cite{E-FardGK}
and~\cite{E-FardGKbis}.

\subsection*{Acknowledgments}

This manuscript collects and expands for the most part a series of
lectures delivered by the second-named author in the framework of the
joint mathematical physics seminar of the Universit\'es d'Artois and
Lille 1, as a guest of the first named institution, from late January
till mid-February 2003. We thank Amine El-Gradechi and the
Universit\'es d'Artois and Lille~1 for the excellent occasion to
lecture on a subject close to our hearts. JMG-B is very grateful to
his `students' on that occasion for much friendliness, and for the
hospitality of the Theory Division of~CERN, where a draft was written
prior to the lectures.

Also we thank Michel Petitot for teaching us the double series, Li
Guo, Michiel Hazewinkel and Fr\'ed\'eric Patras for helpful remarks,
and Christian Brouder, Kurusch Ebrahimi-Fard, Amine El-Gradechi, Dirk
Kreimer and Joseph C. V\'arilly for careful reading of prior versions
of the manuscript and providing insights and welcome advice. We are
also indebted to Lo\"{\i}c Foissy and Dominique Manchon for forwarding
us very valuable material. Joseph V\'arilly generously helped us with
the figures. We owe to the referees, whose comments greatly helped to
improve the presentation. HF acknowledges support from the
Vicerrector\'{\i}a de Investigaci\'on of the Universidad
de~Costa~Rica.

\end{document}